\documentclass[11pt,a4paper]{article}
\pdfoutput=1
\usepackage{euscript,amsmath,amsthm,amsfonts,amssymb}
\usepackage{mathtools}
\usepackage{float}
\usepackage[margin=1in]{geometry}
\usepackage{graphicx}
\usepackage{outlines}
\usepackage[dvipsnames]{xcolor}
\usepackage{jheppub}
\usepackage{subfigure}
\usepackage{adjustbox}
\usepackage{longtable}

\usepackage{makecell}
\usepackage[table]{colortbl}
\usepackage{color}

\usepackage{tabularx}

\def\5{{(5)}}

\newcommand{\ket}[1]{|{#1}\rangle}

\newcommand{\be}{\begin{equation}}
\newcommand{\ee}{\end{equation}}
\newcommand{\bea}{\begin{eqnarray}}
\newcommand{\eea}{\end{eqnarray}}

\DeclareMathOperator{\area}{area}

\newtheorem{theorem}{Theorem}
\newtheorem{definition}{Definition}
\newtheorem{conjecture}{Conjecture}

\title{\boldmath Perfect tensor hyperthreads}
\author[a,b]{Jonathan Harper}
\affiliation[a]{Martin Fisher School of Physics, Brandeis University, Waltham, Massachusetts 02453, USA}
\affiliation[b]{Center for Gravitational Physics and Quantum Information, Yukawa Institute for Theoretical Physics, Kyoto University, Kitashirakawa Oiwakecho, Sakyo-ku, Kyoto 606-8502, Japan}
\preprint{BRX-TH-6705}
\emailAdd{jharper@brandeis.edu}
\abstract{Bit threads, a dual description of the Ryu-Takyanagi formula for holographic entanglement entropy (EE), can be interpreted as a distillation of the quantum information to a collection of Bell pairs between different boundary regions. In this article we discuss a generalization to hyperthreads which can connect more than two boundary regions leading to a rich and diverse class of convex programs. By modeling the contributions of different species of hyperthreads to the EEs of perfect tensors we argue that this framework may be useful for helping us to begin to probe the multipartite entanglement of holographic systems. Furthermore, we demonstrate how this technology can potentially be used to understand holographic entropy cone inequalities and may provide an avenue to address issues of locking.

}
\begin{document}
\maketitle
\flushbottom

\section{Introduction}
The purpose of this article is to begin to address ways of characterizing the multipartite entanglement of holographic states. Given a holographic state $\ket{\psi}$ with a dual classical bulk geometry $M$ we can consider a static time slice $\Sigma$ and partition the boundary $\partial \Sigma$ into a boundary region $A$ along with its purifier the complement $O$. In such a set up it is well known that the bipartite entanglement between $A$ and $O$ can be quantified by the entanglement entropy (EE) $S_A$. In the boundary theory this is calculated as the von Neumann entropy of the reduced density matrix after a partial trace of one of the two boundary regions. One way of understanding this quantity is that it determines the number of Bell pairs which can be distilled from the asymptotic limit of many copies of the holographic state by a quantum channel constructed from local unitaries (LU)
\be
\ket{\psi} \longrightarrow \ket{AO}^{\otimes S_A}
\ee
where the arrow here represents the appropriate distillation protocol.

From the bulk perspective the entanglement entropy can be calculated using the Ryu-Takayanagi (RT) formula \cite{2006JHEP...08..045R} which asks for the minimal area surface homologous to $A$
\be
S_A=\min_{m\sim A} \area(m)
\ee
we call this minimizing surface $m_A$.

\begin{figure}[H]
\centering
\includegraphics[page=8,width=.40\textwidth]{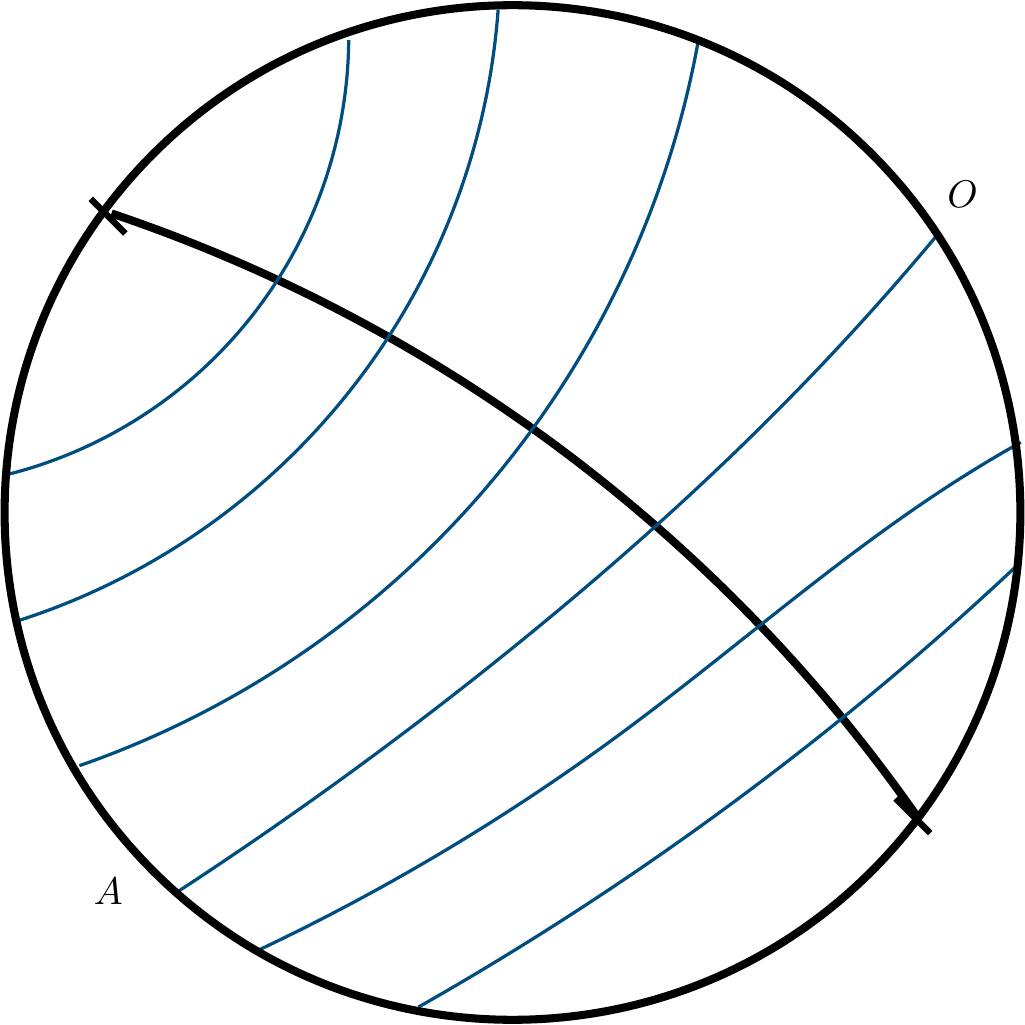}
\caption{\label{fig:BT} The RT surface $m_A$ along with a maximal configuration of bit threads. The area of $m_A$ and the number of bit threads both calculate the holographic entanglement entropy $S_A$.}
\end{figure}

Alternatively, the entanglement entropy is given a maximal configuration of bit threads \cite{2017CMaPh.352..407F,2018CQGra..35j5012H}: simple curves of constant thickness connecting $A$ to $O$ subject to a local density bound. Tools from the theory of convex optimization can be used to show that these two descriptions: minimal RT surfaces and maximal bit thread configurations are in fact the same (see figure \ref{fig:BT}).

\begin{figure}[H]
\centering
\includegraphics[page=9,width=.45\textwidth]{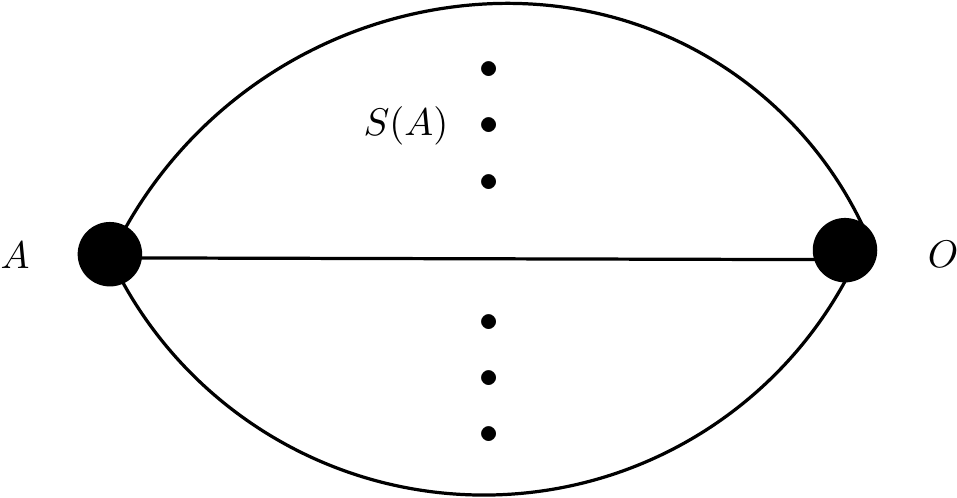}\quad \quad
\raisebox{3.8em}{\includegraphics[page=10,width=.45\textwidth]{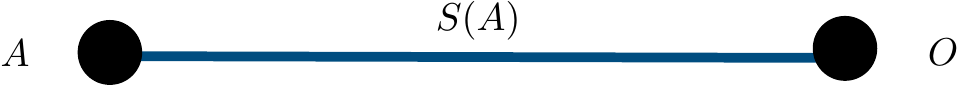}}
\caption{\label{fig:dessication} Given a configuration of bit threads we can course grain or desiccate the geometry keeping only the geometry which the threads pass. This is essentially a collection of $S_A$ wormholes between the boundary regions $A$ and $O$. In this way each bit thread can be viewed as a distilled Bell pair realizing ER=EPR.}
\end{figure}

The bit threads are often represented as a geometrical avatar of the distilled Bell pairs. That is given a configuration of bit threads we can consider a course-graining or desiccation of the geometry where we only keep the portions of $\Sigma$ which bit threads cross. Such a geometry can be viewed as a collection of wormholes, one for each thread. This is in turn equivalent to a simple graph consisting of a single edge of weight $S_A$ which is equivalent to $S_A$ Bell pairs (see figure \ref{fig:dessication}). This perspective provides a realization of the connection between geometry and entanglement.

A natural question to then ask is what if we consider more than one boundary region. Given a partition of $\partial \Sigma$ into $N$ regions along with a purifier $O$: $\partial \Sigma =\{A_1\cdots A_N,O\}$ there are $2^{N}-1$ independent entanglement entropies one can consider. These include single party EEs (e.g. $S_{A_1}$) as well as multiple party EEs consisting of the union of a number of boundary regions (e.g. $S_{A_1A_2}$). It is useful to organize these into an entropy vector space with each EE corresponding to a different orthogonal direction. However, holographic states only comprise a subset of allowed vectors; these form the holographic entropy cone (HEC) \cite{r2,2019ForPh..6700011H, 2018ForPh..6600067H, r3}. This is because there are nontrivial entropy inequalities which constrain the allowed values of various EEs. For example in the case $N=3$ there are two well known classes of such inequalities: subadditivity (SA) and monogamy of mutual information (MMI)
\be
\begin{split}
I_2(A:B)&=S_A+S_B-S_{AB}\geq 0\\
-I_3(A:B:C)&=S_{AB}+S_{BC}+S_{AC}-S_A-S_B-S_C-S_{ABC} \geq 0.
\end{split}
\ee
These have been independently proven using RT surfaces \cite{2007PhRvD..76j6013H,2013PhRvD..87d6003H} and bit threads \cite{2017CMaPh.352..407F,Cui:2018aa}. Holographic entropy cone inequalities are know explicitly for up to $N=5$ \cite{r3} and many, but not all for $N=6$ \cite{N6rays}. However, their structure remains mysterious and elucidating the general properties and features of holographic entropy inequalities has been the subject of current research \cite{N6rays,2022PhRvD.105h6008F,2021arXiv210207535A,2020JHEP...07..245H,2022arXiv220400075H,2021arXiv211200763C}. 

While bit threads are capable of correctly reproducing all single party entanglement entropies, in general beyond $N=2$ bit thread configuration can not correctly reproduce the full entropy vector. This is due to geometric obstacles which prevent the locking, or simultaneous, saturation of the necessary RT surfaces. This indicates that the bipartite distillations corresponding to such thread configurations are too coarse grained as they do not contain the necessary information to correctly reproduce the full entropy vector. So far bit threads have been unable to prove holographic entropy inequalities beyond $N=3$.

The main innovation of this article is the definition of perfect tensor hyperthreads. These consist of a single internal vertex along with an even number of strands each of which connects to a unique boundary region. They are modeled so that their contributions to entanglement entropies match that of perfect tensor states. As such, in analogy with standard bit threads, they should be considered as avatars of perfect tensor states distilled from the full CFT state. In doing so we make extensive use of the $K$-basis construction of \cite{r1} which provides an alternative description of entropy vectors in terms of perfect tensor states.

We construct a procedure for defining an appropriate convex program which, we believe, has the capability to lock the full entropy vector. This allows us to then relate different species of perfect tensor hyperthreads to well known information quantities. For example, for $N=3$ threads which connect two regions (2-threads) are units of mutual information while threads that connect four (4-threads) are units of negative tripartite information\footnote{Assuming the locking of the full entropy vector for any $N$ the information quantities $I_{N}$ can always be written as a linear combination of perfect tensor hyperthread species with positive coefficients. However, beyond $N=3$ these are generally not sign definite as they are no longer facets of the HEC.}. We provide several examples on graphs up to $N=5$ of our construction. We further speculate and provide evidence that this can be extended to general holographic states. These considerations are summarized in our primary conjecture \ref{mainconj}.

For $N\geq 4$ it is necessary for us to introduce negative threads which contribute negatively to the density bound and objective. Holographic entropy cone inequalities become essential and must be explicitly implemented in our programs. This provides an alternate interpretation of the structure of the holographic entropy cone in that the entropy inequalities can be understood as necessary constraints between different species of perfect tensor hyperthreads.

The organization of the rest of the article is as follows: In section \ref{sec:prelim} we provide a quick introduction to a number of preliminary, but necessary topics. These include: tools of convex optimization such as convex programs, convex duality and complementary slackness; bit threads for one and more boundary regions, duality to RT, and locking properties; the holographic entropy cone and entropy inequalities in both the $S$ and $K$-basis. In section \ref{sec:PTHT} we define perfect tensor hyperthreads and then apply this framework to two and three boundary regions in section \ref{sec:N2} and \ref{sec:N3}. In section \ref{sec:N4} we introduce the notion of negative threads which contribute negatively to both the objective and density bound. We show how the holographic entropy cone inequalities can be used to place necessary constraints between different species of perfect tensor hyperthreads. This technology then permits us to describe the case of four region and subsequently five regions in section \ref{sec:N5}. Finally, in section \ref{sec:dis} we conclude with some discussion of a general conjecture of the locking properties of perfect tensor hyperthreads and their possible relation to multipartite distillations of holographic states. Appendix \ref{appex:N5ER} contains explicit perfect tensor hyperthread configurations for many of the $N=5$ extremal rays of the holographic entropy cone. These configurations correctly reproduce the full entropy vector.

\section{Preliminaries}\label{sec:prelim}
\subsection{Tools of convex optimization}
Here we review some key results from the theory of convex optimization \cite{boyd2004convex}.\footnote{For an in-depth introduction see section 2 of \cite{2018CQGra..35j5012H}.}

\paragraph{Convex duality}
A convex program is an optimization problem consisting of a convex objective $f_0$ along with a number of convex inequality constraints $\{f_i\leq0\}$ and affine equality constraints $\{h_i=0\}$. We write the program as
\be
        P_{max}=\left\{
        \begin{aligned}
            & & & \max f_0(x) \\
            & \text{s.t.} & & \forall i,\; f_i(x)\leq 0\\
             & \text{and} & & \forall i,\; h_i(x)=0.
        \end{aligned}
    \right.  
\ee
Given such a program it is always possible to determine an equivalent minimization program by dualizing. This is done in two steps: First the program is written as a single function with Lagrange multipliers imposing the constraints
\be
L(x, \{\lambda\}, \{\gamma\}) = f_{0}(x) + \lambda_{i}f_{i}(x) + \gamma_{i}h_{i}(x), \quad \lambda_{i} \geq 0.
\ee
Next, the roles of the original variables and the Lagrange multipliers are switched and the Lagrangian is optimized with respect to the original variables. Doing so results in a dual objective $\tilde{f}_0$ and a number of dual constraints $\tilde{f}_i$ and $\tilde{h}_i$. Using these we can define the dual minimization program with
\be
       P_{min}= \left\{
        \begin{aligned}
            & & & \min \tilde{f}_0(\lambda,\gamma) \\
            & \text{s.t.} & & \forall i,\; \tilde{f}_i(\lambda,\gamma)\geq 0\\
             & \text{and} & & \forall i,\; \tilde{h}_i(\lambda,\gamma)=0.
        \end{aligned}
    \right.  
\ee
Convex duality is the demand that these two programs are in fact equivalent
\be
    P_{max}=P_{min}.
\ee

\paragraph{Complementary Slackness}
Given such a program an important concept is complementary slackness (CS). Given the Lagrangian of a convex optimization program with an inequality constraint $f_i$, and Lagrange multipliers $\lambda_i$ for any optimal configuration it is true that
\be
\lambda_{i}^*f_i(x^*)=0 \quad \text{(no sum)}
\ee
which implies one of the two constraints $\lambda_i\geq 0$ or $f_{i}(x)\leq 0$ is saturated. As we will see CS is extremely useful for diagnosing properties of optimal configurations and can also be used in many cases to simplify the evaluation of programs given knowledge about the saturation of constraints for particular setups.

\subsection{Bit thread configurations}
As an application of convex duality we consider the calculation of the entanglement entropy in holography. Given a static time slice $\Sigma$ of a holographic state we choose a division of the boundary into a region $A$ along with a purifier $O$. The entanglement entropy $S_A$ is given by the Ryu-Takayanagi (RT) formula \cite{2006JHEP...08..045R} which asks for the minimal area surface homologous to $A$
\be
S_A=\min_{m\sim A}\area(m)=\area(m_A).
\ee
We make use of the following maximization program: Let $P$ be the set of all simple curves with one endpoint on $A$ and the other on $O$. We maximize the number of such objects which can be placed on the time slice $\Sigma$ with the added condition that they take up a finite amount of space in the geometry. We refer to these curves as bit threads between $A$ and $O$. From these we have
\be
        \left\{
        \begin{aligned}
            & & & \max 2\mu(P) \\
            & \text{s.t.} & & \forall x\in \Sigma,\; \int_{P}d\mu(p)\Delta(x,p)\leq 1.
        \end{aligned}
    \right. 
\ee
Here $\mu$ is a measure on the space $P$ and $\Delta(x,p)$ is a delta function which is nonzero at the location of a bit thread $p\in P$. The factor of two is a normalization which we choose for convenience. We refer to a feasible measure (one which satisfies the density bound) as a \emph{thread configuration}. An optimal thread configuration will be denoted as $\mu^*$.

This program can be dualized as follows
\be\label{eq:BT_lan}
\begin{split}
L(d\mu,d\nu)&= 2\int_{P}d\mu(p)-\int_{\Sigma}d\nu(x)\left(\int_{P}\left(d\mu(h)\Delta(x,p)\right)- 1\right)\\
&=\int_{P}d\mu(p) \left(2-\int_{\Sigma}d\nu(x)\Delta(x,p)\right)+\int_{\Sigma}d\nu(x)
\end{split}
\ee
resulting in the equivalent dual minimization program
\be\label{eq:BTmin}
        \left\{
        \begin{aligned}
            & & & \min \nu(x) \\
            & \text{s.t.} & & \forall p\in P,\; \int_{\Sigma}d\nu(x)\Delta(x,p) \geq 2.
        \end{aligned}
    \right.  
\ee

\begin{figure}[H]
\centering
\includegraphics[page=97,width=.40\textwidth]{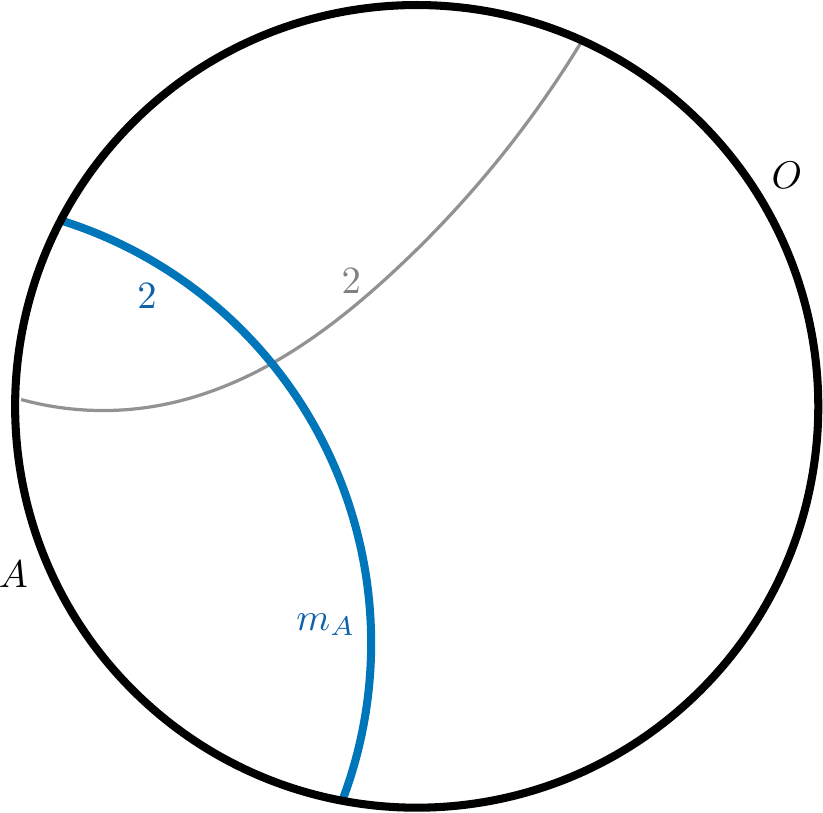}
\caption{\label{fig:BTmin} The optimal barrier configuration $\nu^*$ of \eqref{eq:BTmin} consists of  placing a barrier of two on the RT surface $m_A$. This is the smallest configuration possible such that every 2-thread from $A$ to $O$ will cross a barrier of at least two. The constraint is equivalent to the homology condition of the RT formula.}
\end{figure}

\noindent The measure $\nu$ should be thought of as a required barrier in the manifold. The constraint requires that every bit thread in $P$ must cross a minimum barrier of two in order to be feasible. We refer to such a measure as a \emph{barrier configuration}. This constraint is a natural realization of the usual homology constraint: any barrier which does not connect to the entangling surface $\partial A$ or separate fully $A$ from $O$ will not meet this condition. As such, because we wish for the smallest possible barrier, the correct location for the optimal barrier configuration is precisely the minimal RT surface $m_A$. That is it can be shown that this program has an optimal value of $2S_A$ such that it is equivalent to the RT formula (see figure \ref{fig:BTmin}).

\begin{figure}[H]
\centering
\includegraphics[page=101,width=.40\textwidth]{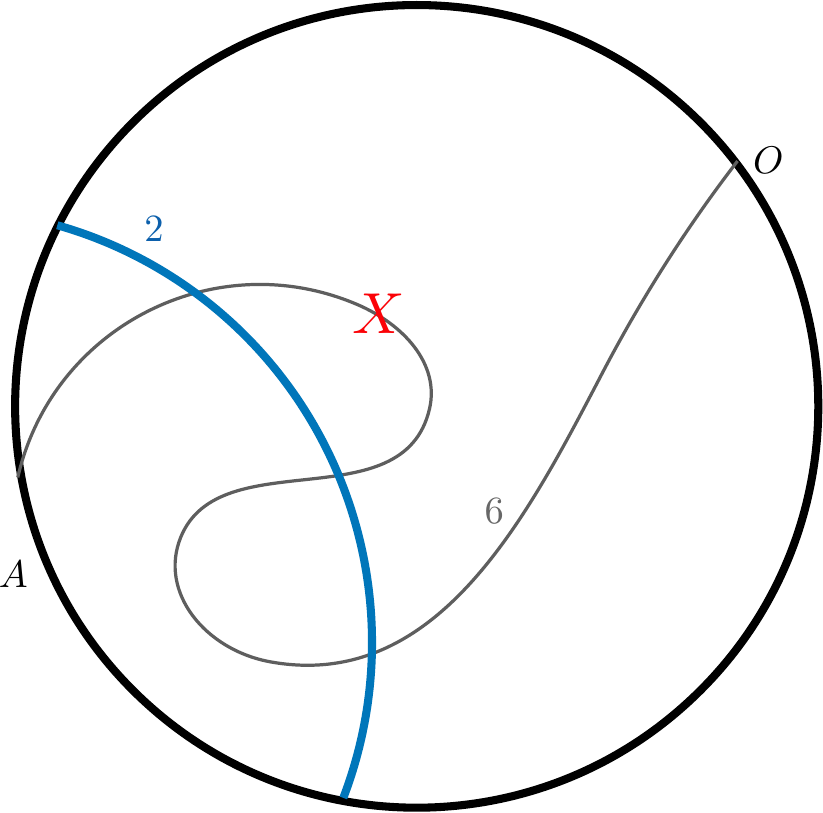}
\caption{\label{fig:CSthread} The following thread can not contribute to an optimal thread configuration as it crosses a total barrier of six. This can be understood as the thread is highly inefficient with each additional crossing of the barrier further preventing other threads from being placed.}
\end{figure}

The following theorem is an immediate consequence of applying CS to \eqref{eq:BT_lan}:
\begin{theorem}[Optimal thread configurations are efficient]\label{T1}
 A bit thread can contribute to an optimal thread configuration only if it crosses a barrier of exactly two.
\begin{proof}
For a given thread $p$ by CS we have either
\be
d\mu^*(p) =0 \text{ or } \left(2-\int_{\Sigma}d\nu^*(x)\Delta(x,p)\right)=0.
\ee
As such, if we define the space of threads which cross a barrier greater than two as $P_0$
\be
P_0=\{p\in P, \text{ s.t. } \int_{\Sigma}d\nu^*(x)\Delta(x,p)> 2 \} 
\ee
Then we are guaranteed
\be
\mu^*(P_0)=0.
\ee
\end{proof}
\end{theorem}

\noindent This can be understood intuitively as the location of an optimal barrier configuration acts as a bottleneck to the thread configuration. Were a thread to cross a barrier greater than two then it would necessarily be preventing other threads from crossing (see figure \ref{fig:CSthread}).

\paragraph{Multiple regions}

\begin{figure}[H]
\centering
\includegraphics[page=98,width=.40\textwidth]{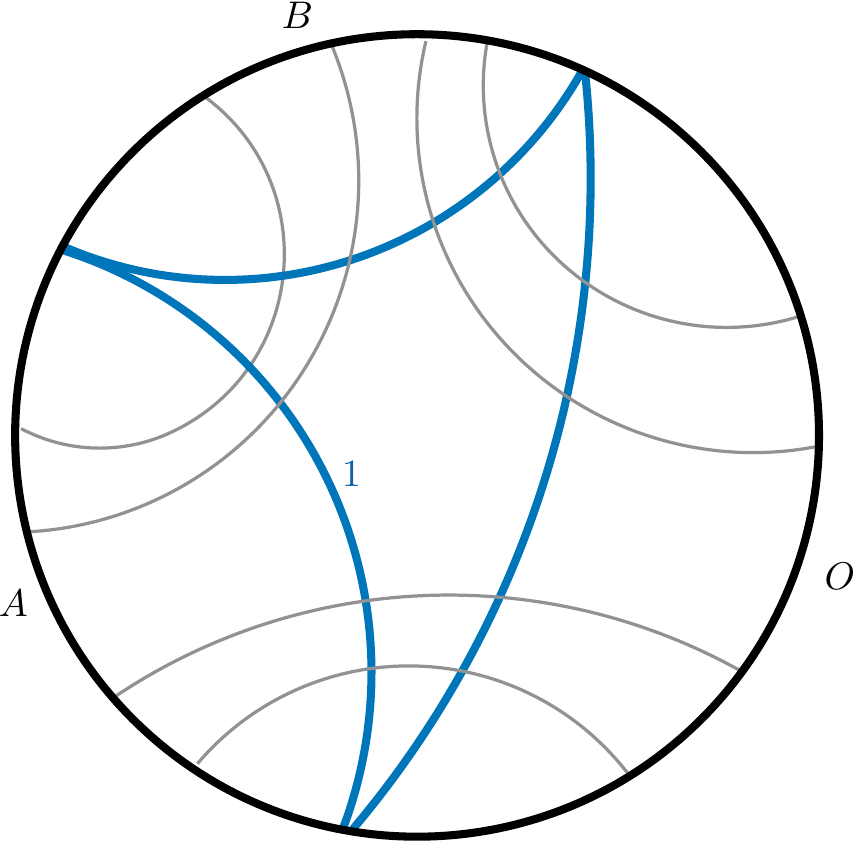}
\caption{\label{fig:multi}When we have multiple boundary regions we include a species of bit thread for each pair of boundary regions. The minimal barrier configuration $\nu^*$ consists of placing a barrier of one on each of the single party RT surfaces such that each thread crosses a barrier of two.}
\end{figure}

We can also consider the following generalization. Let the boundary consist of $N$ regions along with a purifier $O$: $\mathcal{A}=\{A_1,\cdots, A_N,O\}$. To each pair of regions $A_i,A_j$ including the purifier we define a \emph{species} of bit threads $P_{A_i:A_j}$ which is the set of all simple curves with one endpoint on $A_i$ and the other on $A_j$. We take union of these $\frac{1}{2}N(N+1)$ species to be the full space of bit threads $P$ (see figure \ref{fig:multi}). The dual programs remain unchanged except for the difference in the choice of the space $P$
\be\label{eq:BT}
\begin{split}
        BT(\mathcal{A})=&\left\{
        \begin{aligned}
            & & & \max 2\mu(P) \\
            & \text{s.t.} & & \forall x\in \Sigma,\; \int_{P}d\mu(p)\Delta(x,p)\leq 1
        \end{aligned}
    \right. \\
        =&\left\{
        \begin{aligned}
            & & & \min \nu(x) \\
            & \text{s.t.} & & \forall p\in P,\; \int_{\Sigma}d\nu(x)\Delta(x,p) \geq 2.
        \end{aligned}
    \right. 
    \end{split}
\ee
To understand the optimal configurations it is useful to introduce the notion of locking:
\begin{definition}[Locking]
A thread configuration $\mu$ is said to lock a set of surfaces if the density bound is saturated on all of them simultaneously.
\end{definition}

The following theorem of \cite{Cui:2018aa} states the locking capabilities of bit threads\footnote{Also see \cite{locking} for additional details and attempts to increase the locking capabilities of bit threads.}
\begin{theorem}\label{thm:locking}
For the program \eqref{eq:BT} there exists an optimal thread configuration $\mu^*$ such that
\be
2\mu^*(P)=\sum_i S_{A_i}.
\ee
and all single party entropies are locked.
\end{theorem}

\subsection{The holographic entropy cone and the $K$-basis}\label{sec:kcone}

In this section we review some basic facts about the holographic entropy cone \cite{r2,2019ForPh..6700011H, 2018ForPh..6600067H, r3} as well as the $K$-basis construction of \cite{r1}\footnote{For a current in depth introduction to the holographic entropy cone see for example \cite{2021arXiv210207535A,2022arXiv220400075H}}. Given a holographic state with $N$ boundary regions a natural question to ask is among all such states what are the allowed values for the various $2^N-1$ different entanglement entropies. For a given state these can be arranged as an entropy vector
\be
\Vec{\mathcal{S}}^N=\sum_J {S}^N_J \hat{e}^J.
\ee
However, not all entropy vectors are allowed. The holographic entropy cone (HEC) describes the space of allowed vectors as a series of positivity constraints on entropy quantities $Q$
\be
Q=\sum_J\alpha_J \mathcal{S}_J\geq 0.
\ee
As such, there are non-trivial relations between the various entanglement entropies. For example, we can consider the case $N=3$ where we take three regions $A,B,C$ along with a purifier $O$. In this case the entropy space is seven dimensional
\be
\mathcal{S}^3=\{S_A,S_B,S_C,S_{AB},S_{AC},S_{BC},S_{ABC}\}.
\ee
The entropy constraints are given by subadditivty (SA) and monogamy of mutual information (MMI) \cite{2013PhRvD..87d6003H} which correspond to positivity of mutual information and positivity of the \emph{negative} tripartite information
\be
\begin{split}
I(A:B)&=S_A+S_B-S_{AB}\geq 0\\
-I_3(A:B:C)&=S_{AB}+S_{BC}+S_{AC}-S_A-S_B-S_C-S_{ABC} \geq 0.
\end{split}
\ee
These hold for each choice of $A,B,C,O$ giving rise to seven unique inequalities\footnote{These correspond to $\binom{4}{2}=6$ mutual informations, but only one instance of MMI. This is because even though there are $\binom{4}{3}=4$ choices these all give the same inequality due to $I_3$ being secretly symmetric with respect to all three regions and their complement.}.

\begin{figure}[H]
\centering
\begin{tabular}{cc}
\centering
\includegraphics[width=.45\textwidth,page=2]{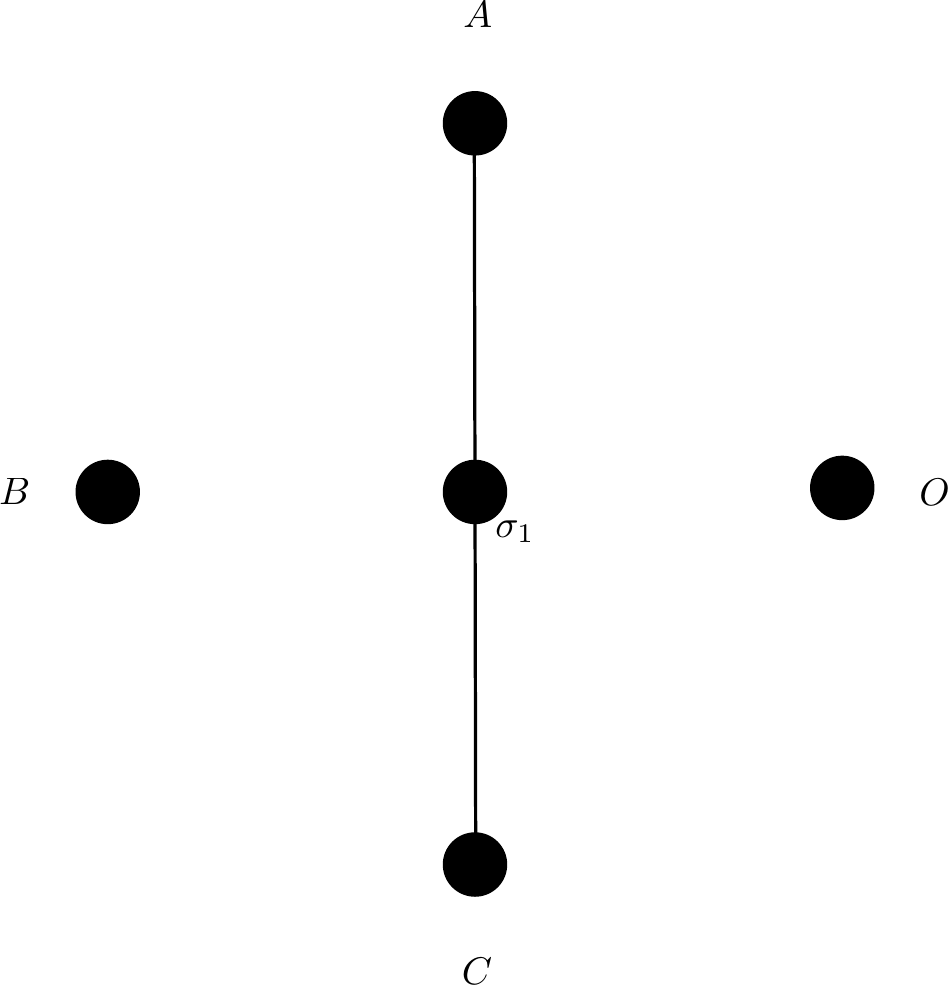}&
\includegraphics[width=.45\textwidth,page=3]{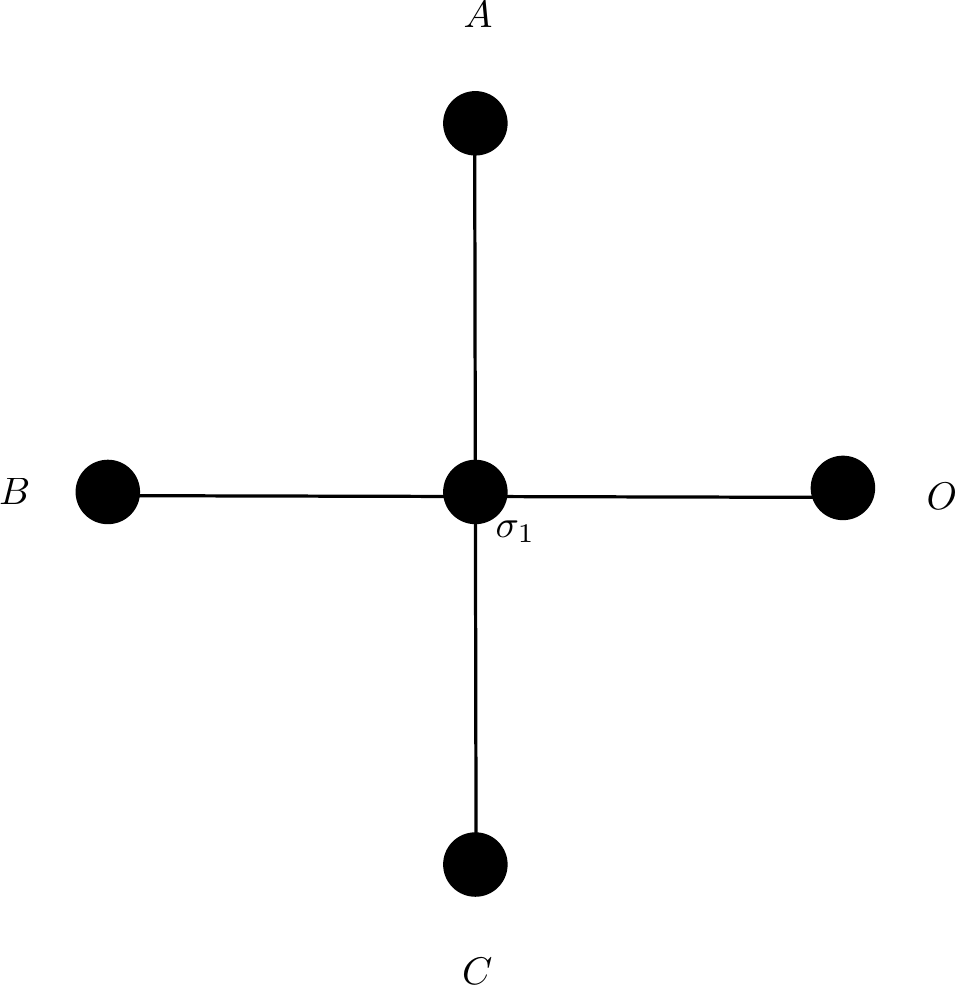}
\end{tabular}
\caption{\label{fig:star} Star graphs for $N=3$ whose entropy vectors calculated using min cuts correspond to $PT_{AC}$ and $PT_{ABCO}$. The coefficients relating the $S$ and $K$ basis are determined by the entanglement entropy of the corresponding star graph. For example $S_{AB}$ contains $1K_{AC}$ and $2K_{ABCO}$ as $S_{AB}(PT_{AC})=1$ and $S_{AB}(PT_{ABCO})=2$.}
\end{figure}

In this article we will make frequent use of the $K$-basis first described by the authors of \cite{r1}. The key observation is that perfect tensors states can be used to define an alternate basis for the holographic entropy cone. A 2s-perfect tensor, $PT_{2s}$, is a 2s-party pure state such that for any positive integer $s$ the reduced density matrix involving any $s$ parties is maximally mixed. The entropy vector of such a state is realized by a $2s$ star graph which consists of $2s$ boundary vertices each connected to a single internal vertex by an edge with a capacity of one (see figure \ref{fig:star}).

For $N$ regions we include $\binom{N+1}{2s}$ terms $K_{2s}$ for each choice of $s$ up to $\lfloor\frac{N+1}{2}\rfloor$ corresponding to each possible even combination of boundary regions \emph{including} the purifier
\be
\Vec{\mathcal{S}}^N=\sum_J \mathcal{K}^N_J \hat{g}^J.
\ee
Explicitly for $N=3$ 
\be
\mathcal{K}^3=\{K_{AB},K_{AC},K_{AO},K_{BC},K_{BO},K_{CO};K_{ABCO}\}
\ee
where the change of basis is implemented by the linear equations
\be
\begin{split}
    S_A&=K_{AB}+K_{AC}+K_{AO}+K_{ABCO}\\
    S_B&=K_{AB}+K_{BC}+K_{BO}+K_{ABCO}\\
    S_C&=K_{AC}+K_{BC}+K_{CO}+K_{ABCO}\\\\
    S_{AB}&=K_{AC}+K_{AO}+K_{BC}+K_{BO}+2K_{ABCO}\\
    S_{AC}&=K_{AB}+K_{AO}+K_{BC}+K_{CO}+2K_{ABCO}\\
    S_{BC}&=K_{AB}+K_{BO}+K_{AC}+K_{AO}+2K_{ABCO}\\\\
    S_{ABC}&=K_{AO}+K_{BO}+K_{CO}+K_{ABCO}\\
\end{split}
\ee
or more succinctly\footnote{Note the three party entropy $S_{ABC}$ by purity can be calculated using the first formula as $S_{O}$.}
\be\label{eq:3LTprelim}
\begin{split}
S_1&=\sum_{j=2}^4K_{1j}+K_{1234}\\
S_{12}&=\sum_{j=3}^4\left(K_{1j}+K_{2j}\right)+2K_{1234}.
\end{split}
\ee
The coefficients in these equations can be directly calculated by determining the corresponding entanglement entropy of the perfect tensor state (see figure \ref{fig:star}). This procedure can be used to generate the correct set of linear equations for any $N$.

Positivity constraints on entropy quantities can also be expressed in the $K$-basis
\be
Q=\sum_J\alpha_J \mathcal{S}_J=\sum_J\beta_J\mathcal{K}_J\geq 0
\ee
however, they have the added property that all coefficients in the constraints will be positive $\forall_J\;\beta_J\geq 0$. For example, the $N=3$ inequalities can be written as
\be
\begin{split}
    I(A:B)=2K_{AB}\geq0\\
    -I_3(A:B:C)=2K_{ABCO}\geq0
\end{split}
\ee
which is simply positivity of the components of $\mathcal{K}^3$.

\subsection*{Graphs and extremal rays}

\begin{figure}[H]
\centering
\begin{adjustbox}{center}
\begin{tabularx}{\textwidth}{Xcc}
  \begin{minipage}{.44\textwidth}\includegraphics[page=92,width=\textwidth]{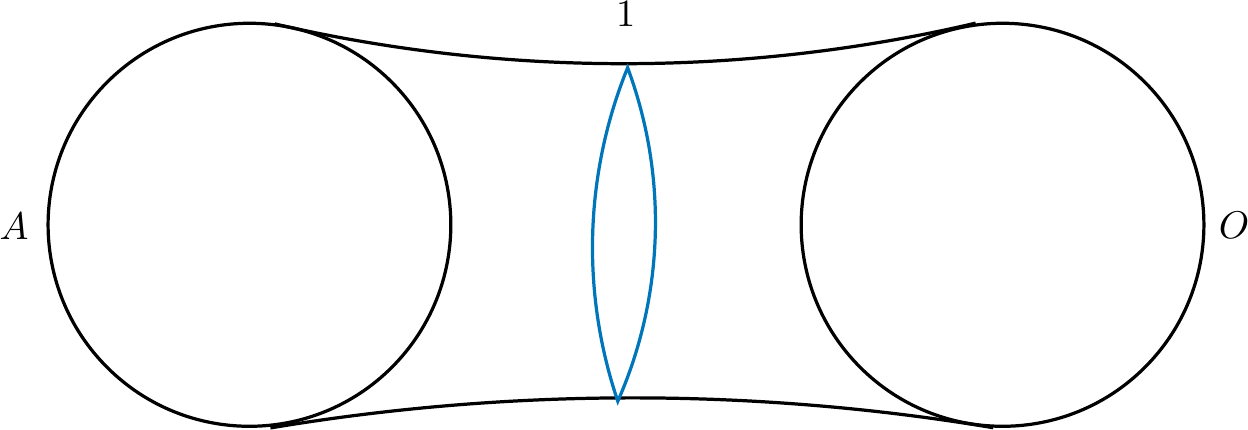}\end{minipage} &$\Longleftrightarrow$&
\includegraphics[page=91,width=.42\textwidth]{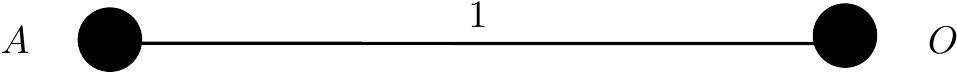}
\end{tabularx}
\end{adjustbox}
\caption{\label{fig:gstate} A bulk geometry consisting of a wormhole between the two boundary regions $A$ and $O$. This particular state is realized by a bulk metric which is AdS-Schwarzschild. The area of the minimal surface or throat is taken to have a value of one. This geometry can be viewed as a graph with a single edge connecting two boundary vertices $A$ and $O$. The capacity of the edge is one.}
\end{figure}

For this paper we will be primarily focused on a particular class of asymptotically $AdS$ multiboundary wormhole geometries. These states can be represented schematically as a graph where the capacity of an edge is equal to the minimal area surface of the corresponding throat (see figure \ref{fig:gstate}). An essential detail is that given any entropy vector of the holographic entropy cone there is a graph and consequently a holographic state which realizes that vector \cite{r2}.

Graphs are particularly convenient for our purposes as it is usually straightforward to construct explicit thread and barrier configurations (compared to the task on a Riemannian manifold). As such, these serve as a useful testing ground of examples for understanding the key properties of our construction.

\begin{figure}[H]
\centering
\includegraphics[page=93,width=.50\textwidth]{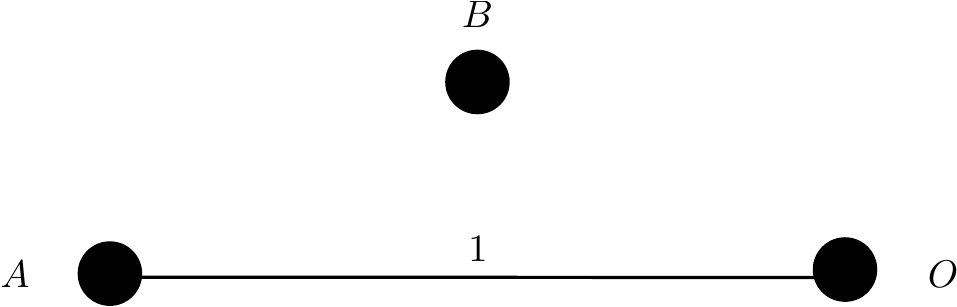}
\caption{\label{fig:eray} An example of an extremal ray for $N=2$. The three entropy constraints consist of $I(A:B)\geq 0, \quad I(A:O)\geq 0, \quad I(B:O)\geq 0$. For this state $I(A:B)=I(B:O)=0, \quad I(A:O)=2$. }
\end{figure}

Among the space of graphs there special graphs called \emph{extremal rays} of the holographic entropy cone\footnote{Knowledge of the extremal rays is equivalent to that of the entropy inequalities. This is because the extremal rays are the facets of the dual cone of the holographic entropy cone.}. These have the property that they saturate the maximum possible number of entropy inequalities. An example of an extremal ray is shown in figure \ref{fig:eray}. We will see that extremal rays are particularly important examples for our purposes and the saturation of the entropy inequalities will lead to tight constraints between thread species.

\section{Perfect tensor hyperthreads}\label{sec:PTHT}
The primary conceptual innovation of this article is the utility of the $K$-basis when considering entanglement entropies from the perspective of bit threads. To each component of the $K$-basis entropy vector we associate a class of thread-like objects which connect on the boundary in the regions specified by the particular $K$.   

\begin{figure}[H]
\centering
\begin{tabular}{cc}
\centering
\includegraphics[width=.45\textwidth,page=99]{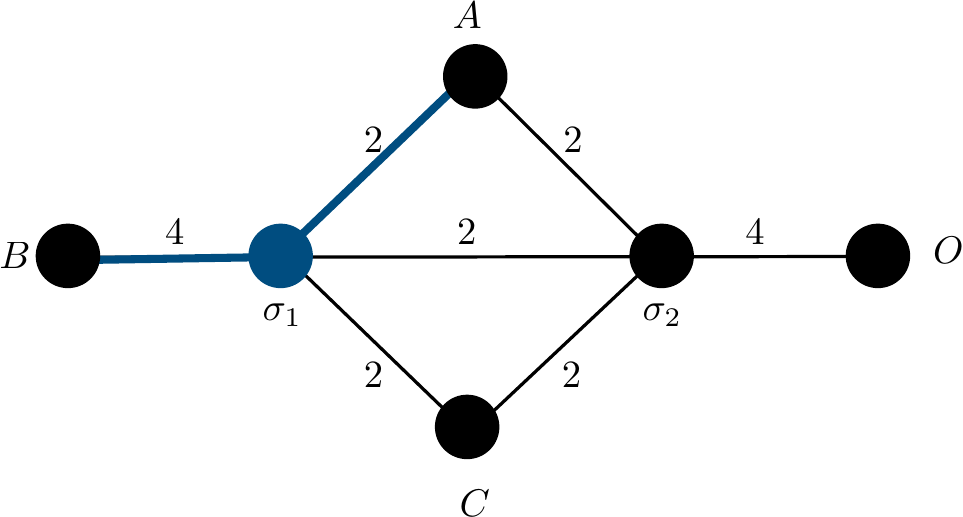}&
\includegraphics[width=.45\textwidth,page=100]{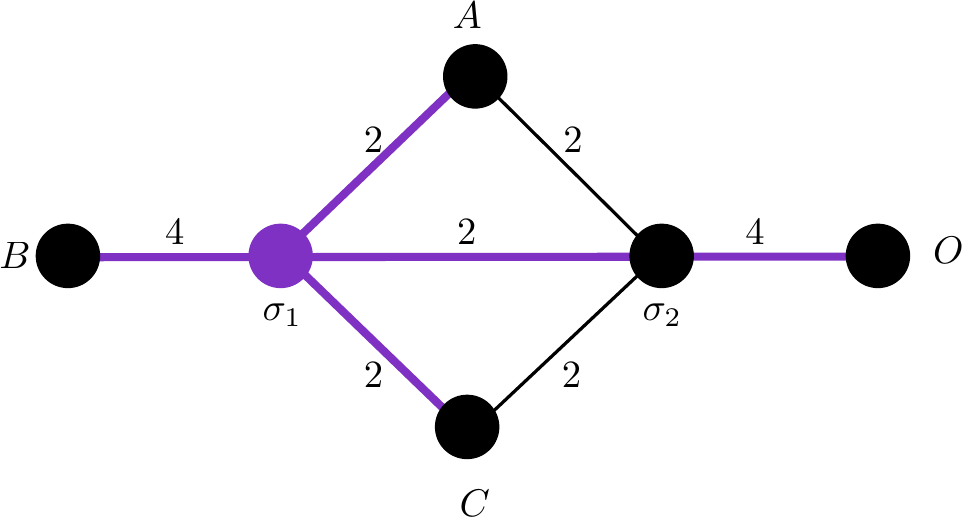}
\end{tabular}
\caption{\label{fig:PTex}L: An $AB$ 2-thread. R: An $ABCO$ 4-thread.}
\end{figure}

\noindent A perfect tensor $k$-hyperthread or ``$k$-thread" is the union of an even number of simple curves\footnote{We will keep this general for the application to manifolds, but in what follows we will work primarily on graphs. In this context the simple curves are paths: a collection of edges connecting two vertices.} connecting different boundary regions to a single internal vertex\footnote{Previous work \cite{2021JHEP...09..118H} defined hyperthreads for GHZ states and allowed for these hyperthreads to have multiple internal vertices. It is essential to our current construction that the perfect tensor hyperthreads have only a \emph{single} internal vertex. This is so that the different species of perfect tensor hyperthreads will cross RT surfaces the appropriate number of times.} (see figure \ref{fig:PTex}). The space of all such perfect tensor hyperthreads $H$ can be split up by the number of regions a perfect tensor hyperthread connects and further into species determined by the exact boundary regions it connects
\be
\begin{split}
    H_2&=H_{A_1:A_2}\cup...\cup H_{A_{n-1}:A_n}\\
    H_4 &= H_{A_1:A_2:A_3:A_4}\cup...\cup H_{A_{n-3}:A_{n-2}:A_{n-1}:A_{n}}\\
     &\;\;\vdots  \\
    H_n &= H_{A_1:...:A_n}.
\end{split}
\ee
When designing a convex program for the perfect tensor hyperthreads our goal will be that maximizing a measure $\mu$ over the space of such objects should correctly reproduce the full entropy vector. As such, we \emph{define} the contribution of the perfect tensor hyperthreads to be the same as that of the corresponding perfect tensor. For example for $N=3$ we would have following directly from \eqref{eq:3LTprelim}
\be
\begin{split}
S_1&=\sum_{j=2}^4\mu_{1j}+\mu_{1234}\\
S_{12}&=\sum_{j=3}^4\left(\mu_{1j}+\mu_{2j}\right)+2\mu_{1234}
\end{split}
\ee
where $\mu_{I}\equiv\mu(H_{I})$ with $\mu$ a measure on the set $H$. That is each 2-thread which connects the boundary regions to \emph{another} boundary region counts for one to that entanglement entropy. While the 4-threads contribute one to the single party entropies and two to the two party entropies.

In order for a configuration of perfect tensor hyperthreads to correctly reproduce all entanglement entropies we must consider an objective which is the sum of all components of the $S$-basis entropy vector
\be\label{eq:PTHTmax}
        \left\{
        \begin{aligned}
            & & & \max \sum\mathcal{S}^N(\mu) \\
            & \text{s.t.} & & \forall x\in \Sigma,\; \int_{H}d\mu(h)\Delta(x,h)\leq 1.
        \end{aligned}
    \right.  
\ee

\noindent Because summing over the entropy vector is symmetric with respect to all boundary regions in general all species of $k$-threads will contribute identically. That is the program will always take the form

\be\label{eq:PTHTmax2}
        \left\{
        \begin{aligned}
            & & & \max \sum_{s=1}^{\lfloor\frac{N}{2}\rfloor}a_{2s}\mu(H_{2s}) \\
            & \text{s.t.} & & \forall x\in \Sigma,\; \int_{H}d\mu(h)\Delta(x,h)\leq 1.
        \end{aligned}
    \right.  
\ee
The coefficients $a_{2s}$ determine the contribution of each species of perfect tensor hyperthread as well as the corresponding required barrier in the dual. From this we can generalize theorem \ref{T1}:
\begin{theorem}[Optimal configurations of perfect tensor hyperthreads are efficient]\label{T3}
 A perfect tensor hyperthread $h\in H_{2s}$ which contributes $a_{2s}$ to an objective can contribute to an optimal configuration of perfect tensor hyperthreads only if it crosses a barrier of exactly $a_{2s}$.
 \end{theorem}

In what follows we will show several application of this procedure to various geometries. We start with $N=2$ and show that in this case \eqref{eq:PTHTmax} is equivalent to \eqref{eq:BT}. That is 2-threads and bit threads are equivalent. We then proceed to $N=3$ where we introduce for the first time 4-threads. As we move to $N=4,5$ a major complication arises because of the structure of the entropy inequalities namely that components of the $K$-basis can be negative. We resolve this by introducing negative threads which contribute negatively to the density bound and objective. As we will see, it is necessary to explicitly implement the entropy inequalities in our convex programs as these provide the essential constraints which relate different species of positive and negative threads. In all cases we provide explicit examples of optimal configurations of perfect tensor hyperthread and optimal barrier configurations. Together these allow us to demonstrate the ability of the perfect tensor hyperthreads to lock the full entropy vector.

\section{2 regions}\label{sec:N2}
To serve as a warm-up we begin with two regions $A$ and $B$ along with the purifier $O$. In this case the entropy vector is three dimensional and in the $K$ basis consists of three $PT_2$s
\be
\mathcal{S}^2=\{S_A,S_B;S_{AB}\}
\ee
\be
\mathcal{K}^2=\{K_{AB},K_{AO},K_{BO}\}.
\ee
The relation between the two is given explicitly by
\be
\begin{split}
S_A&=K_{AB}+K_{AO}\\
S_B&=K_{AB}+K_{BO}\\
S_{AB}&=K_{AO}+K_{BO}
\end{split}
\ee
so that the entanglement entropy is given by the sum of the two $K$s which share its region. A short calculation shows that these are precisely up to a factor the mutual information
\be
\begin{split}
  I(A:B)&=2K_{AB}\\
 I(A:O)&=2K_{AO}\\
 I(B:O)&=2K_{BO}.
\end{split}
\ee
The inequalities of the holographic entropy cone are given by subadditivity which corresponds to positivity of the mutual information and thus positivity of the $K$'s
\be
K_{AB},K_{AO},K_{BO}\geq0.
\ee
Following our procedure we define a class of perfect tensor hyperthreads to each $K_{I}$. Here, we have three species of 2-threads $H_{A:B}\cup H_{A:O}\cup H_{B:O} =H$. Taking the sum of $S$-basis entropy vector
\be
\sum \mathcal{S}^2= 2(K_{AB}+K_{AO}+K_{BO})
\ee
it follows that our objective is given by
\be
2(\mu_{AB}+\mu_{AO}+\mu_{BO})=2\mu(H).
\ee
Imposing the density bound we arrive at the convex program
\be
        \left\{
        \begin{aligned}
            & & & \max 2\mu(H) \\
            & \text{s.t.} & & \forall x\in \Sigma,\; \int_{H}d\mu(h)\Delta(x,h)\leq 1
        \end{aligned}
    \right. 
\ee
which is dual to
\be
        \left\{
        \begin{aligned}
            & & & \min \nu(x) \\
            & \text{s.t.} & & \forall h\in H,\; \int_{\Sigma}d\nu(x)\Delta(x,h) \geq 2.
        \end{aligned}
    \right.  
\ee
This is the same as \eqref{eq:BT} which we know by theorem \ref{thm:locking} has the optimal value $S_A+S_B+S_O=\sum\mathcal{S}^2$ and locks each of the RT surfaces. As such the entropy vector is correctly reproduced.

\section{3 regions}\label{sec:N3}
As we proceed to three regions $A,B,C$ with purifier $O$ we encounter for the first time 4-threads. The entropy vector is seven dimensional
\be
\mathcal{S}^3=\{S_A,S_B,S_C;S_{AB},S_{AC},S_{BC};S_{ABC}\}
\ee
so that the corresponding vector in the $K$-basis consists of six $PT_2$s and a single $PT_4$
\be
\mathcal{K}^3=\{K_{AB},K_{AC},K_{AO},K_{BC},K_{BO},K_{CO};K_{ABCO}\}.
\ee
with the change of basis given by
\be
\begin{split}\label{eq:LT3}
S_1&=\sum_{j=2}^4K_{1j}+K_{1234}\\
S_{12}&=\sum_{j=3}^4\left(K_{1j}+K_{2j}\right)+2K_{1234}.
\end{split}
\ee
The inequalities of the holographic entropy cone in the $K$- basis  correspond to the  positivity of all of the $K$s
\be
\begin{split}
   K_{AB},K_{AC},K_{AO},K_{BC},K_{BO},K_{CO}\geq 0\\
   K_{ABCO} \geq 0.
\end{split}
\ee
Taking the sum of the $S$-basis entropy vector we find
\be
\sum \mathcal{S}^3= 4( K_{AB}+K_{AC}+K_{AO}+K_{BC}+K_{BO}+K_{CO}) + 10 K_{ABCO} 
\ee
so that assigning to each $K$ a species of perfect tensor hyperthreads our objective is given by
\be
4\mu(H_2) + 10\mu(H_4).
\ee
Imposing the density bound we arrive at the convex program
\be\label{eq:max3}
        \left\{
        \begin{aligned}
            & & & \max 4\mu(H_2)+10\mu(H_4) \\
            & \text{s.t.} & & \forall x\in \Sigma,\; \int_{H}d\mu(h)\Delta(x,h)\leq 1
        \end{aligned}
    \right.  
\ee
which can be dualized
\be
\begin{split}
L(d\mu,d\nu)&= 4\int_{H_2}d\mu(h)+10\int_{H_4}d\mu(h)-\int_{\Sigma}d\nu(x)\left(\int_{H}\left(d\mu(h)\Delta(x,h)\right)- 1\right)\\
&=\int_{H_2}d\mu(h) \left(4-\int_{\Sigma}d\nu(x)\Delta(x,h)\right)+\int_{H_4}d\mu(h) \left(10-\int_{\Sigma}d\nu(x)\Delta(x,h)\right)+\int_{\Sigma}d\nu(x)
\end{split}
\ee
resulting in the equivalent dual minimization program
\be
        \left\{
        \begin{aligned}
            & & & \min \nu(x) \\
            & \text{s.t.} & & \forall h\in H_2,\; \int_{\Sigma}d\nu(x)\Delta(x,h) \geq 4,\\
             & \text{and} & & \forall h\in H_{4},\; \int_{\Sigma}d\nu(x)\Delta(x,h) \geq 10
        \end{aligned}
    \right.  
\ee
which asks a for the minimal barrier configuration such that each 2-thread crosses a barrier of at least 4 and each 4-thread a barrier of at least 10.

\subsection*{Example}
\begin{figure}[H]
\centering
\includegraphics[page=88,width=.60\textwidth]{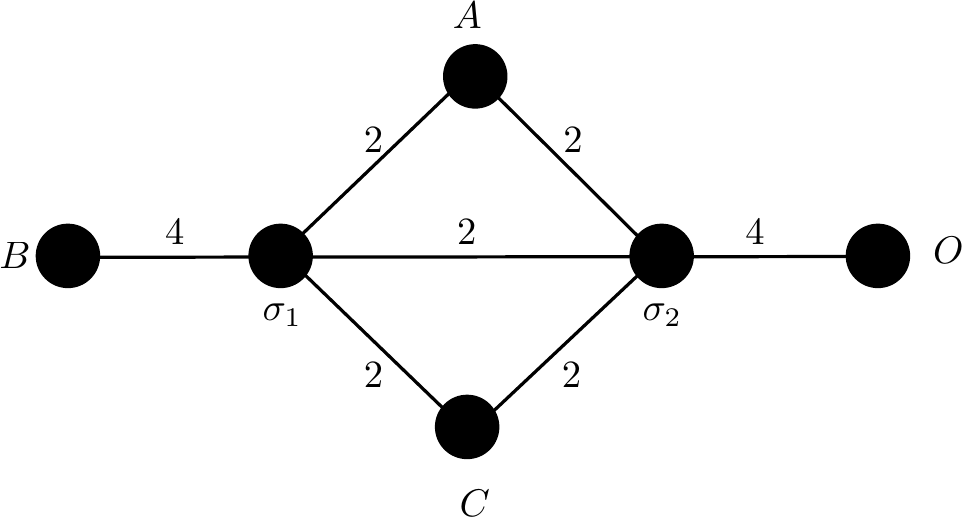}
\caption{\label{fig:3g}The graph $g$ consists of three boundary vertices $A,B,C$ and purifier $O$ along with two internal vertices $\sigma_1$ and $\sigma_2$. Edge capacities are labeled for each edge and give the maximum number of threads which can be placed on it. Viewing the graph as a multi boundary wormhole geometry, the edge capacities also gives the area of the minimal area surface or throat between the connecting vertices.}
\end{figure}
As an example we consider the graph $g$ shown in figure \ref{fig:3g} which has $S$ and $K$-basis entropy vectors
\be
\begin{split}
\mathcal{S}^3_g&=\!\{S_A=4,S_B=4,S_C=4;S_{AB}=6,S_{AC}=8,S_{BC}=6;S_{ABC}=4\}\\
\mathcal{K}^3_g&=\!\{K_{AB}=1,K_{AC}=0,K_{AO}=1,K_{BC}=1,K_{BO}=0,K_{CO}=1;K_{ABCO}=2\}.
\end{split}
\ee

\begin{figure}[H]
\centering
\includegraphics[width=.6\textwidth,page=89]{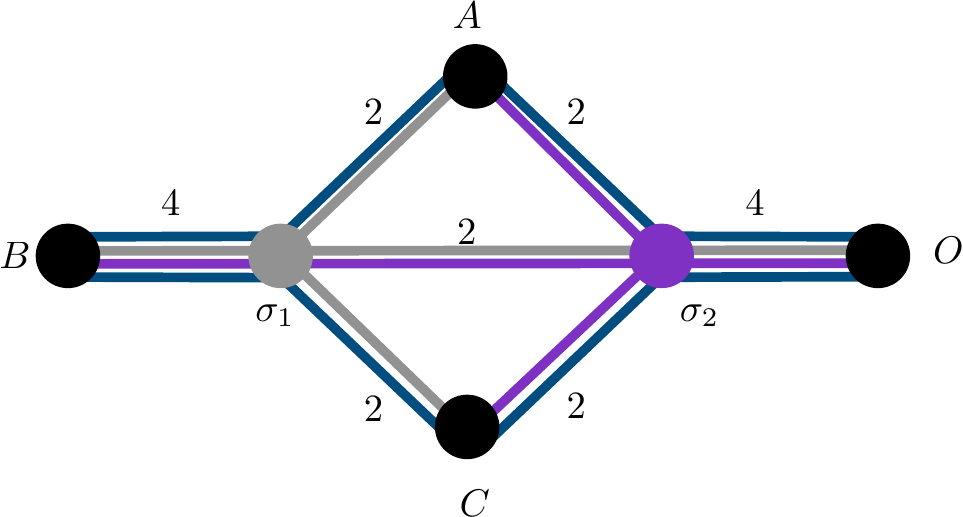}
\caption{\label{fig:graph4} A maximal configuration of perfect tensor hyperthreads on the 3 boundary graph $g$ for the program \eqref{eq:max3}. The configuration consists of two 4-threads one of which splits on each of the two internal vertices as well as four 2-threads shown in blue. The entropy vector in the $K$-basis is given by $\mathcal{K}^3_g=\{1,0,1,1,0,1;2\}$ which matches the number of perfect tensor hyperthreads of each species. The objective obtains a value of 4*4+10*2=36. The dual obtains its minimum with a barrier configuration consisting of 2 placed on each edge. This corresponds to a barrier of 1 being placed on each of the RT surfaces corresponding to each of the seven entanglement entropies which make up the entropy vector in the $S$-basis. Taking into account the capacities, the objective has the optimal value 5*2*2+2*4*2=36. The equality of the two is demanded by the duality and guarantees these configurations are in fact optimal.}
\end{figure}

\begin{figure}[H]
\centering
\includegraphics[page=90,width=.6\textwidth]{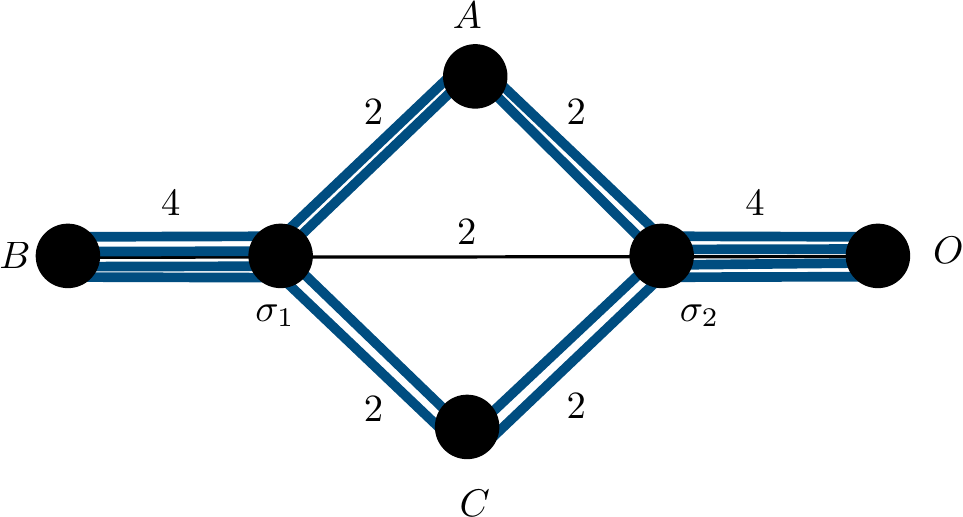}
\caption{\label{fig:g3BT} An optimal 2-thread configuration $\mu^*$ of $g$ for the program \eqref{eq:BT}. The corresponding entropy vector of the configuration is
$\mathcal{S}^3_\mu=\{4,4,4;\textcolor{red}{4},8,\textcolor{red}{4};4\}$. The entropies $S_{AB}$ and $S_{BC}$ shown in red do not match $\mathcal{S}^3_g$. This is a consequence of theorem \ref{thm:locking} which states only \emph{single party} entropies are guaranteed to be locked. Note the corresponding $K$-basis vector of the thread configuration is $\mathcal{K}^3_\mu=\{2,0,2,2,0,2;0\}$ which matches the number of each thread species.}
\end{figure}

\noindent For graphs the program \eqref{eq:max3} can be straightforwardly applied by explicit construction of both maximal configurations of perfect tensor hyperthread and minimal barrier configurations. An example of such an analysis is given in figure \ref{fig:graph4} which demonstrates the ability of the perfect tensor hyperthreads to lock the full entropy vector. This should be compared with a bit thread configuration figure \ref{fig:g3BT} which is optimal for \eqref{eq:BT}.

More generally one would wish to know if a maximal configuration of perfect tensor hyperthreads exists for \emph{any} graph with $N=3$. We state this as the following conjecture:

\begin{conjecture}
For any graph with $N=3$ the program
\be
        \left\{
        \begin{aligned}
            & & & \max 4\mu(H_2)+10\mu(H_4)  \\
            & \text{s.t.} & & \forall x\in \Sigma,\; \int_{H}d\mu(h)\Delta(x,h)\leq 1
        \end{aligned}
    \right.  
\ee
locks the entropy vector $\mathcal{S}^3$ s.t. $K_{I}=\mu^*(H_I).$
\end{conjecture}
\noindent As supporting evidence we have performed similar analysis on a large number of random graphs with random capacities. As of yet no contradictions have been found.

\begin{figure}[H]
\centering
\begin{tabular}{cc}
\centering
\includegraphics[width=.4\textwidth,page=11]{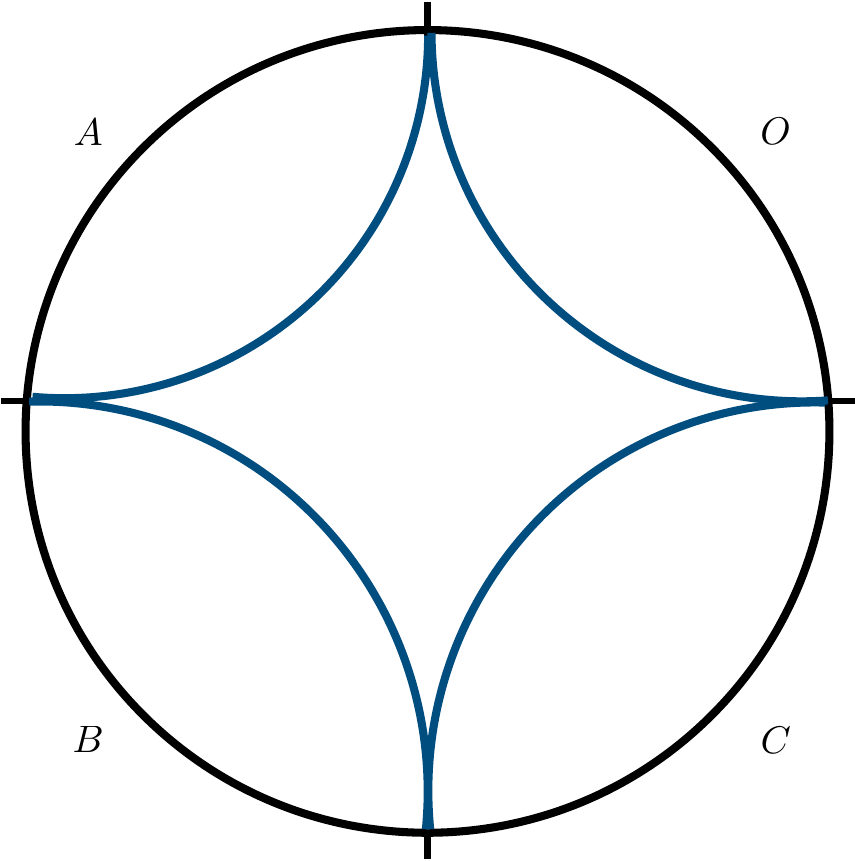}&
\includegraphics[width=.4\textwidth,page=12]{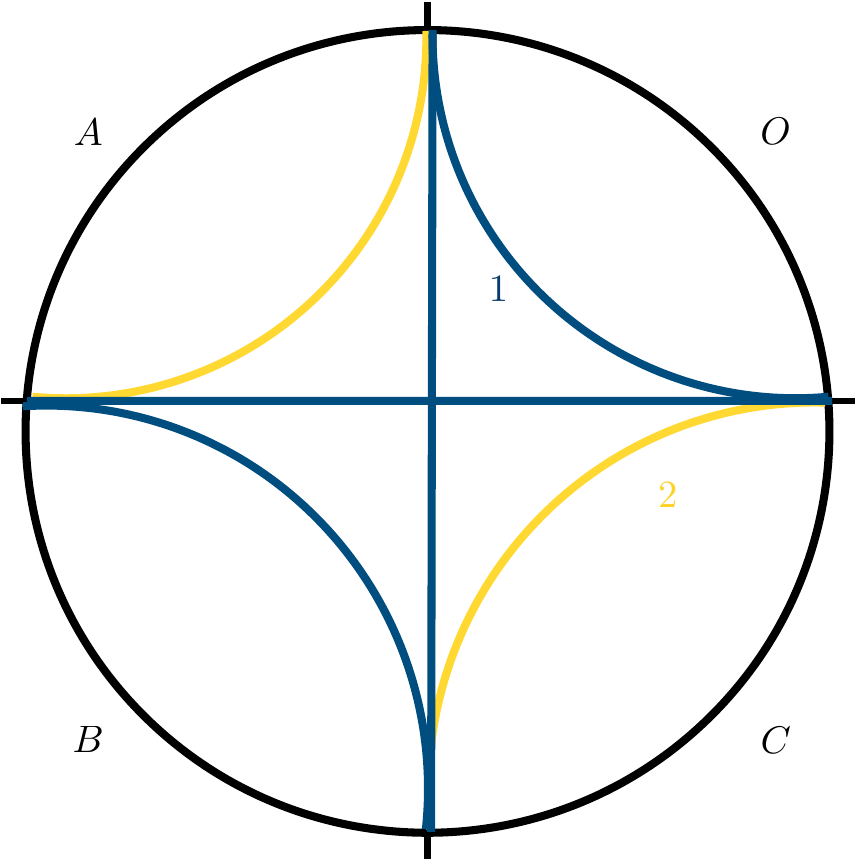}
\end{tabular}
\caption{\label{fig:ex_MMI} We consider a static time slice of vacuum $AdS_3$ with three regions $A,B,C$ along with the purifier $O$. First we place a barrier in the manifold corresponding to each of the RT surfaces whose area gives the entanglement entropy. In order to show this configuration is minimal it is necessary to construct a corresponding maximal configuration. This is hard to do in the context of a Riemannian manifold. As such we emphasize that the barrier configuration we construct is not necessarily optimal and only provides an upper bound. L: First we place barriers on each of the single party RT surfaces $m_A, m_B, m_C, m_{ABC}$. R: We place barrier on the two party RT surfaces $m_{AB},m_{BC},m_{AC}$. Note here that $m_{AC}=m_{A}\cup m_{C}$ so that the total barrier on these surfaces is 2 emphasized here in yellow.  }
\end{figure}

\begin{figure}[H]
\centering
\begin{tabular}{cc}
\centering
\includegraphics[width=.4\textwidth,page=94]{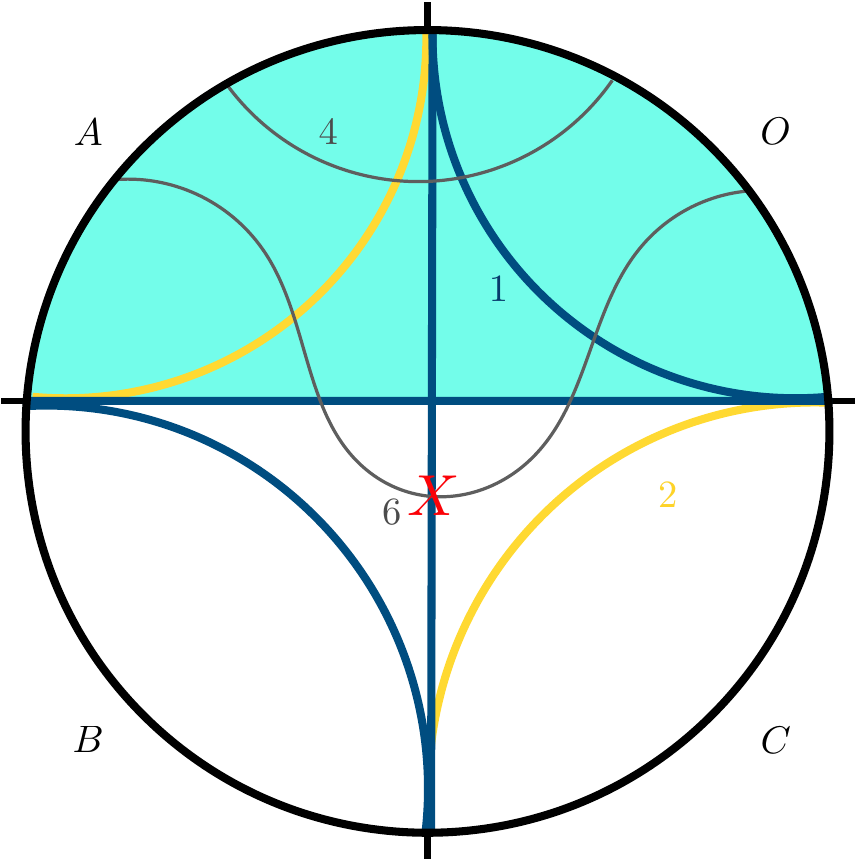}&
\includegraphics[width=.4\textwidth,page=95]{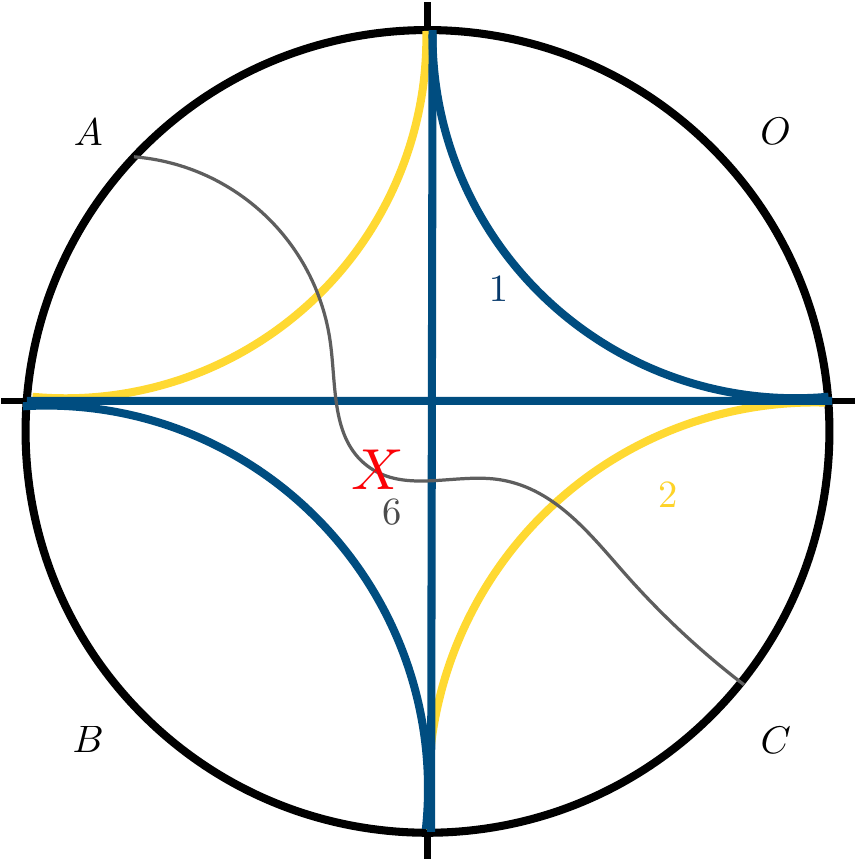}
\end{tabular}
\caption{\label{fig:ex_MMI2}L: A valid $AO$ 2-thread is shown as it crosses a total barrier of four. By theorem \ref{T3} it is able to contribute to a maximal perfect tensor hyperthread configuration. Shown in light blue is the region that such 2-threads can inhabit. This is because any such thread which extended past $m_{BC}$ would necessarily see more than the minimum required barrier of four. This essentially combs the $AO$ 2-threads closer to $A$ and $O$. R: The barrier configuration explicitly forbids $AC$ 2-threads as there is no way for such a thread to cross exactly a barrier of four. This is consistent with $I(A:C)=2K_{AC}=2\mu^{*}_{AC}=0$.}
\end{figure}

\begin{figure}[H]
\centering
\begin{tabular}{cc}
\centering
\includegraphics[width=.4\textwidth,page=1]{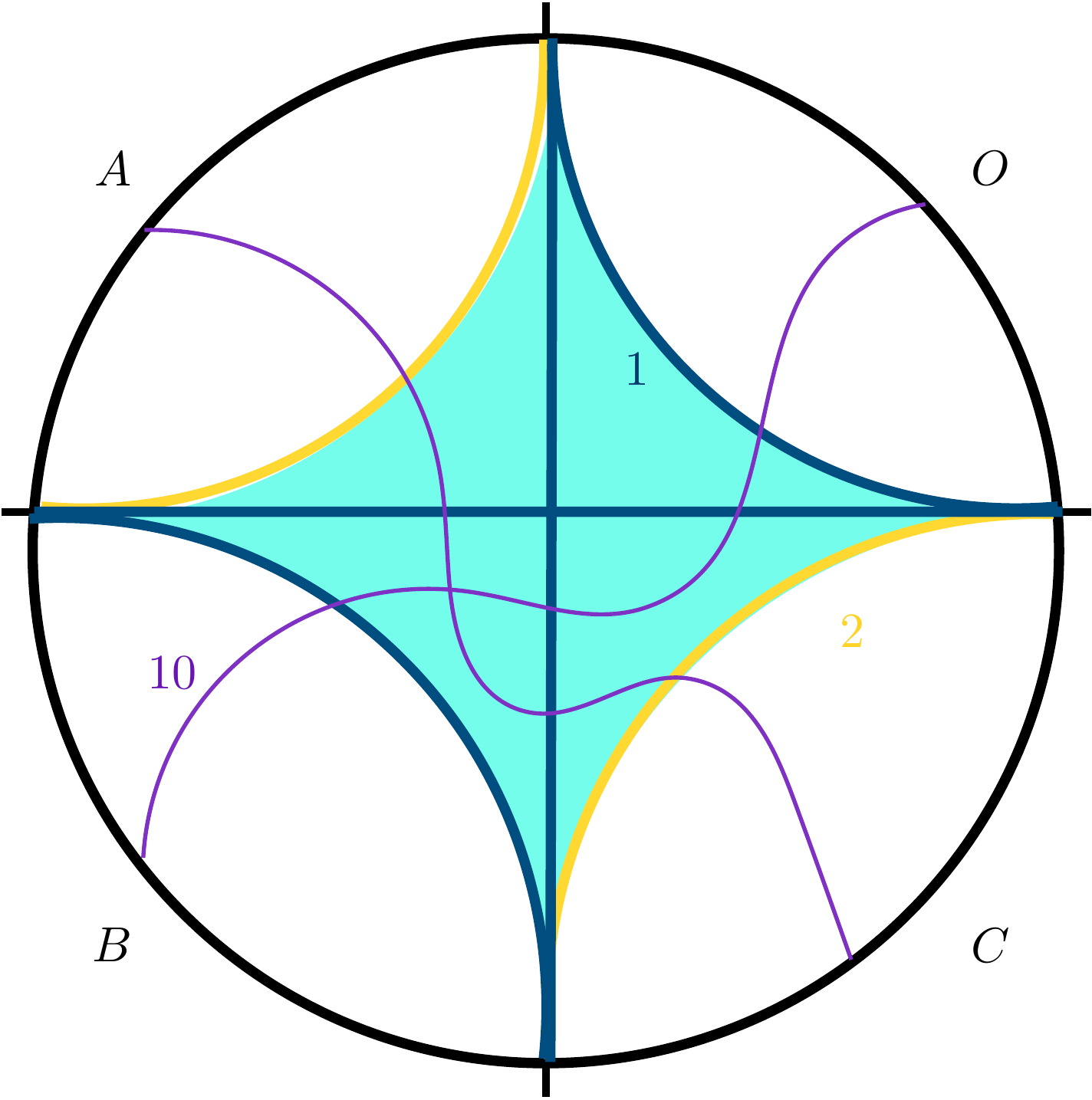}&
\includegraphics[width=.4\textwidth,page=96]{figs/PT_BTc.pdf}
\end{tabular}
\caption{\label{fig:ex_MMI3}L: Shown is a 4-thread which crosses exactly a barrier of ten and can contribute to an optimal perfect tensor hyperthread configuration as a result of theorem \ref{T3}. Such threads must have their vertex in the central light blue region as can be verified by comparing with the figure on the right. As such 4-threads are combed to the center of the geometry.}
\end{figure}

One is also interested in the application of perfect tensor hyperthreads directly to general holographic states. Typically, it is very difficult to explicitly construct hyperthread configurations as one must consider the placement of an infinite number of threads. In general hyperthreads with fractional weight can be used which forces one to also consider threads which cross one another in the geometry. Such issues make it more likely that there are potential geometric constraints which limit the ability of the hyperthreads to lock. Still, it is possible to construct valid barrier configurations which place upper bounds on the objective. Theorem \ref{T3} can also be used to understand the general behavior and location of particular species of hyperthreads (see figures \ref{fig:ex_MMI}, \ref{fig:ex_MMI2}, \ref{fig:ex_MMI3}). If such obstacles do prevent locking it is possible that this may be fixed by considering an alternate definition of the perfect tensor hyperthreads perhaps by considering a different density bound.

\section{4 regions}\label{sec:N4}
When we move to the case of four boundary regions $A,B,C,D$ with purifier $O$, the entropy cone is 15 dimensional consisting of
\be
\mathcal{S}=\{S_A,S_B,S_C,S_D;S_{AB},S_{AC},S_{AD},S_{BC},S_{BD},S_{CD};S_{ABC},S_{ABD},S_{ACD},S_{BCD};S_{ABCD}\}.
\ee
In the $K$-basis this corresponds to ten $PT_2$s and five $PT_4$s
\be
\begin{split}
\mathcal{K}=\{&K_{AB},K_{AC},K_{AD},K_{AO},K_{BC},K_{BD},K_{BO},K_{CD},K_{CO},K_{DO};\\
&K_{ABCD},K_{ABCO},K_{ABDO},K_{ACDO},K_{BCDO}\}.
\end{split}
\ee
with change of basis is given by
\be\label{eq:LT4}
\begin{split}
S_1&=\sum_{j=2}^5K_{1j}+\sum_{1<j<k<l}^5K_{1jkl}\\
S_{12}&=\sum_{j=3}^5\left(K_{1j}+K_{2j}\right)+\sum_{2<j<k<l}^5\left(K_{1jkl}+K_{2jkl}\right)+2\sum_{2<j<k}^5K_{12jk}.\\
\end{split}
\ee
The inequalities of the holographic entropy cone are again given by subadditivity and MMI, however in the $K$-basis they now take the form
\be
\begin{split}
    K_{12}\geq 0\\
   K_{1234}+K_{1235}\geq 0
\end{split}
\ee
which consists of 10 and 10 total constraints respectively corresponding to the $\binom{5}{2}$ and $\binom{5}{3}$ ways of choosing boundary regions.

Taking the sum of the $S$-basis entropy vector we find
\be
\sum \mathcal{S}^4= 8\sum K_2 + 20 \sum K_4
\ee
so that assigning to each $K$ a species of perfect tensor hyperthreads our objective is given by
\be
8\mu(H_2)+20\mu(H_4).
\ee
As such, we should consider the program
\be
        \left\{
        \begin{aligned}
            & & & \max 8\mu(H_2)+20\mu(H_4) \\
            & \text{s.t.} & & \forall x\in \Sigma,\; \int_{H}d\mu(h)\Delta(x,h)\leq 1.
        \end{aligned}
    \right.  
\ee

\begin{figure}[H]
\centering
\includegraphics[width=.5\textwidth,page=33]{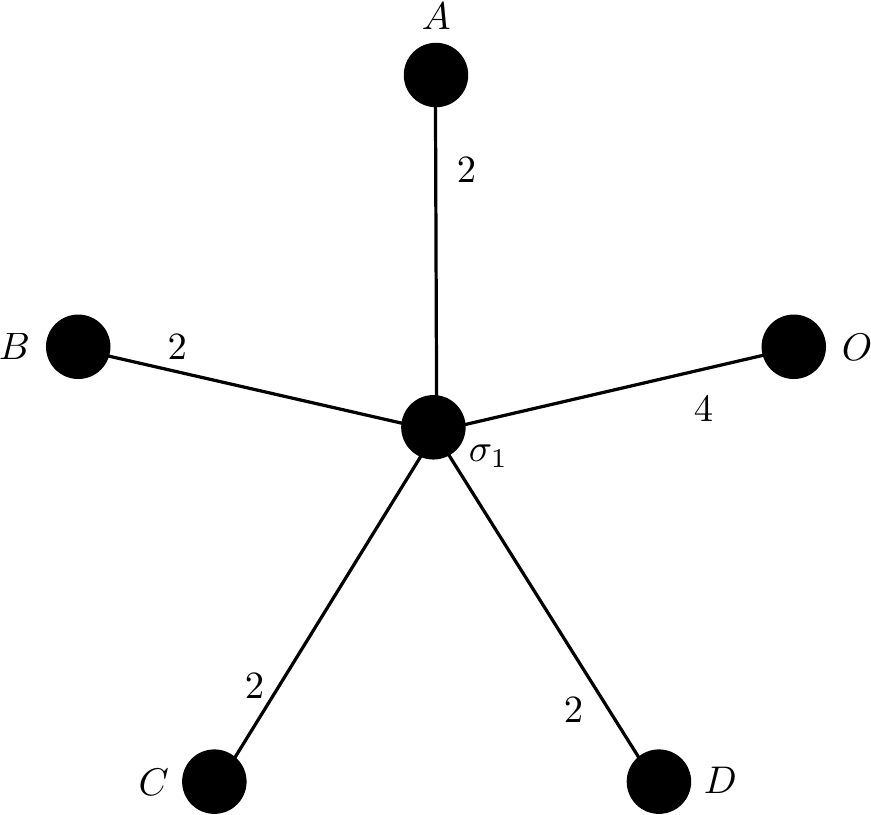}
\caption{\label{fig:4example}A graph with four boundary vertices $A,B,C,D$ and purifier $O$ along with one internal vertices $\sigma_1$. Edge capacities are labeled for each edge.
}
\end{figure}

We are immediately presented with a problem caused by the form of the inequalities of the holographic entropy cone. We are no longer guaranteed that all of the $K$s will be explicitly positive. Instead MMI only guaranteed that the sum of any two $K_4$s will be. This is highlighted by the following example: we consider the graph shown in figure \ref{fig:4example}, which has the entropy vector in the $S$ and $K$-basis
\be
\mathcal{S}=2*\{1111;222222;3333;2\}
\ee
\be
\mathcal{K}=\{0000000000;-11111\}.
\ee
Note this has $K_{ABCD}=-1$ which is negative. As constructed our program is not capable of reproducing the necessary entropy vector with a maximal perfect tensor hyperthread configuration. To proceed it is necessary for us to adapt our program to allow for the existence of negative threads and also explicitly implement inequalities of the holographic entropy cone which constrain the relationships between different species. We tackle these each in turn.

\subsection{Intermission: Negative threads}
To illustrate the new technology of negative threads we first consider for simplicity the case of $N=1$. Here, the entropy cone is one dimensional given in the $S$ and $K$-basis by $S_A$ or $K_{AO}$.

We start with the program
\be
        \left\{
        \begin{aligned}
            & & & \max 2\mu(H) \\
            & \text{s.t.} & & \forall x\in \Sigma,\; \int_{H}d\mu(h)\Delta(x,h)\leq 1.
        \end{aligned}
    \right.  
\ee
where we take $H$ to be the set of positive and negative 2-threads between $A$ and its complement $O$. The Hahn decomposition theorem states that given such a signed measure it is always possible to decompose the set into a positive and negative part
\be
H = H^+ \cup H^-
\ee
on which we have separately the (non-signed) measures $\mu^+$ and $\mu^-$. The measure on $H$ is then given by
\be
        \left\{
        \begin{aligned}
            & & & \max \; 2\mu_{+}(H^+)-2\mu_-(H^-) \\
            & \text{s.t.} & & \forall x\in \Sigma,\; \int_{H^+}d\mu_+(h)\Delta(x,h)-\int_{H^-}d\mu_-(h)\Delta(x,h)\leq 1.
        \end{aligned}
    \right.  
\ee
Since generically a positive and negative 2-thread can be used to cancel one another what we are really interested in is maximal thread configurations which contain the fewest threads possible. That is there are no ``extra" negative threads. The total number of threads for a given measure $\mu$ can be expressed as
\be
N=\mu^+(H^+)+\mu^-(H^-)
\ee
which can be used to perturb the program towards a solution with the fewest total threads (i.e. max $-N$). Using this we have
\be
        \left\{
        \begin{aligned}
            & & & \max \; (2-\epsilon)\mu_+(H^+)-(2+\epsilon)\mu_-(H^-) \\
            & \text{s.t.} & & \forall x\in \Sigma,\; \int_{H^+}d\mu_+(h)\Delta(x,h)-\int_{H^-}d\mu_-(h)\Delta(x,h)\leq 1
        \end{aligned}
    \right.  
\ee
which can be dualized. Imposing the constraints we have the Lagrangian
\be
\begin{split}
L(d\mu,d\nu)=&(2-\epsilon)\int_{H^+}d\mu_+(h)-(2+\epsilon)\int_{H^-}d\mu_-(h)\\
-&\int_{\Sigma}d\nu(x)\left(\int_{H^+}\left(d\mu_+(h)\Delta(x,h)\right)-\int_{H^-}\left(d\mu_-(h)\Delta(x,h)\right)- 1\right)\\
=&\int_{\Sigma}d\nu(x)+\int_{H^+}d\mu_+(h) \left((2-\epsilon)-\int_{\Sigma}d\nu(x)\Delta(x,h)\right)\\
+&\int_{H^-}d\mu_-(h)\left(-(2+\epsilon)+\int_{\Sigma}d\nu(x)\Delta(x,h)\right)
\end{split}
\ee
from which we derive the dual minimization program
\be\label{eq:epsilonmin}
        \left\{
        \begin{aligned}
            & & & \min \nu(\Sigma) \\
            & \text{s.t.} & & \forall h\in H^+,\; \int_{\Sigma}d\nu(x)\Delta(x,h) \geq 2-\epsilon,\\
             & \text{and} & & \forall h\in H^-,\; \int_{\Sigma}d\nu(x)\Delta(x,h) \leq 2+\epsilon.
        \end{aligned}
    \right.  
\ee
   
\noindent That is positive threads must cross a barrier of \emph{at least} $2-\epsilon$ while the negative threads can cross a barrier \emph{at most} $2+\epsilon$. The optimal configuration is given by a barrier of $2-\epsilon$ on $m_A$. From this we can conclude from theorem \ref{T3} that no negative threads will contribute. 

This can be generalized easily: The barrier condition for all positive threads is decreased by $\epsilon$ while that for negative threads is increased by $\epsilon$.

\begin{figure}[H]
\centering
\includegraphics[width=.4\textwidth,page=4]{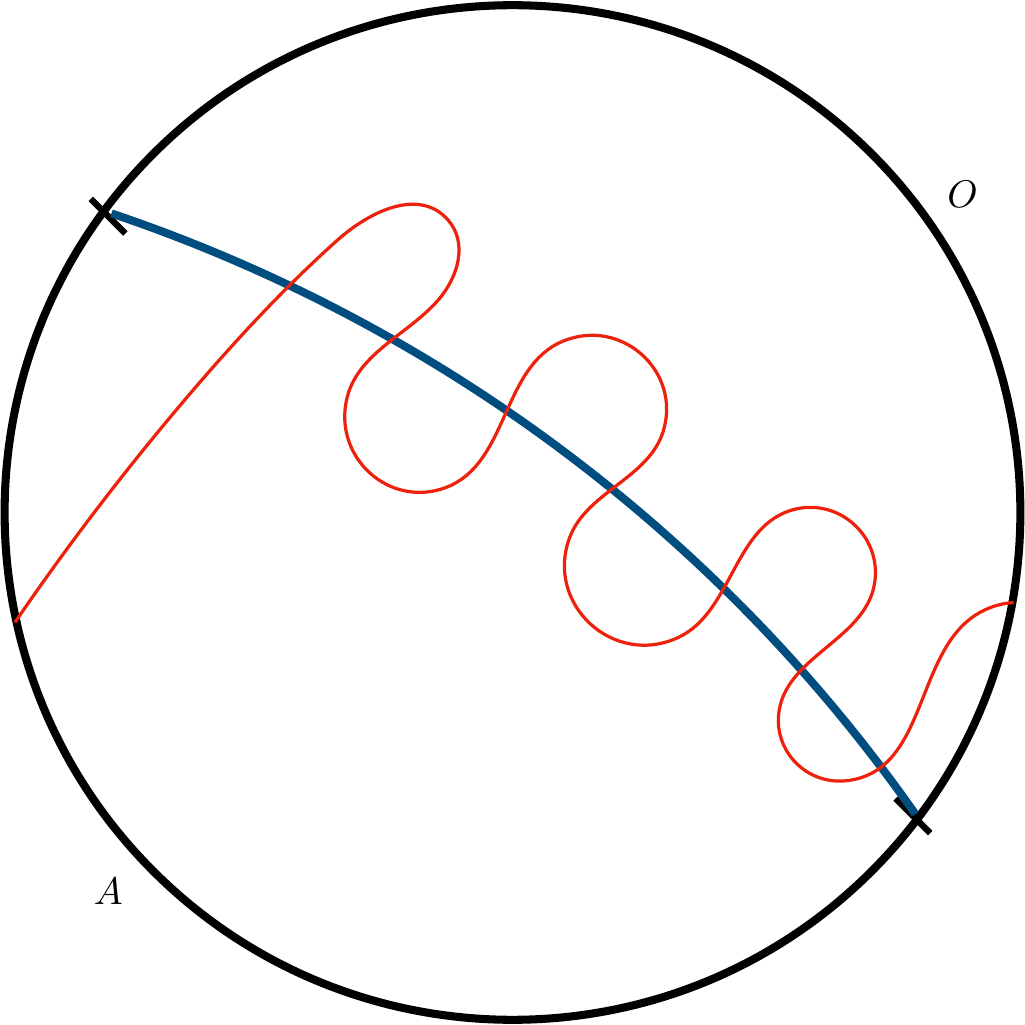}
\caption{\label{fig:setupneg} Given a potential  barrier it is always possible to construct a negative thread which crosses it an arbitrary number of times. Since the minimization program puts an upper bound on the barrier a negative thread can cross it is necessary to restrict the space of negative threads. This can be done for example by limiting the multiplicity of threads across RT surfaces. For this paper we will only work with negative threads on graphs. The figure here is shown on a manifold to emphasize that the negative thread crosses $m_A$ many times and as such without this restriction placing the barrier on $m_A$ would not be a feasible.}
\end{figure}

With an upper bound imposed on the space of negative threads, it is possible to have threads which will cross a chosen barrier any number of times (see figure \ref{fig:setupneg})\footnote{Moving forward positive threads will always be shown in blue while negative threads will be shown in red.}. For the positive threads this is not an issue since the lower bound and theorem \ref{T3} imply that such threads will not contribute. As such, in order to construct a sensible program it is necessary for us to limit the types of negative threads which can appear. This could potentially be done in many different ways, but here we present one resolution which uses the RT surfaces to define a set of threads whose path is the most direct.

Starting with a partition of the boundary into regions we consider the set of RT surfaces. A perfect tensor hyperthread is said to be \emph{straight} if each strand crosses these surfaces with multiplicity $0$ or $1$. For the rest of this paper we will always work with the set of straight perfect tensor hyperthread and denote it simply as $H$\footnote{Ideally, with more work the set of allowed negative perfect tensor hyperthread could be defined without reference to the $RT$ surfaces. For the graph examples we consider in this paper this can be done as it is enough for us to restrict the space of perfect tensor hyperthread to those for which each strand crosses every edge with multiplicity 0 or 1.}.

\subsection{Adding entropy inequality constraints}
Suppose we demand that our program additionally satisfy a constraint of the form
\be
\sum_J\beta_J\mu_J\geq 0
\ee
where we take $\beta\geq0$. We think of this as a potential inequality of the holographic entropy cone. For simplicity we start with $N=1$ and impose the constraint $\mu(H)\geq0$. Then the program becomes

\be
        \left\{
        \begin{aligned}
            & & & \max \; \mu_{+}(H^+)-\mu_-(H^-) \\
            & \text{s.t.} & & \forall x\in \Sigma,\; \int_{H^+}d\mu_+(h)\Delta(x,h)-\int_{H^-}d\mu_-(h)\Delta(x,h)\leq 1\\
           &\text{and} & &  \mu(H)\geq0.
        \end{aligned}
    \right.  
\ee
We can dualize as

\be
\begin{split}
&\int_{H^+}d\mu_+(h)-\int_{H^-}d\mu_-(h)-\int_{\Sigma}d\nu(x)\left(\int_{H^+}\left(d\mu_+(h)\Delta(x,h)\right)-\int_{H^-}\left(d\mu_-(h)\Delta(x,h)\right)- 1\right)\\-&\int_{\Sigma}d\alpha(x)\left(-\int_{H^+}d\mu_+(h)+\int_{H^-}d\mu_-(h)\right)
\end{split}
\ee
where we have introduced the new Lagrange multiplier $\alpha$. This can be rewritten as

\be
\begin{split}
\int_{\Sigma}d\nu(x)+\int_{H^+}d\mu_+(h) \left(1-\int_{\Sigma}d\nu(x)\Delta(x,h)+\alpha\right)+\int_{H^-}d\mu_-(h)\left(-1+\int_{\Sigma}d\nu(x)\Delta(x,h)-\alpha\right)
\end{split}
\ee
which gives the minimization program
\be
        \left\{
        \begin{aligned}
            & & & \min_{\nu,\alpha} \nu(\Sigma) \\
            & \text{s.t.} & & \forall h\in H^+,\; \int_{\Sigma}d\nu(x)\Delta(x,h) \geq 1+\alpha\\
             & \text{and} & & \forall h\in H^-,\; \int_{\Sigma}d\nu(x)\Delta(x,h) \leq 1+\alpha
        \end{aligned}
    \right.  
\ee
so the barrier requirement can be raised a positive number. In general the barrier for positive and negative threads $H_{I}$ will be increased by $\beta_{I}$ from the imposed constraint. Each such constraint imposed introduces a new measure on $\Sigma$ which can independently raise the required barrier of a set of species.

Since positive and negative threads are paired, for any path there will be both a corresponding positive and negative thread, it is necessary for both constraints to be satisfied at the same time. As such this program can be more simply written using equality constraints\footnote{If we wish we can always use the same trick as \eqref{eq:epsilonmin} to obtain an optimal configuration with the minimum number of threads and to distinguish between the positive and negative threads. For simplicity and notational clarity we omit such a step in what follows.}
\be
        \left\{
        \begin{aligned}
            & & & \min_{\nu,\alpha} \nu(\Sigma) \\
            & \text{s.t.} & & \forall h\in H,\; \int_{\Sigma}d\nu(x)\Delta(x,h) = 1+\alpha.
        \end{aligned}
    \right.  
\ee

Note that in the dualization the entropy inequality $-Q_i\leq0$ is imposed by the Lagrange multiplier $\alpha_i\geq0$. As such CS applied to these variables leads us to the following observation:
\begin{theorem}\label{thm:QCS}
An entropy inequality $Q_i$ affects the optimization of a configuration of perfect tensor hyperthreads only if the entropy vector is such that $Q_i=0$ i.e. it is saturated.
\end{theorem}
\noindent As a consequence entropy vectors corresponding to extremal rays are of particular interest as they have the maximum number of saturated entropy inequalities. These states have optimal perfect tensor hyperthread configurations which have particularly exacting constraints between the positive and negative threads of different species. Extremal rays thus serve as a particularly useful testing ground for constructing optimal perfect tensor hyperthread configurations and searching for potential counter examples to the locking of the full entropy vector.

Using what we have learned let us generalize and consider the case for $N=2$. We choose two boundary regions $A,B$ along with the purifier $O$. The $K$-basis is given by 3 $PT_2$s: $K_{AB},K_{AO},K_{BO}$. First we consider the space of both positive and negative 2-threads and consider the program \emph{without} imposing the holographic entropy cone inequalities $K_{AB},K_{AO},K_{BO}\geq0$

\be\label{eq:negmax}
        \left\{
        \begin{aligned}
            & & & \max 2\mu(H) \\
            & \text{s.t.} & & \forall x\in \Sigma,\; \int_{H}d\mu(h)\Delta(x,h)\leq 1
        \end{aligned}
    \right.  
\ee
which is dual to a barrier configuration where positive threads cross at least 2 and negative threads cross at most 2. Simplifying we have 
\be
        \left\{
        \begin{aligned}
            & & & \min \nu(\Sigma) \\
            & \text{s.t.} & & \forall h\in H,\; \int_{\Sigma}d\nu(x)\Delta(x,h) = 2
        \end{aligned}
    \right.  
\ee

so that every thread must cross a barrier of exactly 2.
\begin{figure}[H]
\centering
\begin{tabular}{cc}
\centering
\includegraphics[width=.3\textwidth,page=102]{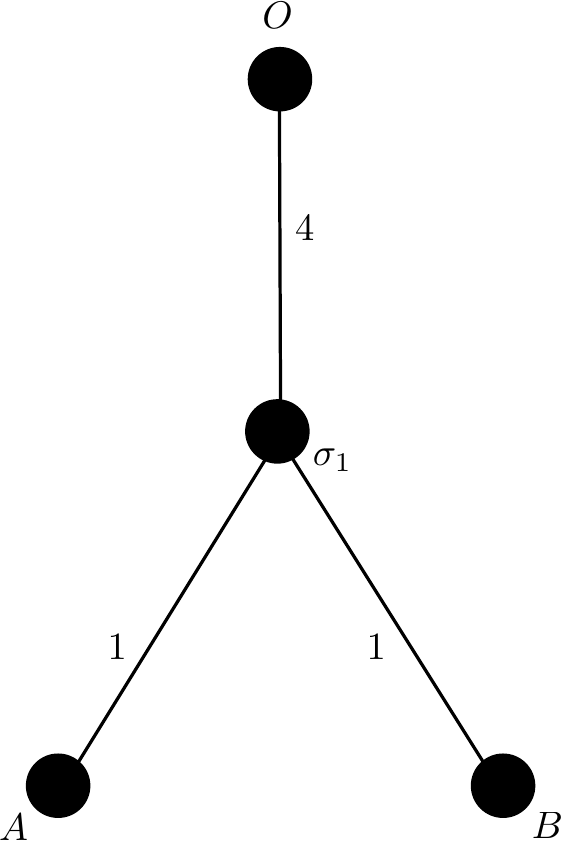}&
\includegraphics[width=.3\textwidth,page=103]{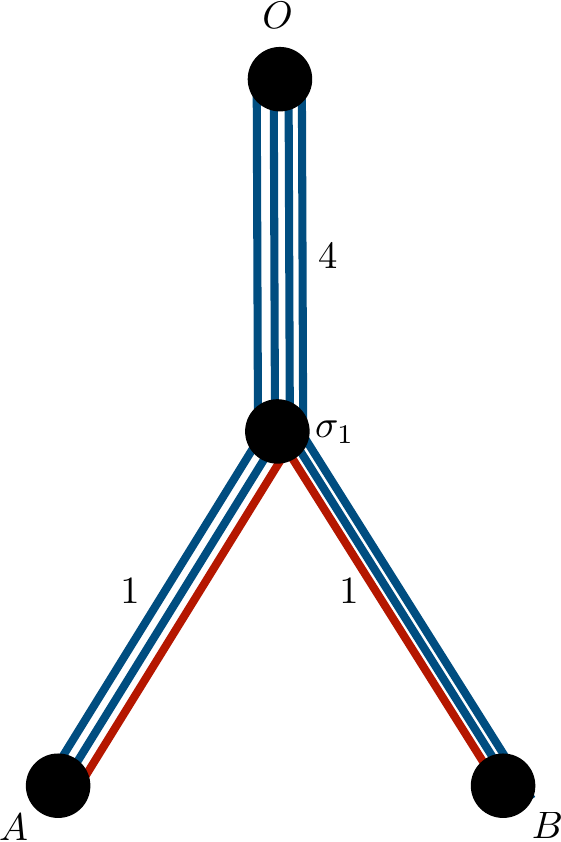}
\end{tabular}
\caption{\label{fig:n2} L: We consider the following graph which has entropy vector $\mathcal{S}^2_g=\{1,1;2\}$ or $\mathcal{K}_g^2=\{0,1,1\}$. R: The optimal thread configuration of \eqref{eq:negmax} consists of a single negative 2-thread between $A$ and $B$ as well as four positive 2-threads which obtains an objective value of $(4-1)*2=6$. The barrier configuration is given by a barrier of one on each edge for a value of $4*1+1*2=6$. This is constructed so that every 2-thread crosses a barrier of \emph{exactly} 2. The entropy vector of the thread configuration is $\mathcal{S}^2_\mu=\{1,1;4\}$ or $\mathcal{K}_\mu^2=\{-1,2,2\}$ which is not the same as the graph. In fact, it is not a valid entropy vector of the holographic entropy cone as it explicitly breaks subadditivity.}
\end{figure}

\begin{figure}[H]
\centering
\includegraphics[width=.3\textwidth,page=104]{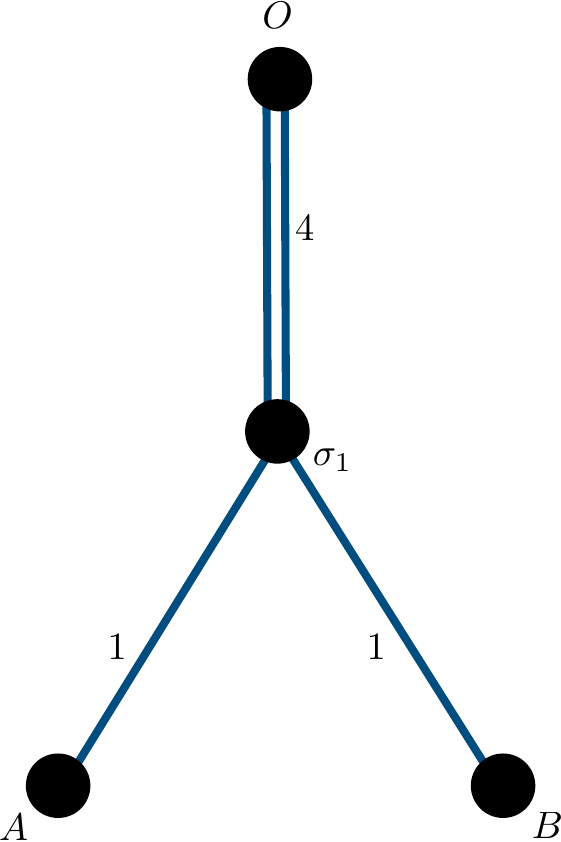}
\caption{\label{fig:withei} The maximal thread configuration of \eqref{eq:negmaxc} for the graph $g$ has an entropy vector $\mathcal{S}^2_\mu=\{1,1;2\}$ or $\mathcal{K}_\mu^2=\{0,1,1\}$ which matches that of the graph and obtains a value of $2*2=4$. The barrier configuration is given by a barrier of two on the edges $e_{A\sigma_1}$ and $e_{B\sigma_1}$ for a value of $2*2=4$. Here $\alpha_{2}=\alpha_{3}=0$ while $\alpha_{1}=2$ so that each $AB$ 2-thread is required to cross an increased barrier of exactly 4. This is consistent as the only saturated entropy inequality is $K_{AB}=0$ so that by theorem \ref{thm:QCS} only $\alpha_{1}$ is nonzero.}
\end{figure}

\noindent We are immediately confronted with a problem as \emph{this program without imposing constraints does not know about subadditivity} (see figure \ref{fig:n2}). As such in general it will not reproduce the correct entropies.

Now, we explicitly impose the three entropy constraints
\be\label{eq:negmaxc}
        \left\{
        \begin{aligned}
            & & & \max 2\mu(H) \\
            & \text{s.t.} & & \forall x\in \Sigma,\; \int_{H}d\mu(h)\Delta(x,h)\leq 1\\
            & \text{and} & & \mu_{AB}\geq 0\\
                & \text{and} & & \mu_{AO}\geq 0\\
                    & \text{and} & & \mu_{BO}\geq 0\\
        \end{aligned}
    \right.  
\ee
which is dual to 
\be
        \left\{
        \begin{aligned}
            & & & \min \nu(\Sigma) \\
            & \text{s.t.} & & \forall h\in H_{A:B},\; \int_{\Sigma}d\nu(x)\Delta(x,h) = 2+\alpha_1\\
             & \text{and} & & \forall h\in H_{A:O},\; \int_{\Sigma}d\nu(x)\Delta(x,h) = 2+\alpha_2\\
              & \text{and} & & \forall h\in H_{B:O},\; \int_{\Sigma}d\nu(x)\Delta(x,h) = 2+\alpha_3.
        \end{aligned}
    \right.  
\ee
As a result optimal thread and barrier configuration will obey the entropy inequalities (see figure \ref{fig:withei}).

\subsection{4 regions revisited}
Equipped with knowledge of how to include negative threads we are now prepared to tackle the case of $N=4$ we return to our example figure \ref{fig:4ex}:

\begin{figure}[H]
\centering
\includegraphics[width=.4\textwidth,page=33]{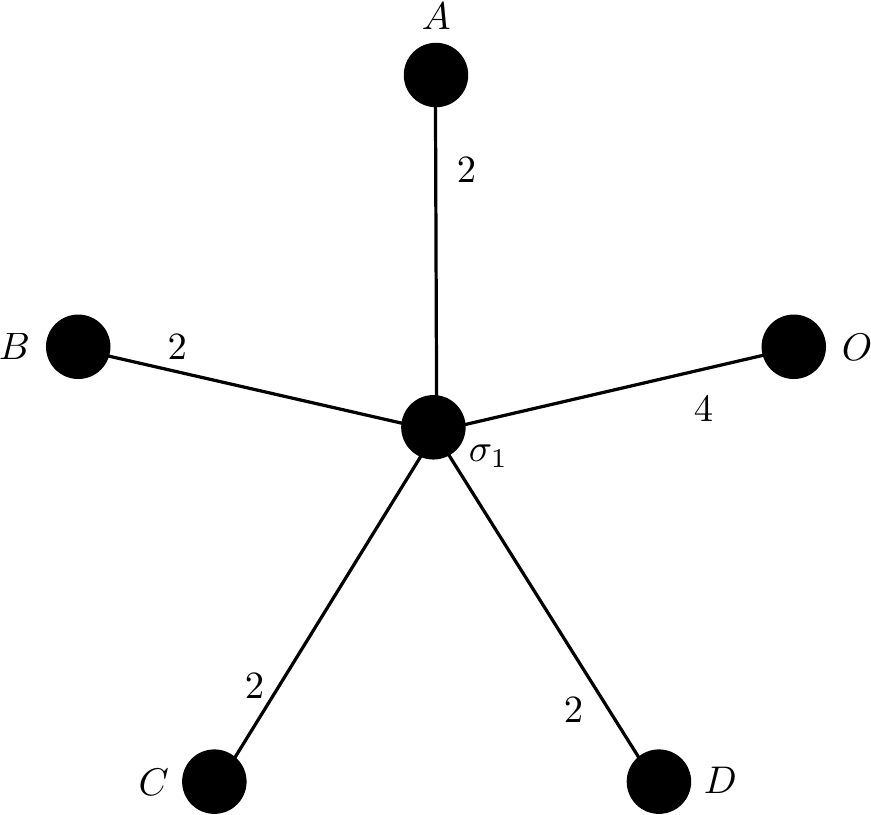}
\caption{\label{fig:4ex}A graph with four boundary vertices $A,B,C,D$ and purifier $O$ along with one internal vertices $\sigma_1$. Edge capacities are labeled for each edge.}
\end{figure}
which has $S$ and $K$-basis vectors
\be
\mathcal{S}=2*\{1111;222222;3333;2\}
\ee
\be
\mathcal{K}=\{0000000000;-11111\}.
\ee
We now consider the program
\be
        \left\{
        \begin{aligned}
            & & & \max \sum\mathcal{S}^4(\mu) \\
            & \text{s.t.} & & \forall x\in \Sigma,\; \int_{H}d\mu(h)\Delta(x,h)\leq 1,\\
             & \text{and} & & \forall_i Q^4_i(\mu)\geq 0
        \end{aligned}
    \right.  
\ee
where $H$ is the full space of both positive and negative perfect tensor hyperthreads and we explicitly implement the holographic entropy cone inequalities $Q^4_i\geq0$ as constraints in the program. We proceed by explicitly constructing a maximal configuration of perfect tensor hyperthreads as well as a minimal barrier configuration. Together these show optimality of the chosen configurations.

\paragraph{Maximizing}
Using knowledge of the chosen graph we start with some simplifications. First, because of subadditivity we choose to implicitly assume $\mu_-(H_2)=0$ that is there are no negative 2-threads. Next, making use of theorem \ref{thm:QCS} we only explicitly implement those entropy constraints which are saturated for this state. Besides subadditivity these correspond to four instances of MMI:
\be
\begin{split}
\mu_{ABCD}+\mu_{ABCO}\geq 0\\
\mu_{ABCD}+\mu_{ABDO}\geq 0 \\
\mu_{ABCD}+\mu_{ACDO}\geq 0 \\
\mu_{ABCD}+\mu_{BCDO}\geq 0 
\end{split}
\ee
as such we are interested in the maximization program
\be
        \left\{
        \begin{aligned}
            & & & \max 8\mu(H_2)+20\mu(H_4) \\
            & \text{s.t.} & & \forall x\in \Sigma,\; \int_{H}d\mu(h)\Delta(x,h)\leq 1,\\
             & \text{and} & & \mu_{ABCD}+\mu_{ABCO}\geq 0\\
             & \text{and} & & \mu_{ABCD}+\mu_{ABDO}\geq 0 \\
             & \text{and} & & \mu_{ABCD}+\mu_{ACDO}\geq 0 \\
             & \text{and} & & \mu_{ABCD}+\mu_{BCDO}\geq 0. 
        \end{aligned}
    \right.  
\ee
Our task now is to explicitly construct a perfect tensor hyperthread configuration satisfying the constraints which we believe will be maximal. To do so we use the $K$-basis entropy vector as an ansataz for the number of each species we expect. Note, that if the maximal configuration has the same numbers of each thread species as the corresponding $K$, then it necessarily satisfies all of the entropy inequalities (since the vector in the $K$-basis \emph{is} a valid entropy vector).

We choose the following configuration which consists of 5 total threads:

\begin{figure}[H]
\centering
\begin{tabular}{cc}
\centering
\includegraphics[width=.4\textwidth,page=34]{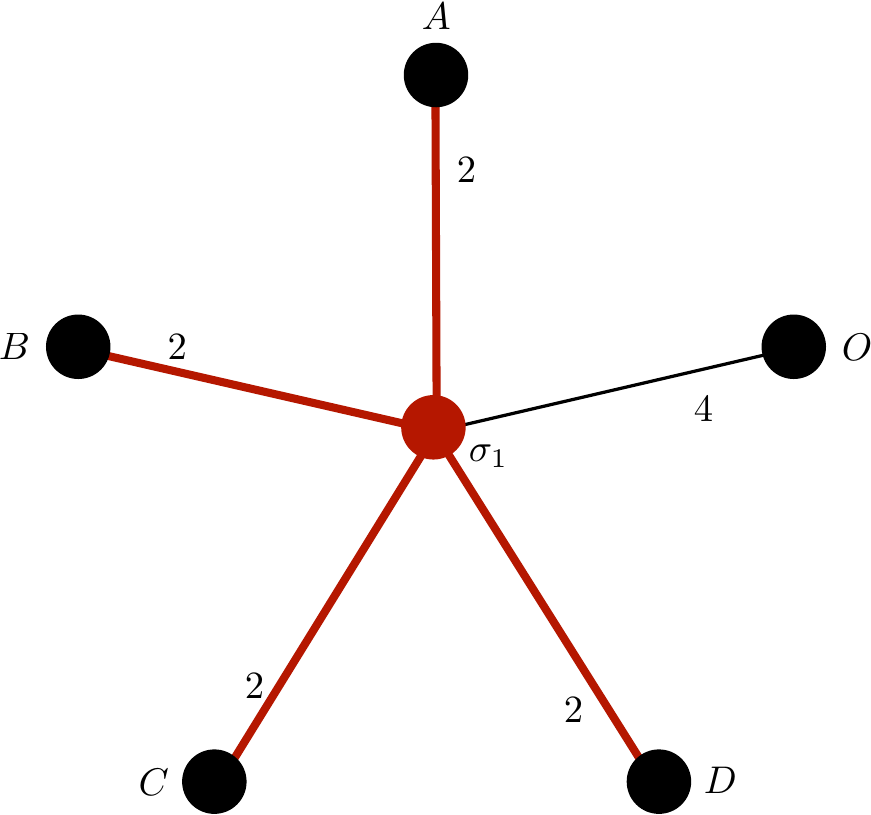}&
\includegraphics[width=.4\textwidth,page=35]{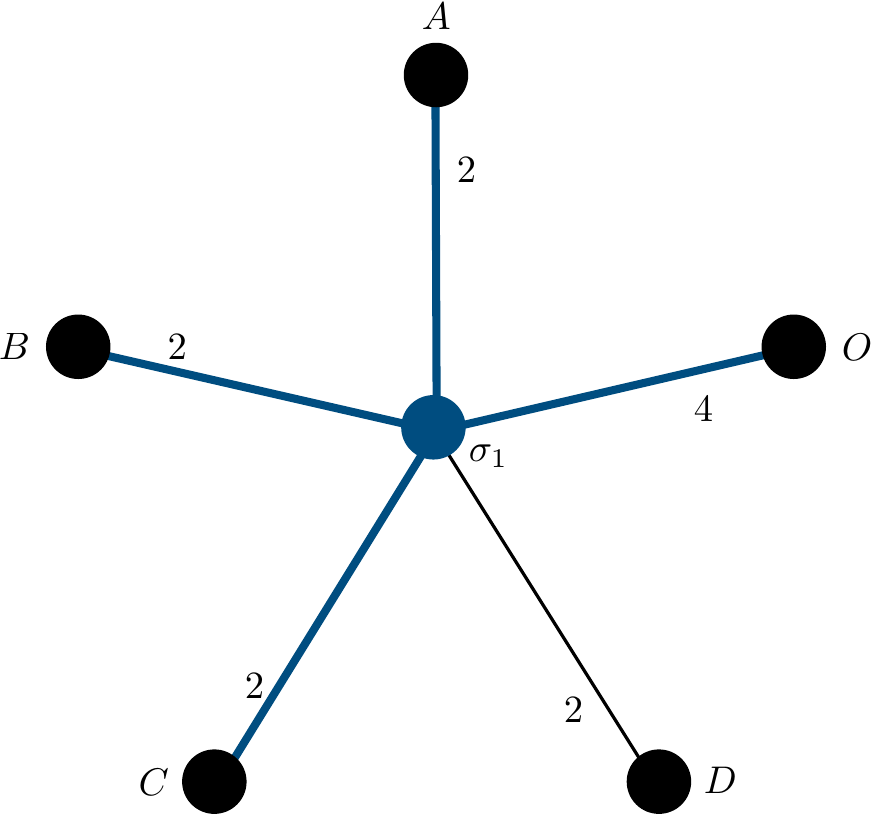}\\
$h_1$ & $h_2$ \\[6pt]
\end{tabular}
\caption{The thread configuration consists of two classes of threads $h_1$ and $h_2$ which are collections of thread species which respect a symmetry of the graph. $h_1$ consists of a single negative thread between $ABCD$. $h_2$ consists four positive 4-threads one between each of $ABCO,ABDO,ACDO$ and $BCDO$. We write this collection of regions more succinctly as $(ABCD)_3O$. Shown is the $ABCO$ 4-thread as a representative of the class.}
\end{figure}

\noindent The important information about this configuration can be summarized in the following table:
\begin{table}[H]
\begin{center}
\scriptsize
\begin{tabular}{| c |c | c | c | c | c | c | }
\hline
Type & Target & Contributing Species & Vertex  & Example hyperthread & $e_{\sigma_1 A}$ & $e_{\sigma_1 O}$  \\ 
\hline
\hline
$h_1$ & -1 & $ABCD$ & $\sigma_1$ &$\{\sigma_1;A;B;C;D\}$ & 1 & 0 \\
\hline
$h_2$ & 1 & $(ABCD)_3O$ &$\sigma_1$ &$\{\sigma_1;A;B;C;O\}$ & 3 & 4\\
\hline
\end{tabular}
\end{center}
\label{tab:ray3_intext}
\end{table}
\noindent Each class of perfect tensor hyperthreads lists the contributing species to that class as well as the target number of perfect tensor hyperthreads \emph{for each contributing species}. We use the notation $(A_1,A_2,\cdots)_k$ to mean ``Any combination of $k$ elements". Here the class $h_1$ consists of a single species while the class $h_2$ consists of four:
\be
\begin{split}
 h_1: \quad &ABCD\\
h_2: \quad &(ABCD)_3O=\{ABCO,ABDO,ACDO,BCDO\}
\end{split}
\ee
 Also shown is an example of a perfect tensor hyperthread for that class. Lastly, the total number of times a collection of one perfect tensor hyperthread from each species of the class crosses each edge are listed. Due to the symmetry of the graph a number of edges are suppressed as these will give rise to identical constraints. Using this information we can explicitly verify that this perfect tensor hyperthread configuration satisfies all of the density bound constraints and in fact uses up all available space in the graph if we place a number of perfect tensor hyperthreads equal to the target of each species
\be
\begin{split}
 e_{\sigma_1 A}: \quad &h_1+3h_2=2\\
 e_{\sigma_1 O}: \quad &4h_2=4\\
&h_1 \rightarrow -1, \; h_2 \rightarrow 1.
\end{split}
\ee
As such for this chosen configuration the objective obtains a value of 
\be
(4-1)*20=60
\ee
which is equal to the sum of the entries of the $S$-basis entropy vector. Necessarily then this configuration locks the full entropy vector as each individual entanglement entropy is correctly calculated by the counting of the appropriate species of perfect tensor hyperthread using \eqref{eq:LT4} and the identification $\mu^*_I=K_I$.

\paragraph{Minimizing}
Dualizing, we have the minimization program
\be
        \left\{
        \begin{aligned}
            & & & \min \nu(x) \\
            & \text{s.t.} & & \forall h\in H_2,\; \int_{\Sigma}d\nu(x)\Delta(x,h) \geq 8,\\
             & \text{and} & & \forall h\in H_{A:B:C:O},\; \int_{\Sigma}d\nu(x)\Delta(x,h) = 20+\alpha_1\\
             &\text{and} & & \forall h\in H_{A:B:D:O},\; \int_{\Sigma}d\nu(x)\Delta(x,h) = 20+\alpha_2 \\
             &\text{and} & & \forall h\in H_{A:C:D:O},\; \int_{\Sigma}d\nu(x)\Delta(x,h) = 20+\alpha_3 \\
             &\text{and} & & \forall h\in H_{B:C:D:O},\; \int_{\Sigma}d\nu(x)\Delta(x,h) = 20+\alpha_4\\
             &\text{and} & &\forall h\in H_{A:B:C:D},\; \int_{\Sigma}d\nu(x)\Delta(x,h) = 20+\alpha_1+\alpha_2+\alpha_3+\alpha_4
        \end{aligned}
    \right.  
\ee
where each $\alpha_i$ is the Lagrange multiplier which implemented one of the entropy inequalities. The symmetry of the particular entropy vector can be used to reduce the number of constraints as it is necessary that all of the $\alpha_i$s have the same value. Redefining this single parameter as $\alpha$ we have
\be
        \left\{
        \begin{aligned}
            & & & \min \nu(x) \\
            & \text{s.t.} & & \forall h\in H_2,\; \int_{\Sigma}d\nu(x)\Delta(x,h) \geq 8,\\
             & \text{and} & & \forall h\in H_{A:B:C:O},\; \int_{\Sigma}d\nu(x)\Delta(x,h) = 20+\alpha\\
             &\text{and} & & \forall h\in H_{A:B:D:O},\; \int_{\Sigma}d\nu(x)\Delta(x,h) = 20+\alpha \\
             &\text{and} & & \forall h\in H_{A:C:D:O},\; \int_{\Sigma}d\nu(x)\Delta(x,h) = 20+\alpha \\
             &\text{and} & & \forall h\in H_{B:C:D:O},\; \int_{\Sigma}d\nu(x)\Delta(x,h) = 20+\alpha\\
             &\text{and} & &\forall h\in H_{A:B:C:D},\; \int_{\Sigma}d\nu(x)\Delta(x,h) = 20+4\alpha.
        \end{aligned}
    \right.  
\ee
This can be further simplified by noting again because of the symmetry whatever barrier is placed on the edges connecting $A,B,C,D$ to the internal vertex $\sigma_1$ must be the same
\be
        \left\{
        \begin{aligned}
            & & & \min 8b_{A}+4b_O \\
             & \text{s.t.} & & 3b_A+b_O = 20+\alpha\\
             &\text{and} & & 4b_A = 20+4\alpha.
        \end{aligned}
    \right.  
\ee
Here $b_O$ is the barrier to be placed on $e_{\sigma_1 O}$ and because of symmetry $b_{A}$ will be the barrier placed on each of the other edges. This program can be explicitly evaluated to find
\be
60, \; b_A \rightarrow 6, \; b_O \rightarrow 3, \; \alpha \rightarrow 1
\ee
which agrees both with the sum of the $S$-basis entropy vector as well as the constructed perfect tensor hyperthread configuration. Together these show that the perfect tensor hyperthread and barrier configurations are optimal with the perfect tensor hyperthread configuration locking the full entropy vector. This whole procedure is summarized in figure \ref{fig:4_config}:

\begin{figure}[H]
\centering
\begin{tabular}{cc}
\centering
\includegraphics[page=7,width=.4\textwidth]{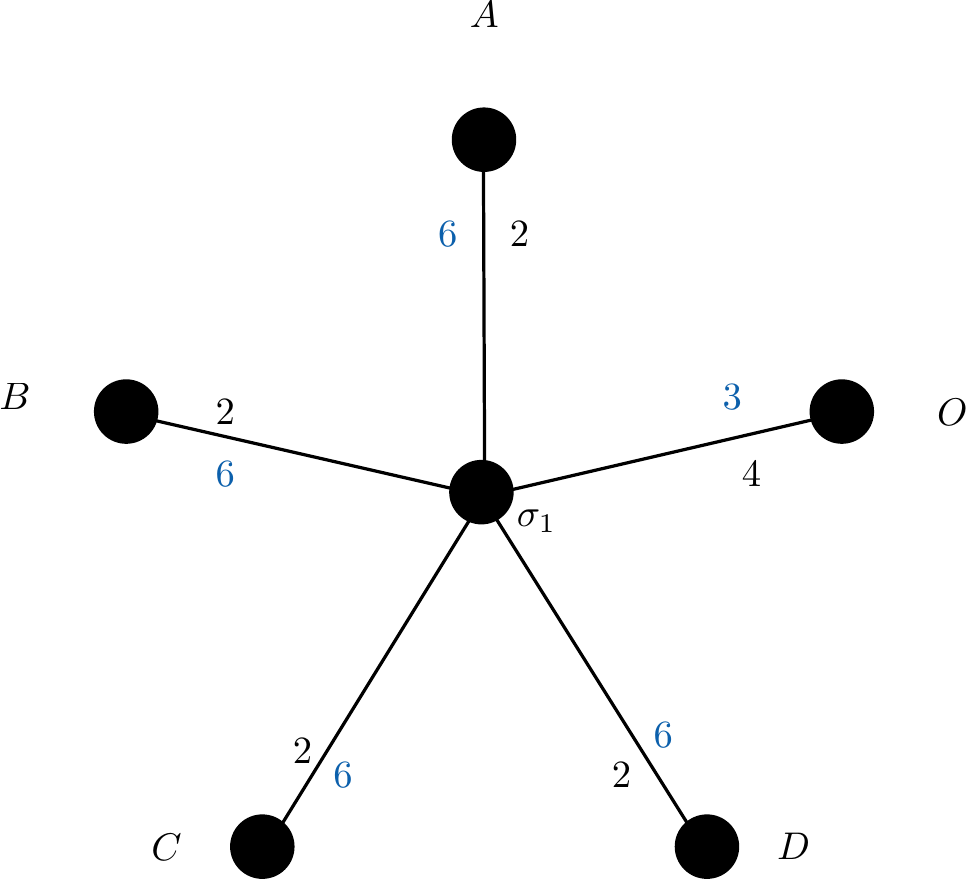}&
\includegraphics[page=13,width=.4\textwidth]{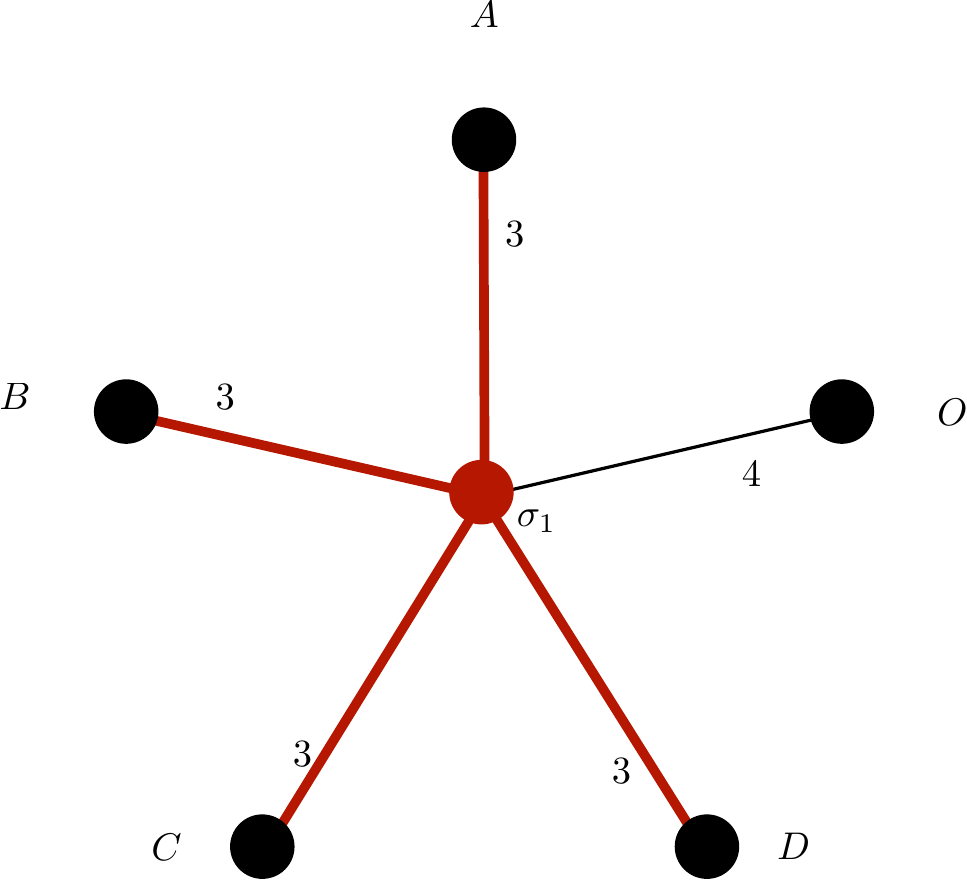}\\
\includegraphics[width=.4\textwidth,page=14]{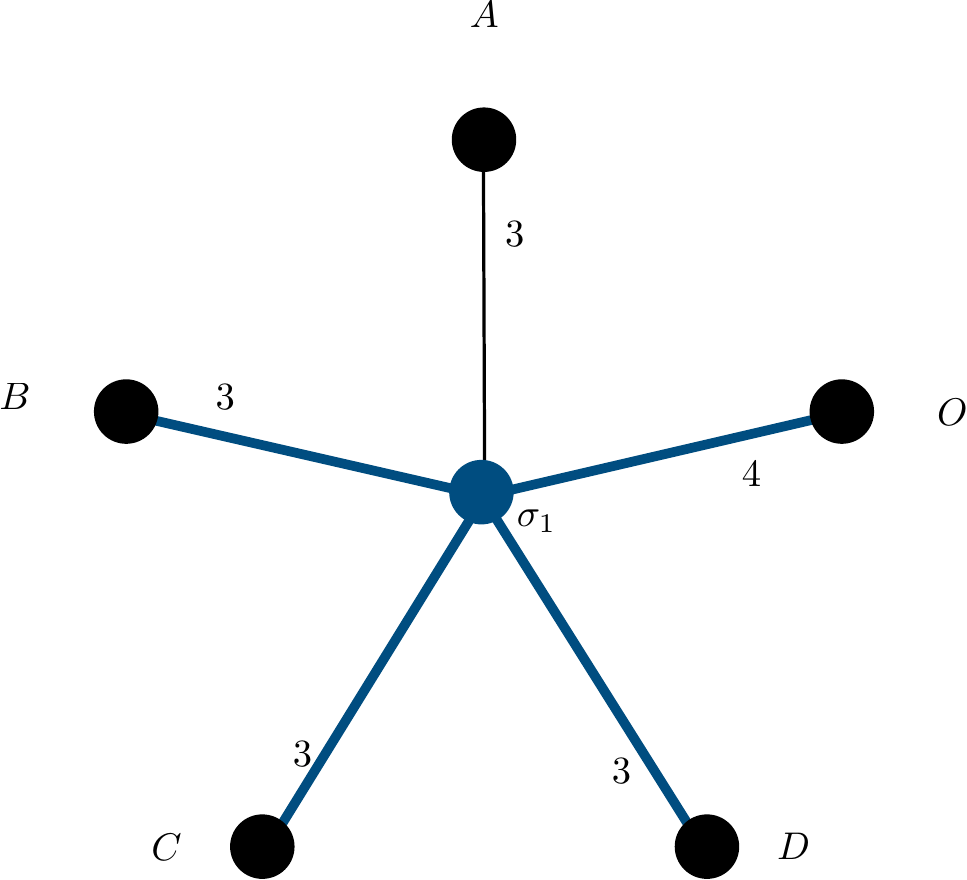}&\includegraphics[width=.4\textwidth,page=15]{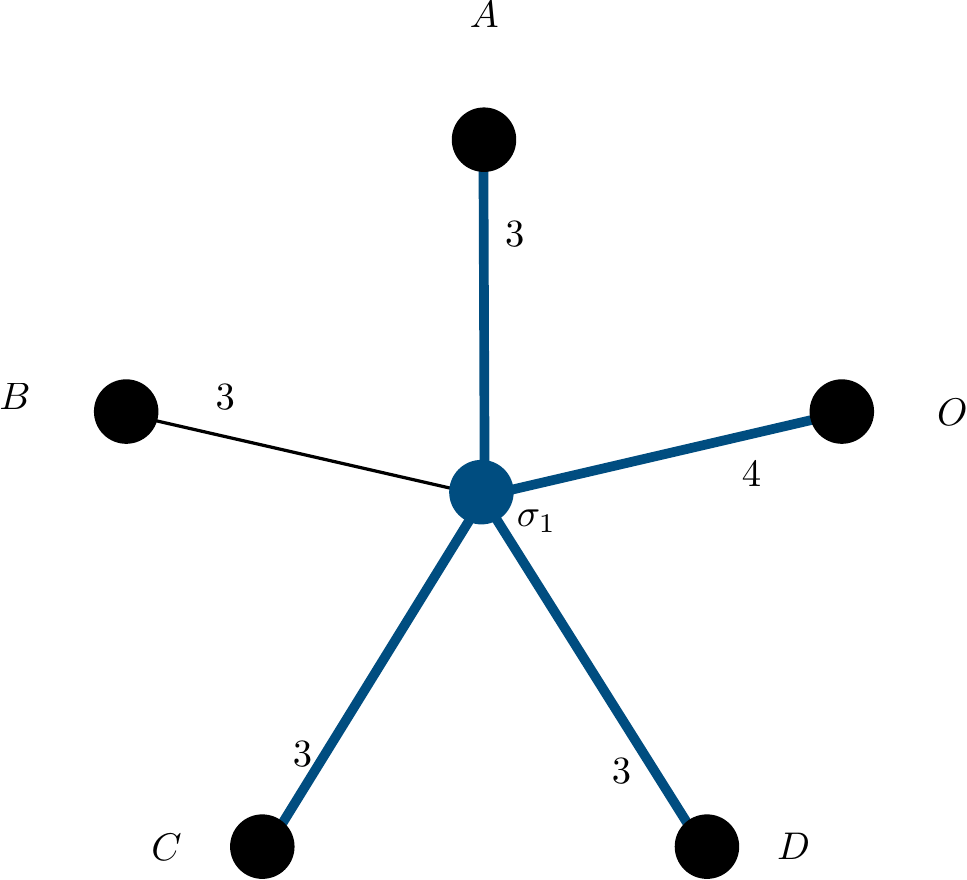}\\
\includegraphics[width=.4\textwidth,page=16]{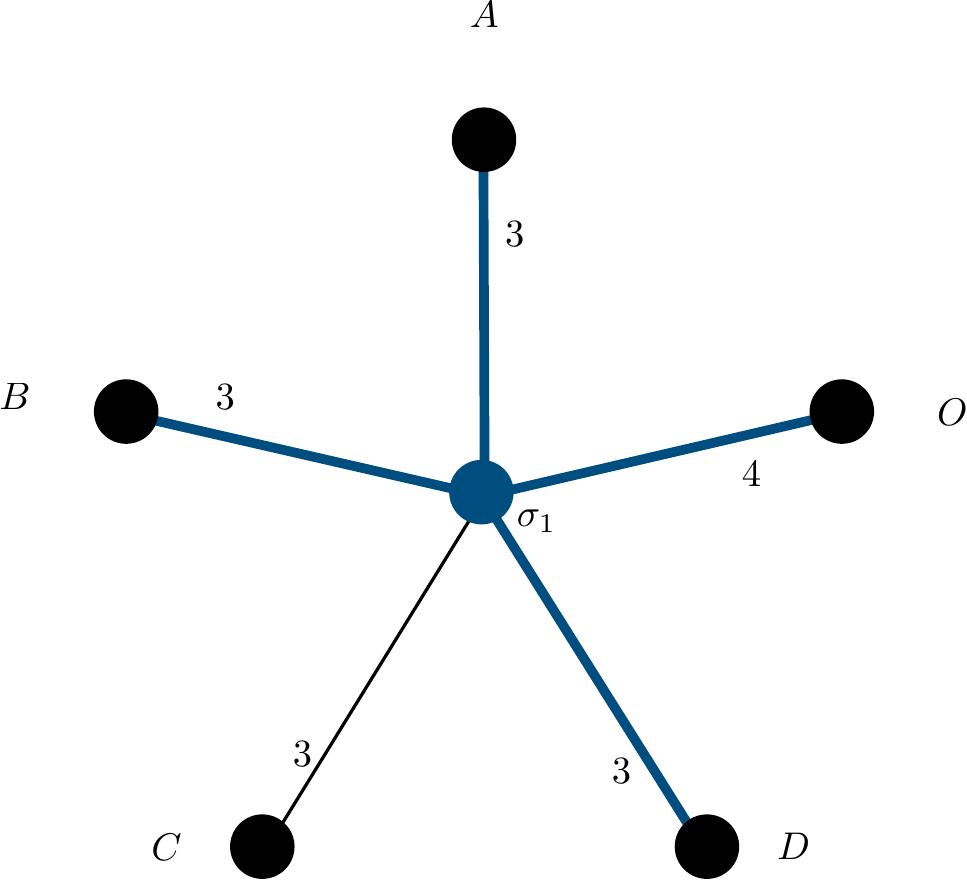}&\includegraphics[width=.4\textwidth,page=17]{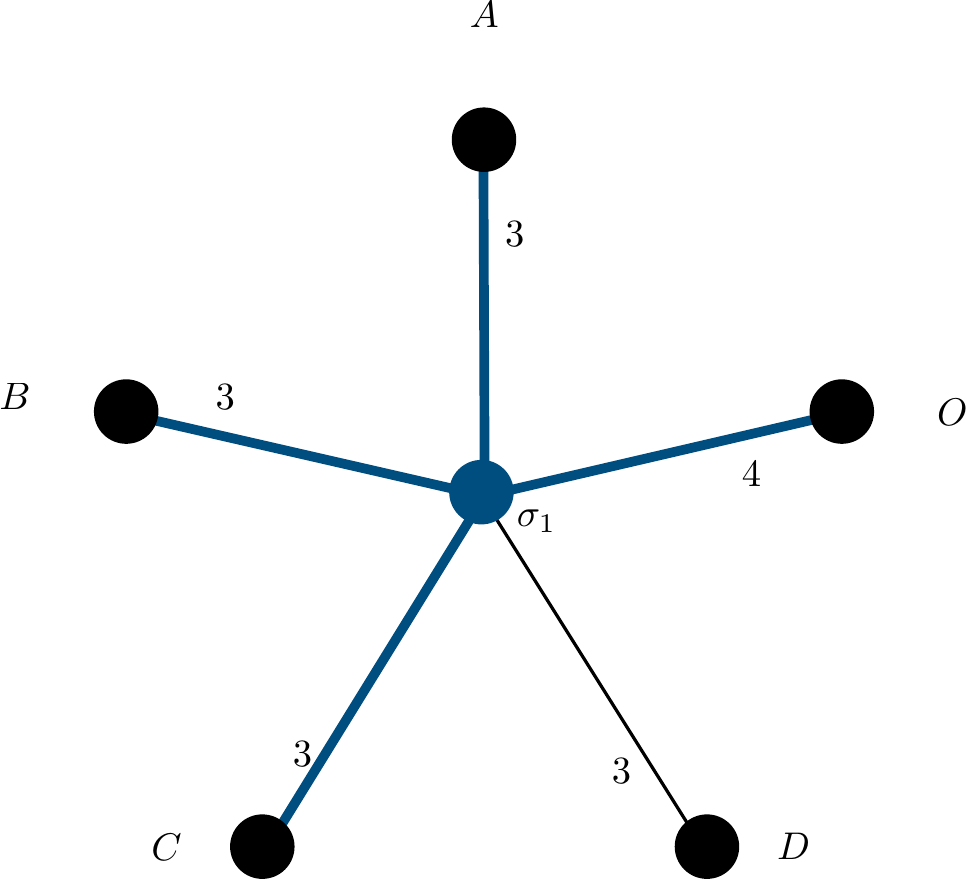}
\end{tabular}
\caption{\label{fig:4_config}TL: The edge capacities are in black while the optimal barrier configuration for each edge is in blue. This configuration corresponds to $\alpha=1$ such that each thread of $H_{(ABCD)_3O}$ crosses a barrier of exactly 21 while $H_{A:B:C:D}$ crosses a barrier of exactly 24. TR: The negative thread $h_2$ corresponding to $\mu^*_{ABCD}=-1$. Shown are the updated edge capacities corresponding to the extra space the positive threads have available in the geometry. The remaining diagrams show the four positive threads comprising $h_1$ corresponding to $\mu^*_{(ABCD)_3O}=1$. Their placement on the graph uses up all of the available capacity such that no additional threads can be placed.}
\end{figure}

\section{5 regions}\label{sec:N5}
We now consider five boundary regions $A,B,C,D,E$ with purifier $O$. The entropy space is 31 dimensional consisting of
\be
\begin{split}
\mathcal{S}= \{ &S_A, S_B, S_C, S_D, S_E ; \\
&S_{AB}, S_{AC}, S_{AD}, S_{AE}, S_{BC}, S_{BD}, S_{BE}, S_{CD}, S_{CE}, S_{DE} ;\\
&S_{ABC}, S_{ABD}, S_{ABE}, S_{ACD}, S_{ACE}, S_{ADE}, S_{BCD},
S_{BCE}, S_{BDE}, S_{CDE} ;\\
&S_{ABCD}, S_{ABCE}, S_{ABDE}, S_{ACDE}, S_{BCDE} ;\\
&S_{ABCDE} \}.
\end{split}
\ee
In the $K$-basis we have fifteen $PT_2$s and $PT_4$s as well as a new feature, a single $PT_6$
\be
\begin{split}
   \mathcal{K}=\{&K_{AB},K_{AC},K_{AD},K_{AE},K_{AO},K_{BC},K_{BD},K_{BE},K_{BO},K_{CD},K_{CE},K_{CO},K_{DE},K_{DO},K_{EO};\\
   &K_{ABCD},K_{ABCE},K_{ABCO},K_{ABDE},K_{ABDO},K_{ABEO},K_{ACDE},K_{ACDO},\\
   & K_{ACEO},K_{ADEO},K_{BCDE},K_{BCDO},K_{BCEO},K_{BDEO},K_{CDEO};\\
   &K_{ABCDEO}\}
\end{split}
\ee

with change of basis is given by
\be
\begin{split}
S_1&=\sum_{j=2}^6K_{1j}+\sum_{1<j<k<l}^6K_{1jkl}+K_{123456}\\
S_{12}&=\sum_{j=3}^6\left(K_{1j}+K_{2j}\right)+\sum_{2<j<k<l}^6\left(K_{1jkl}+K_{2jkl}\right)+2\sum_{2<j<k}^6K_{12jk}+2K_{123456}\\
S_{123}&=\sum_{j=4}^6\left(K_{1j}+K_{2j}+K_{3j}\right)+\sum_{3<j<k<l}^6\left(K_{1jkl}+K_{2jkl}+K_{3jkl}\right)\\
&+2\sum_{3<j<k}^6\left(K_{12jk}+K_{13jk}+K_{23jk}\right)+\sum_{j=4}^6K_{123j}+3K_{123456}.
\end{split}
\ee
Considering the sum of the $S$-basis entropy vector we are interested in the program
\be\label{eq:5max}
        \left\{
        \begin{aligned}
            & & & \max 16\mu(H_2)+40\mu(H_4)+66\mu(H_6) \\
            & \text{s.t.} & & \forall x\in \Sigma,\; \int_{H}d\mu(h)\Delta(x,h)\leq 1,\\
             & \text{and} & & \forall_i \;Q^5_i(\mu)\geq 0.
        \end{aligned}
    \right.  
\ee

\subsection*{Example}

\begin{figure}[H]
\centering
\includegraphics[width=.4\textwidth,page=22]{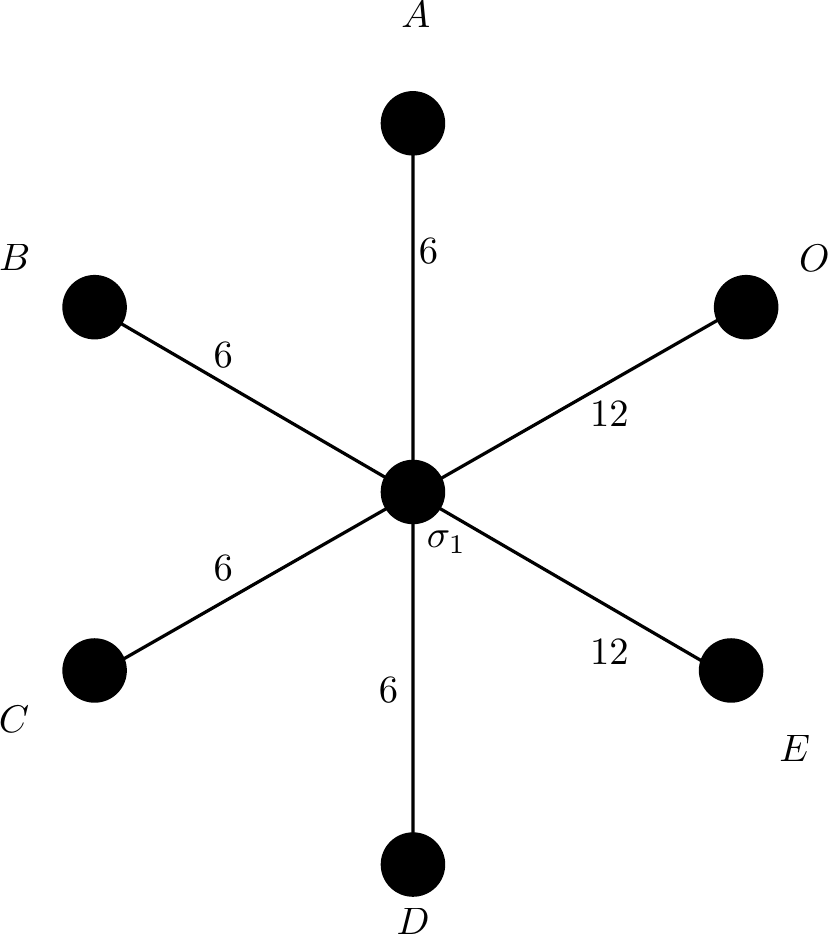}
\caption{\label{fig:ray6main}A graph with five boundary vertices $A,B,C,D,E$ and purifier $O$ along with one internal vertices $\sigma_1$. Edge capacities are labeled for each edge.
 }
\end{figure}

We consider the entropy vector 
\be
\mathcal{S}=6*\{ 11112; 2223223233; 3343443444; 43333; 2\}
\ee
which is realized by the  graph figure \ref{fig:ray6main}. In the $K$-basis this corresponds to
\be
\mathcal{K}=\{000000000000000;2-1-1-1-12-1-122-1-1222;4\}.
\ee
This state has an exchange symmetry between $A,B,C,D$ as well as another between $E$ and $O$.

We consider perfect tensor hyperthread and barrier configurations to establish optimality:

\paragraph{Maximization}
We consider the perfect tensor hyperthread configuration
\begin{table}[H]
\begin{center}
\scriptsize
\begin{tabular}{| c |c | c | c | c | c | c | }
\hline
Type & Target & Contributing Species & Vertex  & Example hyperthread & $e_{\sigma_1 A}$ & $e_{\sigma_1 O}$  \\ 
\hline
\hline
$h_1$ & -1 & $(ABCD)_3(EO)_1$ & $\sigma_1$ &$\{\sigma_1;A;B;C;O)\}$ & 6 & 4 \\
\hline
$h_2$ & 2 & $ABCD,(ABCD)_2EO$ &$\sigma_1$ &$\{\sigma_1;A;B;C;D\}$ & 4 & 6\\
\hline
$h_3$ & 4 & $ABCDEO$ &$\sigma_1$ &$\{\sigma_1;A;B;C;D;E;O\}$ & 1 & 1\\
\hline
\end{tabular}
\end{center}
\label{tab:ray6_intext}
\end{table}
\noindent which matches the $K$-basis vector and has the objective value
\be
(12-6)*40+4*66=504.
\ee
It can be shown that this configuration satisfies the density bound constraints
\be
\begin{split}
 e_{\sigma_1 A}: \quad &6h_1+4h_2+h_3=6\\
 e_{\sigma_1 O}: \quad &4h_1+6h_2+h_3=12\\
&h_1 \rightarrow -1, \; h_2 \rightarrow 2, \; h_3 \rightarrow 4.
\end{split}
\ee
Threads of each class are shown in figure \ref{fig:5_config}:
\begin{figure}[H]
\centering
\begin{tabular}{cc}
\centering
\includegraphics[width=.4\textwidth,page=25]{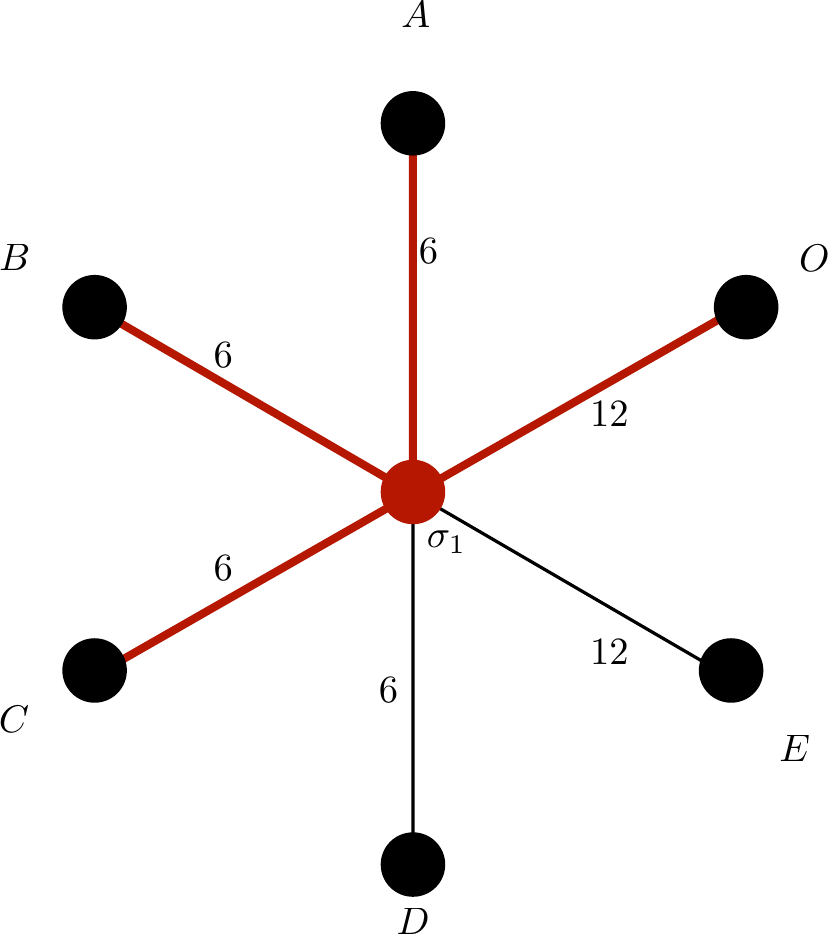}&
\includegraphics[width=.4\textwidth,page=24]{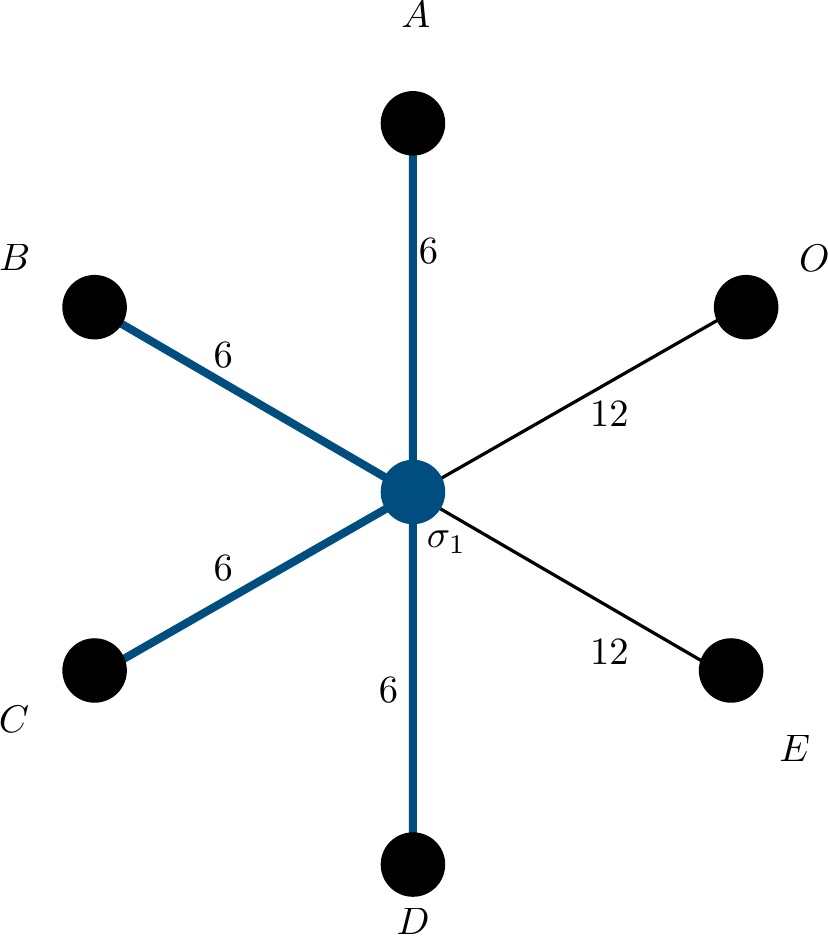}\\
$h_1$ & $h_2$ \\[6pt]
\multicolumn{2}{c}{\includegraphics[width=.4\textwidth,page=23]{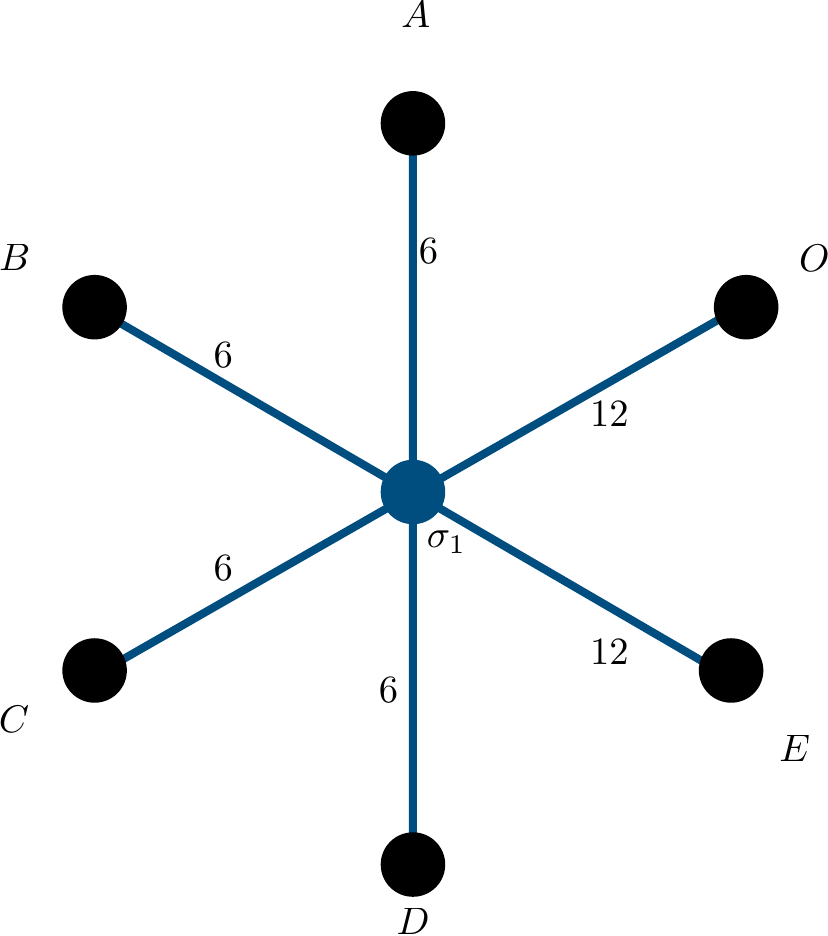}}\\
\multicolumn{2}{c}{$h_3$}
\end{tabular}
\caption{\label{fig:5_config}The thread configuration consists of three classes of threads $h_1$, $h_2$ and $h_3$ which are collections of thread species which respect a symmetry of the graph. $h_1$ consists of eight negative threads one between each of $(ABCD)_3(EO)_1$. $h_2$ consists fourteen positive 4-threads two between each of $ABCD,(ABCD)_2EO$. $h_3$ consist of 4 positive 6-threads. Shown is a representative of each class.}
\end{figure}

\paragraph{Minimization}
For the chosen state besides subadditivity, there are 28 total entropy inequalities which are saturated and by theorem \ref{thm:QCS} must be explicitly implemented. They comprise two classes: MMI between three single-party regions, and MMI between two two-party regions and a single-party region. We will use the wildcard $\bullet$ to indicate any of the regions $A,B,C,D$ where it should be understood that the same region cannot be repeated. The needed inequalities are:
\begin{itemize}
    \item $MMI_{(1,1,1)}$ 16 inequalities excluding those of the form $I_{3}(\bullet:E:O)$. For example the four inequalities which include $\mu_{ABCE}$ are given by
    \be
    \begin{split}
        \mu_{ABCD}+ \mu_{ABCE}+ \mu_{ABCO}&\geq0\\
       \mu_{ABCE}+ \mu_{ABDE}+ \mu_{ABEO}&\geq0\\
        \mu_{ABCE}+ \mu_{ACDE}+ \mu_{ACEO}&\geq0\\
         \mu_{ABCE}+ \mu_{BCDE}+ \mu_{BCEO}&\geq0\\.
    \end{split}
    \ee
    In these 16 inequalities $\mu_{ABCD}$ appear four times and each $\mu_{\bullet\bullet EO}$ two times. Each of $\mu_{\bullet\bullet\bullet E}$ and $\mu_{\bullet\bullet\bullet O}$ appears once with $\mu_{ABCD}$ and three times with $\mu_{\bullet\bullet EO}$.
    \item $MMI_{(1,2,2)}$ 12 inequalities of the form $I_{3}(\bullet:\bullet\bullet:EO)$. For example
    \be
    \mu_{ABDO}+ \mu_{ABDE}+ \mu_{ACDO}+ \mu_{ACDE}+ \mu_{6}\geq0.
    \ee
    Each term $\mu_{\bullet\bullet\bullet E}$ and $\mu_{\bullet\bullet\bullet O}$ occurs in 6 equations while $\mu_6$ occurs in all 12.
\end{itemize}
Taking these into account along with the symmetry of the graph the dual minimization program is
\be
        \left\{
        \begin{aligned}
            & & & \min \nu(x) \\
            & \text{such that} & & \forall h\in H_2,\; \int_{\Sigma}d\nu(x)\Delta(x,h) \geq 16,\\
             & \text{and} & & \forall h\in H_{A:B:C:D},\; \int_{\Sigma}d\nu(x)\Delta(x,h) = 40+4\alpha_1\\
             &\text{and} & & \forall h\in H_{\bullet:\bullet:E:O},\; \int_{\Sigma}d\nu(x)\Delta(x,h) = 40+2\alpha_2 \\
             &\text{and} & & \forall h\in H_{\bullet:\bullet:\bullet:E},H_{\bullet:\bullet:\bullet:O},\; \int_{\Sigma}d\nu(x)\Delta(x,h) = 40+\alpha_1+3\alpha_2+\alpha_3 \\
             &\text{and} & & \forall h\in H_{6},\; \int_{\Sigma}d\nu(x)\Delta(x,h) = 66+2\alpha_3.
        \end{aligned}
    \right.  
\ee
which can simplified to the following linear program
\be
        \left\{
        \begin{aligned}
            & & & \min 24b_{A}+24b_O \\
             & \text{s.t.} & & 4b_A = 40+\alpha_1\\
             &\text{and} & & 2b_A+2b_O = 40+2\alpha_2\\
             &\text{and} & & 3b_A+b_O = 40+\alpha_1+3\alpha_2+\alpha_3\\
                   &\text{and} & & 4b_A+2b_O =66+2\alpha_3. \\
        \end{aligned}
    \right.  
\ee
Here $b_O$ is the barrier to be placed on $e_{\sigma_1 E}$ and $e_{\sigma_1 O}$. Because of symmetry $b_{A}$ will be the barrier placed on each of the other edges. This program can be explicitly evaluated to find
\be
504, \; b_A \rightarrow 13, \; b_O \rightarrow 8, \; \alpha_1 \rightarrow 3,\; \alpha_2 \rightarrow 1,\; \alpha_3 \rightarrow 1.
\ee
which matches the maximization. It can be verified that each perfect tensor hyperthread species will cross exactly the required barrier. 

Together the  perfect tensor hyperthread and barrier configuration demonstrate that these are optimal with the perfect tensor hyperthread configuration explicitly locking the entropy vector.

\section{Discussion}\label{sec:dis}
As shown perfect tensor hyperthreads typically have stronger locking properties than bit threads and are particularly well suited to addressing problems associated with the full entropy vector. This is because of the $K$-basis which provides a linear transformation between the contribution of different species of perfect tensor hyperthread  and the entanglement entropies which make up the $S$-basis. While the ability of perfect tensor hyperthreads to lock the full entropy vector remains an open question, we have presented evidence that this holds up to five boundary regions especially for graphs.

An interesting feature of the holographic entropy cone is that beyond three regions, entropy inequalities do not require all $K$s to be positive. This leads to the notion of negative threads which contribute negatively to the objective and density bound. It was shown for examples with four and five boundary regions that the holographic entropy cone inequalities provide the necessary constraints which relate positive and negative threads. Of particular interest are extremal rays: graphs which saturate the maximum number of entropy inequalities. This is because by theorem \ref{thm:QCS} only saturated inequalities provide active constraints for the optimization programs. As such extremal rays have the strongest constraints relating the different species of negative and positive threads.

\subsection{Main conjecture}
We are now prepared to state the full conjecture for perfect tensor hyperthreads. This states that for any number of boundary regions a maximal configuration can be found such that the number of each species is given by the entropy vector in the $K$-basis so long as the different  perfect tensor hyperthread species are subject to the inequalities of the holographic entropy cone. These provide rules for how different species are related to one another.

\begin{conjecture}\label{mainconj}
For a static time slice $\Sigma$ and a given partition of $\partial\Sigma$ into $N$ regions and purifier $O$ let $H$ be the space of positive and negative perfect tensor hyperthreads and $\{Q^N_i\geq 0\}$ the set of holographic entropy cone inequalities. There exists a suitable definition of ``well-behaved" perfect tensor hyperthreads such that the program

\be
        \left\{
        \begin{aligned}
            & & & \max \sum\mathcal{S}^N(\mu) \\
            & \text{s.t.} & & \forall x\in \Sigma,\; \int_{H}d\mu(h)\Delta(x,h)\leq 1,\\
             & \text{and} & & \forall_i Q^N_i(\mu)\geq 0
        \end{aligned}
    \right.  
\ee
locks the entropy vector $\mathcal{S}^N$ with $K_{I}=\mu^*(H_I).$
\end{conjecture}
We summarize current and future progress:
\begin{itemize}
    \item In this article we have provided explicit analysis of optimal configurations of perfect tensor hyperthreads and barriers for several graphs with $N\leq5$. As of the time of writing this is current as the holographic entropy cone is known completely only up to $N=5$\footnote{See for example \cite{2021arXiv210207535A,2022arXiv220400075H} for some comments on the current progress of $N=6$.} which consists of 372 independent entropy inequalities of 8 distinct types and 2267 extremal rays arranged in 19 orbits \cite{r3}.
    
    In the case of $N=3$ we have successfully performed this analysis on a large number of random graphs with random capacities.
    
    \item In appendix \ref{appex:N5ER} we extend this to many of the extremal rays which make up the $N=5$ holographic entropy cone. We do so by constructing configurations of perfect tensor hyperthreads whose number for each species matches the corresponding $K$ in the $K$-basis while obeying all density bounds. As a result these configurations satisfy all of the entropy inequalities of the holographic entropy cone and correctly reproduce each entropy in the entropy vector\footnote{Let us emphasize this point. Because the $K$-basis entropy vector is a valid entropy vector as long as the number of each species in the configuration of perfect tensor hyperthreads matches the corresponding $K$ we are guaranteed that all of the entropy inequalities are satisfied. What this means is we can use the $K$-basis vector as an ansatz and \emph{without} knowledge of the form of any of the entropy inequalities know that the perfect tensor hyperthread configuration is feasible as long as it satisfies the density bounds.}. We have \emph{not} constructed barrier configurations as this would require analysis similar to that of section \ref{sec:N5}. That is in \eqref{eq:5max} all 372 entropy inequalities would have to be explicitly implemented and dualized or one would have to make use of theorem \ref{thm:QCS} and identify for each ray the entropy inequalities which are saturated and construct the appropriate program from this information. What this means is hypothetically there could be a configuration of perfect tensor hyperthreads which had an objective value higher than that of the sum of the entropies of the entropy vector and still satisfied all of the entropy inequalities. Given how highly constrained extremal rays are this seems unlikely. However, one would need to in each case construct an explicit barrier configuration to provide an upper bound and rule out this possibility.
    
    \item Any entropy vector of the holographic entropy cone can be written as a positive linear combination of entropy vectors of extremal rays. As such knowledge that perfect tensor hyperthreads work for extremal rays is enough to guarantee that for each entropy vector of the holographic entropy cone there exists \emph{a} graph for which our construction holds. This graph is formed from the union of graphs comprised of the extremal rays which the desired entropy vector decomposes into.
    
    \item Ideally, one would be able to prove general locking theorems for perfect tensor hyperthreads on graphs akin to \cite{doi:10.1137/S0895480195287723}. $N=3$ would be the simplest case to consider as these graphs contain only 2 and 4-threads and do not have the added complications of higher party threads, negative threads, and explicit implementation of entropy inequalities. In general more robust tools and methods need to be developed to allow for an easier diagnosing of the locking properties of general convex programs. This seems like an immediate natural next step which we leave to future ongoing work.
    
    \item Once established for graphs the next step would be to prove the conjecture for general holographic states. This would require extending locking theorems to Riemannian manifolds as was done in the case of bit threads \cite{2017CMaPh.352..407F,2018CQGra..35j5012H,Cui:2018aa}. Since perfect tensor hyperthreads will necessarily cross it seems possible that unlike bit threads there may be obstacles to uplifting graph locking theorems. To circumvent this may require alternate definition possibly by using a different density bound. Still the analysis provided in figures \ref{fig:ex_MMI}, \ref{fig:ex_MMI2}, \ref{fig:ex_MMI3} seems to demonstrate that modulo these concerns perfect tensor hyperthreads are capable of providing sensible information about key information quantities as well as phase changes. This also provides an interesting geometric picture where higher party entanglement is generally located (as measured by the perfect tensor hyperthread's internal vertex) deeper in the bulk. While we only provided the one example similar analysis can be performed for other holographic states given information about the location of the RT surfaces.
    \item More care will need to be given to the precise definition of the space of negative threads preferably without reference to specific surfaces. This will be especially true in the context of manifolds.  We have used the term ``well-behaved" in the conjecture to indicate such a definition.
\end{itemize}

\subsection{Multipartite distillation}
A natural picture emerges when one considers the connection between bit threads, information and geometry. Consider a bipartition of a holographic state in $A$ and its purifier $O$. This state can be distilled to a number of Bell pairs equal to the entanglement entropy $S_A$. However, this is also equivalent to the number of bit threads which comprise an optimal bit thread configuration. As such one can view the bit threads \emph{as} the distilled Bell pairs. This is a natural realization of ER=EPR where the bit threads can be viewed as ``building up" a coarse grained version of the geometry (essentially the graph desiccation which only has knowledge of RT surfaces) necessary to support such a structure of entanglement. 

What we would like to do is extend this to perfect tensor hyperthreads. We imagine a resource theory where the resource is given by the full entropy vector of the holographic state and the target states are precisely the perfect tensor states comprising the $K$-basis entropy vector. In this set up it is natural, in analogy with bit threads and Bell pairs, to associate to each perfect tensor hyperthread a distilled perfect tensor state.

Let $\psi$ be a holographic state and $\mathcal{K}$ its $K$-basis entropy vector. If all of the components are strictly positive then we conjecture the existence of quantum channel which enables a distillation of the form:
\be
\ket{\psi}\longrightarrow \prod_{I\in\mathcal{K}}\ket{PT_{I}}^{\otimes K_{I}}.
\ee
For example given the state corresponding to the graph figure \ref{fig:3g}
\be
\ket{\psi}: \quad \mathcal{K}=\{1,0,1,1,0,1;2\}
\ee
would correspond to the distillation 
\be
\ket{\psi}\longrightarrow \ket{AB}\otimes\ket{AO}\otimes\ket{BC}\otimes\ket{CO}\otimes\ket{ABCD}^{\otimes 2}
\ee

 For generic $K$-basis entropy vectors we must interpret the negative components. Potentially, these could be viewed as states which must be provided along with the holographic state in order to enact a distillation to the positive components. That is if we decompose $\mathcal{K}=\mathcal{K}_{+}+\mathcal{K}_{-}$ into its positive and negative components then we wish to consider an assisted distillation of the form
\be
\ket{\psi}\otimes\prod_{I\in\mathcal{K}_{-}}\ket{PT_{I}}^{\otimes K_{I}}\longrightarrow \prod_{I\in\mathcal{K}_{+}}\ket{PT_{I}}^{\otimes K_{I}}.
\ee

The state corresponding to the graph figure \ref{fig:4ex}
\be
\ket{\psi}: \quad \mathcal{K}=\{0000000000;-11111\}
\ee
would result in the distillation
\be
\ket{\psi}\otimes \ket{ABCD}\longrightarrow \ket{ABCO}\otimes\ket{ABDO}\otimes\ket{ACDO}\otimes\ket{BCDO}.
\ee

If such a distillation protocol can be established\footnote{Technical note: Recent work \cite{2020JHEP...04..208A,2021JHEP...10..047H} has determined that holographic states must contain tripartite entanglement as diagnosed by a difference between the mutual information and entanglement wedge cross section \cite{2018NatPh..14..573U,2018JHEP...01..098N} or reflected entropy \cite{2019arXiv190500577D}. Notably, this is in tension with the state decomposition conjecture of \cite{Cui:2018aa} which posits for $N=3$ holographic states are comprised mostly of bipartite and four party perfect tensor entanglement. One possible resolution is that such a decomposition of the holographic state is only possible under a distillation protocol such as the one presented here which, to our knowledge and current understanding, \emph{could not} be comprised of local unitaries. As such the distillation of the holographic state in this manner will in general \emph{not} preserve the reflected entropy. This makes sense as the number of species of perfect tensors matches exactly the entanglement entropies of the entropy vector and as such is too coarse grained to also be capable of generically locking the entanglement wedge cross section. It is possible that the technology presented here may be adapted to describe tripartite entanglement and in doing so be sensitive to both the entropy vector and reflected entropy. We leave this to future ongoing work.} then we can associate perfect tensor hyperthreads with distilled perfect tensor states. This would in effect generalize the notion of ER=EPR to this class of multipartite states and provide additional examples and understanding to the relation between geometry and entanglement in holography.

\subsection{The positive $K$ cone}

For $N$ regions we consider the holographic entropy cone $HEC_N$ along with the entropy inequalities $\{Q^N_i\geq0\}$. Within the cone is a proper subcone which is given by the positive orthant in the $K$-basis. That is we define the positive $K$ cone $HEC^+_N$
\be
HEC^+_N=\{\mathcal{K}^N\} \text{ s.t. } \{K_{I}\geq0\}.
\ee
The two cones are the same for $N\leq3$.

Recall that in the $K$-basis the entropy inequalities take the form
\be
Q=\sum_J\beta_J\mathcal{K}_J\geq 0 \text{ s.t. } \{\beta_J\geq0\}
\ee
As such positivity of the $K$'s is enough to guarantee that all of the holographic entropy cone inequalities are trivially satisfied.

Because it is simplicial the entropy inequalities and extremal rays of  $HEC^+_N$ are particularly easy to describe and are in fact related. The entropy inequalities are precisely $\{K_{I}\geq0\}$ while the set of extremal rays is given by even star graphs. Furthermore, this work suggests that the states of this cone admit unaided distillations to perfect tensor states (i.e. the optimal configurations of perfect tensor hyperthreads contain no negative threads). 

Roughly, we expect states to be in the $HEC^+_N$ when the various boundary regions are relatively similar. One way to understand this is that negative threads act to eliminate bottlenecks and make geometry the uniform. When the regions are of similar sizes there can be no advantage from such an exchange.

Given the notable difference in the complexity of the description of entropy vectors in the positive $K$ cone and generic entropy vectors of the HEC it would be interesting to explore if this can be further related to differences in the geometry between such states.

\acknowledgments
The work of J.H. is supported in part by the Simons Foundation through \emph{It from Qubit: Simons Collaboration on Quantum Fields, Gravity, and Information} and in part by MEXT-JSPS Grant-in-Aid for Transformative Research Areas (A) ``Extreme Universe” No. 21H05187. J.H would like to thank Matthew Headrick for many useful discussion. J.H. would also like to thank Tadashi Takayangi, Sergio Hern\'{a}ndez-Cuenca and Guglielmo Grimaldi for reading an early version of this paper. J.H. is grateful to UC Davis and UT Austin for hospitality where early versions of this work were presented.

\appendix

\section{Locking configurations of perfect tensor hyperthreads for some 5 region extremal rays}\label{appex:N5ER}

In this appendix we provide configurations of perfect tensor hyperthreads whose number of each species matches the $K$-basis entropy vector for many of the $N=5$ extremal rays. Information about these graphs as well as the general structure of the $N=5$ holographic entropy cone can be found in \cite{r3}. For an explicit listing of the entropy inequalities in the $K$-basis see \cite{r1}.

We start with some notation:
\begin{itemize}
    \item Boundary vertices are labeled $A,B,C,D,E,O$. The region $O$ is taken to be the purifier. Disconnected boundary vertices are suppressed.
    \item Internal vertices are labeled $\sigma_i$
    \item An edge connecting two vertices $v_1,v_2$ will be labeled as $e_{v_1v_2}$ with capacity $c_{v_1v_2}$.
    \item A $k-$thread $h$ is determined by a central internal vertex $\sigma_c$ and $k$ strands or paths: one from $\sigma_c$ to each of the boundary vertices the threads connects. For example a simple 4-thread $h\in H_{A:B:C:D}$ on $PT_{ABCD}$ would be given by
    \be
    h=\{\sigma_1;A;B;C;D\}.
    \ee
    \item Often based on the symmetries of the graph different species will contribute similarly. We will use the notation $(A_1,A_2,\cdots)_k$ to mean ``any combination of $k$ elements". For example:
    \be
    (ABC)_2(DE)O=(ABDO,ABEO,ACDO,ACEO,BCDO,BCEO)
    \ee
    \item Positive threads will be shown in blue while negative threads will be shown in red.
    \item Each entropy vector is normalized so that the capacities of the graph are as small as possible while always requiring the solutions presented to have an integer number of each perfect tensor hyperthread species.
\end{itemize}

Before proceeding a few notes are in order about the configurations presented:
\begin{itemize}
    \item A species of positive threads will typically split on the internal vertex which connects directly to the greatest number of boundary vertices to which its strands will connect.
    \item In none of the solutions must negative threads be split on an internal vertex which does not connect to contributing boundary vertices. Similarly it is never necessary for a strand to cross the same edge multiple times. This is a manifestation of the notion of ``straight" threads presented in the main text.
    \item In all cases the perfect tensor hyperthread configurations respect the symmetry of the graph. This often requires the threads to be symmetrized over multiple identical paths or splitting vertices. As a result some collections of threads will have fractional contributions to a particular edge as different contributing threads will cross an edge a different number of times.
    \item Rays 1,2,4 represent the exchange of a single 2,4,6-thread respectively.
       \item The holographic entropy inequalities are superbalenced \cite{2020JHEP...07..245H}. In the $K$-basis this implies that only ray 1 has nonzero $K_2$s. In the context of perfect tensor hyperthreads this is borne out as only ray 1 contains 2-threads for the constructed configurations.
    \item Rays 3 and 6 are the examples presented in the main body of the article. They are reproduced here for completeness.
    \item Interestingly, ray 5 is the only ray which contains negative 6-threads as these are highly constrained by the entropy inequalities.
    \item In addition to those present there are four more extremal rays for a total of 19. These have not yet been explicitly constructed due to the large combinatorial increase in the number of possible contributing perfect tensor hyperthreads. We expect no obstacles to the construction of these configurations which consist entirely of the exchange of negative 4-threads along with positive 4-threads and 6-threads.
\end{itemize}

\newpage
\subsection*{Ray 1}

\be\label{eq:ray1S}
  \mathcal{S}=\{10000; 1111000000; 1111110000; 11110; 1\}\\ 
\ee

\be\label{eq:ray1K}
\mathcal{K}=\{000010000000000;000000000000000;0\}
\ee

\begin{figure}[H]
\centering
\includegraphics[width=.5\textwidth,page=31]{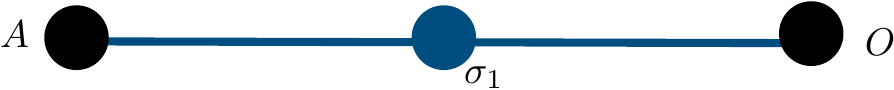}
\caption{\label{fig:ray1}  }
\end{figure}

\subsection*{Ray 2}

\be\label{eq:ray2S}
  \mathcal{S}=\{11100; 2211211110; 1222212211; 11222; 1\}\\ 
\ee

\be\label{eq:ray2K}
\mathcal{K}=\{000000000000000;001000000000000;0\}
\ee

\begin{figure}[H]
\centering
\includegraphics[width=.5\textwidth,page=32]{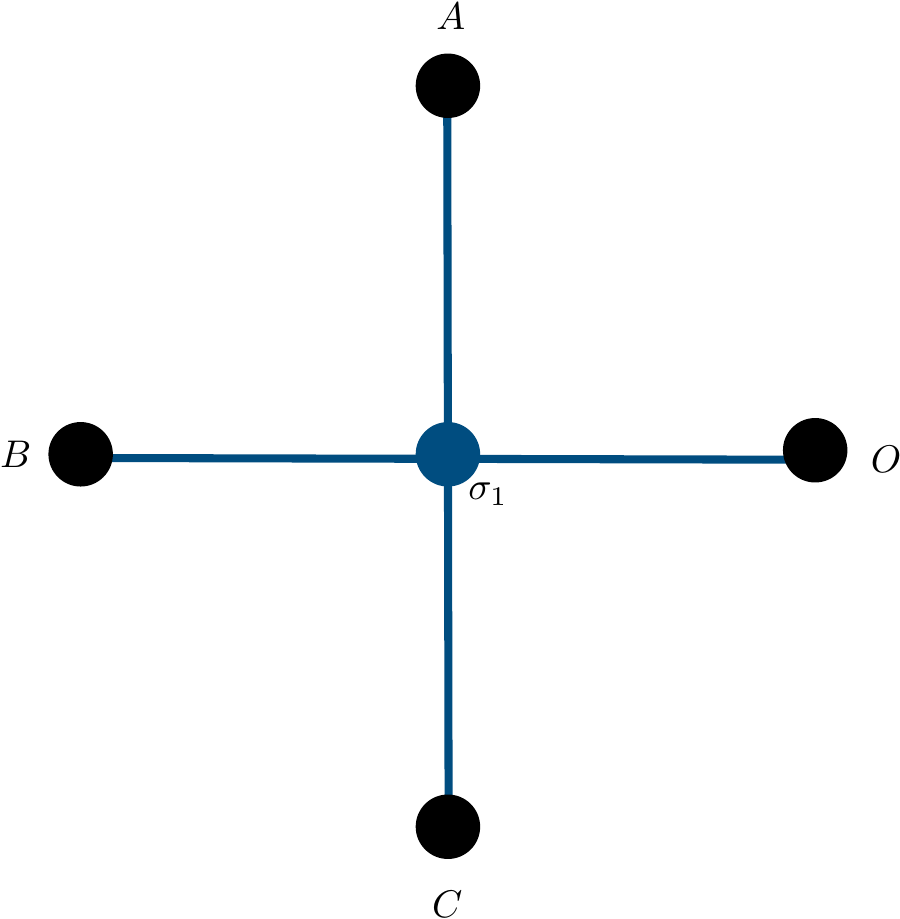}
\caption{\label{fig:ray2}  }
\end{figure}

\newpage
\subsection*{Ray 3}

\be\label{eq:ray3S}
  \mathcal{S}=2\{11110; 2221221211; 3323223222; 23333; 2\}\\ 
\ee

\be\label{eq:ray3K}
\mathcal{K}=\{000000000000000;-101010010001000;0\}
\ee

\begin{figure}[H]
\centering
\includegraphics[width=.5\textwidth,page=33]{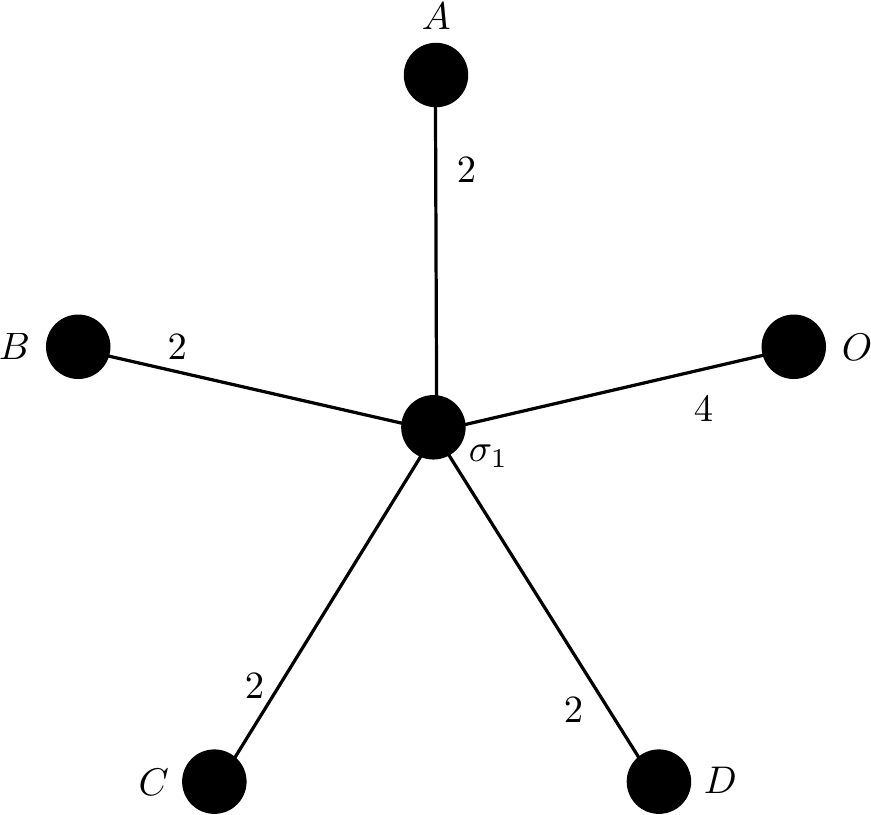}
\caption{\label{fig:ray3}  }
\end{figure}

\begin{table}[H]
\begin{center}
\scriptsize
\begin{tabular}{| c |c | c | c | c | c | c | }
\hline
Type & Target & Contributing Species & Vertex  & Example hyperthread & $e_{\sigma_1 A}$ & $e_{\sigma_1 O}$  \\ 
\hline
\hline
$h_1$ & -1 & $ABCD$ & $\sigma_1$ &$\{\sigma_1;A;B;C;D\}$ & 1 & 0 \\
\hline
$h_2$ & 1 & $(ABCD)_3O$ &$\sigma_1$ &$\{\sigma_1;A;B;C;O\}$ & 3 & 4\\
\hline
\end{tabular}
\end{center}
\label{tab:ray3}
\end{table}

\be
\begin{split}
 e_{\sigma_1 A}: \quad &h_1+3h_2=2\\
 e_{\sigma_1 O}: \quad &4h_2=4\\
&h_1 \rightarrow -1, \; h_2 \rightarrow 1
\end{split}
\ee

\begin{figure}[H]
\centering
\begin{tabular}{cc}
\centering
\includegraphics[width=.45\textwidth,page=34]{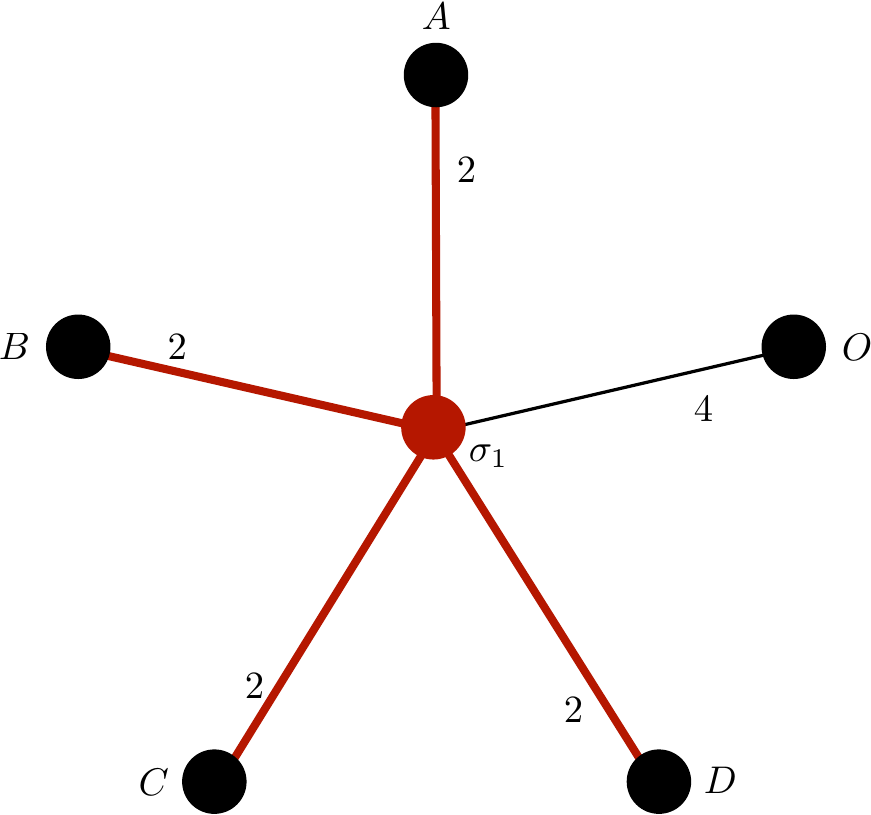}&
\includegraphics[width=.45\textwidth,page=35]{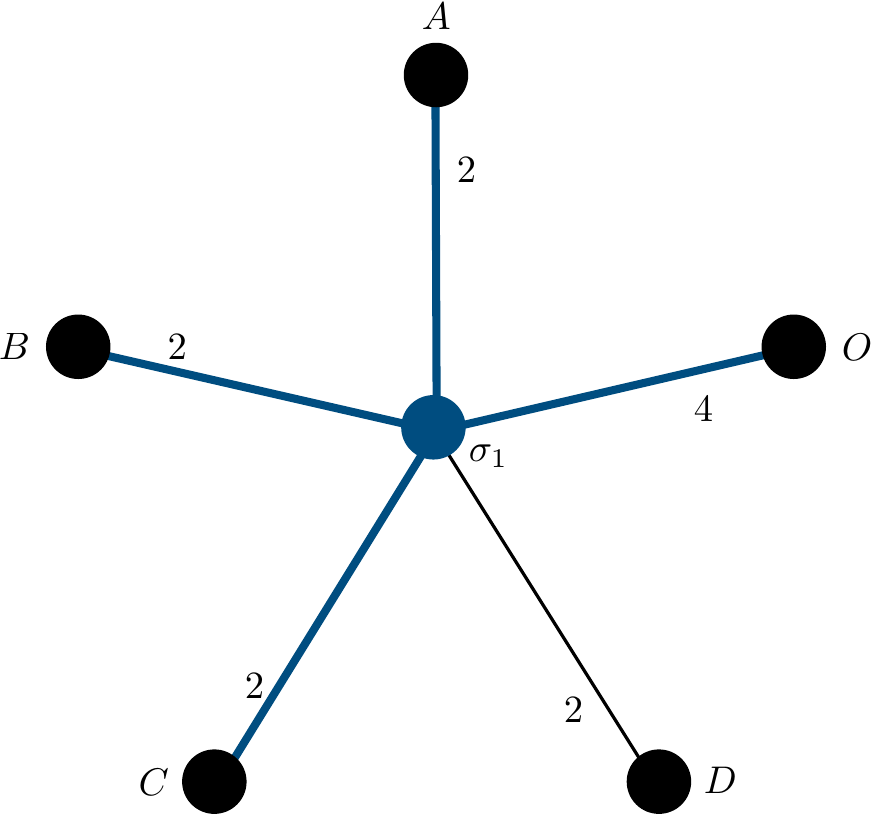}\\
$h_1$ & $h_2$ \\[6pt]
\end{tabular}
\caption{}
\end{figure}

\subsection*{Ray 4}

\be\label{eq:ray4S}
  \mathcal{S}=\{11111; 2222222222; 3333333333; 22222; 1\}\\ 
\ee

\be\label{eq:ray4K}
\mathcal{K}=\{000000000000000;000000000000000;1\}
\ee

\begin{figure}[H]
\centering
\includegraphics[width=.5\textwidth,page=36]{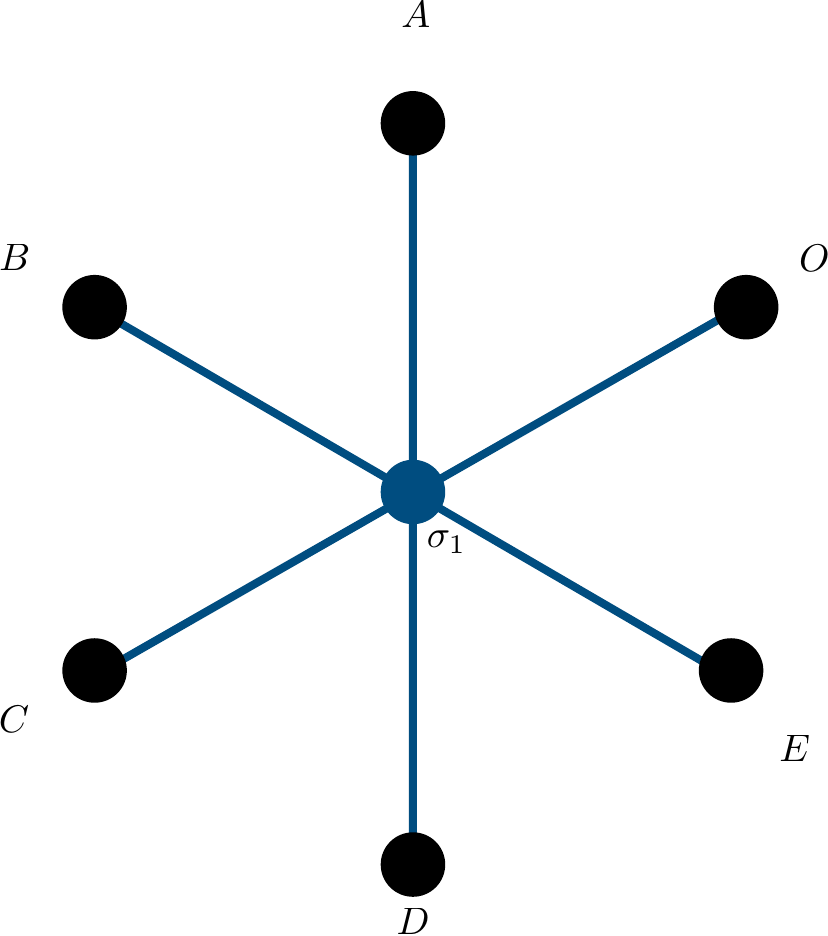}
\caption{\label{fig:ray4}  }
\end{figure}

\newpage
\subsection*{Ray 5}
\be\label{eq:ray5S}
  \mathcal{S}=6\{ 11111; 2222222222; 3333333333; 44444; 3\}\\ 
\ee

\be\label{eq:ray5K}
\mathcal{K}=\{000000000000000;-1-12-122-1222-12222;-2\}
\ee

\begin{figure}[H]
\centering
\includegraphics[width=.5\textwidth,page=18]{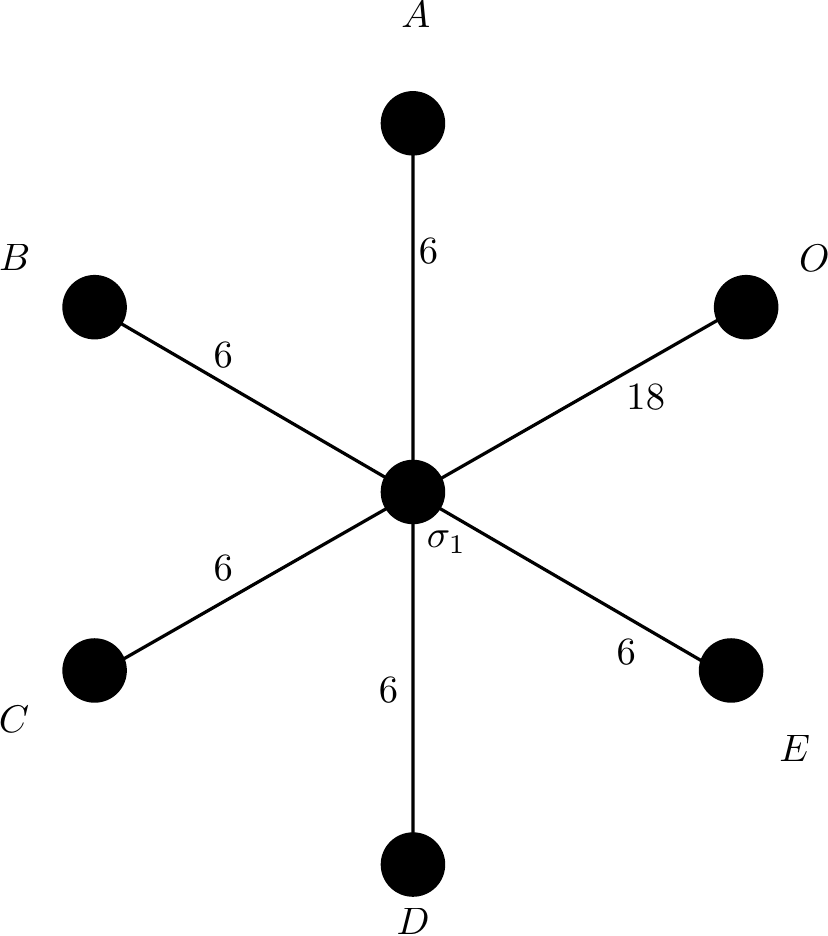}
\caption{\label{fig:ray5}  }
\end{figure}

\begin{table}[H]
\begin{center}
\scriptsize
\begin{tabular}{| c |c | c | c | c | c | c | }
\hline
Type & Target & Contributing Species & Vertex  & Example hyperthread & $e_{\sigma_1 A}$ & $e_{\sigma_1 O}$  \\ 
\hline
\hline
$h_1$ & -1 & $(ABCDE)_4$ & $\sigma_1$ &$\{\sigma_1;A;B;C;D\}$ & 4 & 0 \\
\hline
$h_2$ & 2 & $(ABCDE)_3O$ &$\sigma_1$ &$\{\sigma_1;A;B;C;O\}$ & 6 & 10\\
\hline
$h_3$ & -2 & $ABCDEO$ &$\sigma_1$ &$\{\sigma_1;A;B;C;D;E;O\}$ & 1 & 1\\
\hline
\end{tabular}
\end{center}
\label{tab:ray5}
\end{table}

\be
\begin{split}
 e_{\sigma_1 A}: \quad &4h_1+6h_2+h_3=6\\
 e_{\sigma_1 O}: \quad &10h_2+h_3=18\\
&h_1 \rightarrow -1, \; h_2 \rightarrow 2, \; h_3 \rightarrow -2
\end{split}
\ee

\begin{figure}[H]
\centering
\begin{tabular}{cc}
\centering
\includegraphics[width=.45\textwidth,page=20]{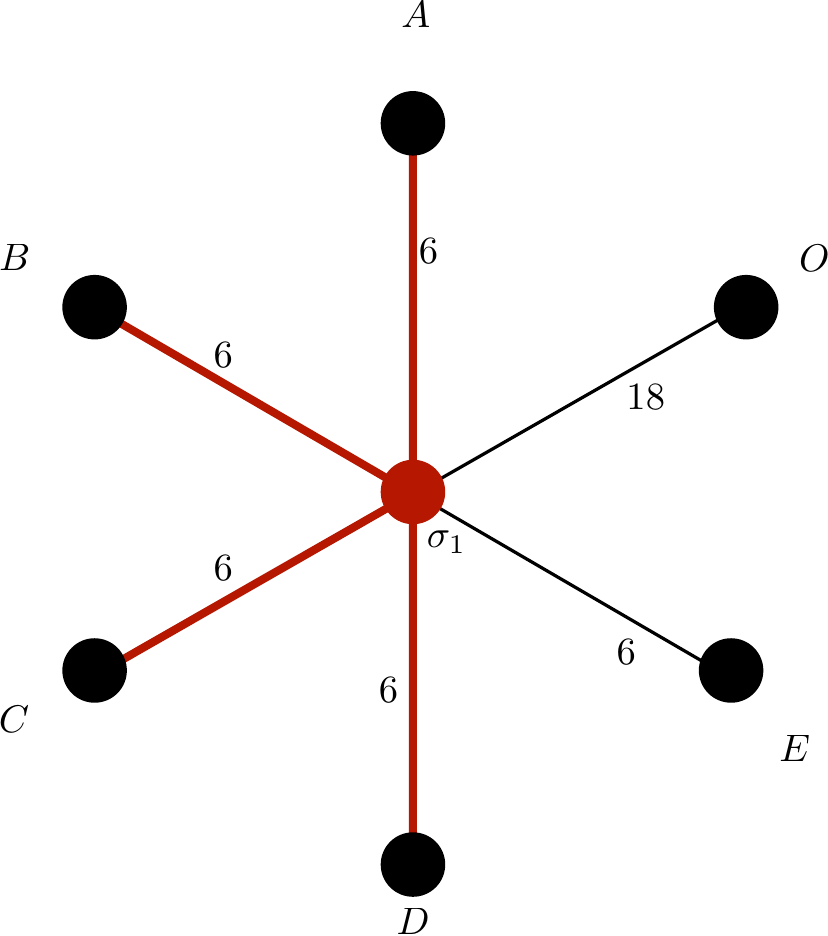}&
\includegraphics[width=.45\textwidth,page=21]{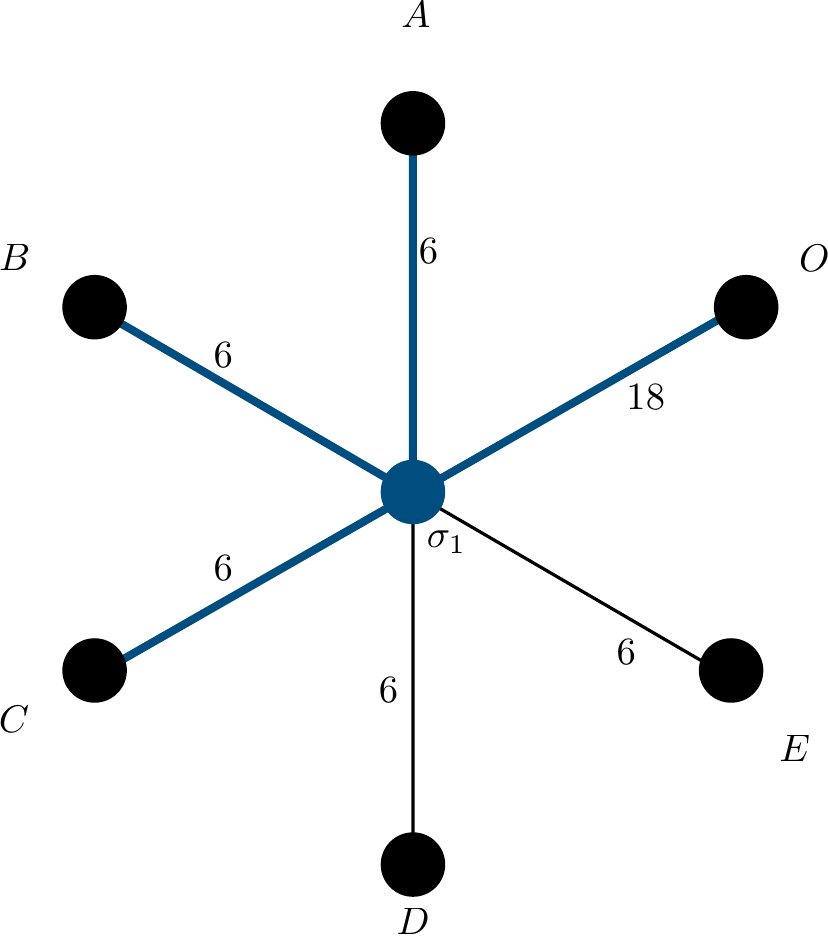}\\
$h_1$ & $h_2$ \\[6pt]
\multicolumn{2}{c}{\includegraphics[width=.45\textwidth,page=19]{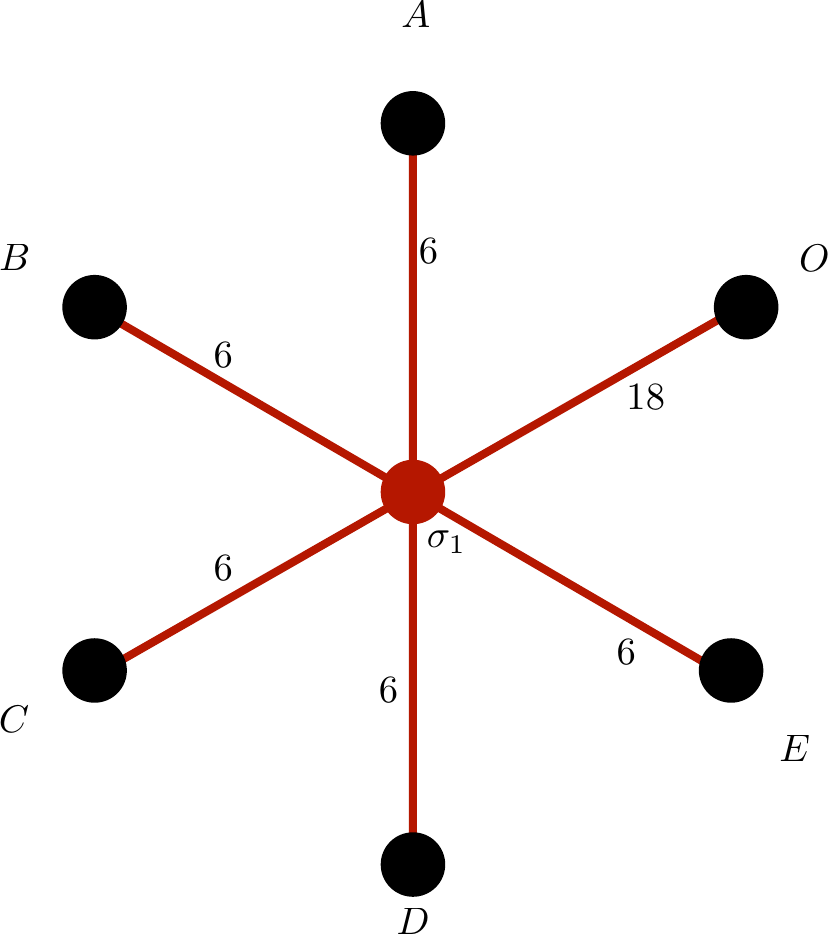}}\\
\multicolumn{2}{c}{$h_3$}
\end{tabular}
\caption{}
\end{figure}

\newpage
\subsection*{Ray 6}
\be\label{eq:ray6S}
\mathcal{S}=6*\{ 11112; 2223223233; 3343443444; 43333; 2\}
\ee

\be\label{eq:ray6K}
\mathcal{K}=\{000000000000000;2-1-1-1-12-1-122-1-1222;4\}
\ee

\begin{figure}[H]
\centering
\includegraphics[width=.5\textwidth,page=22]{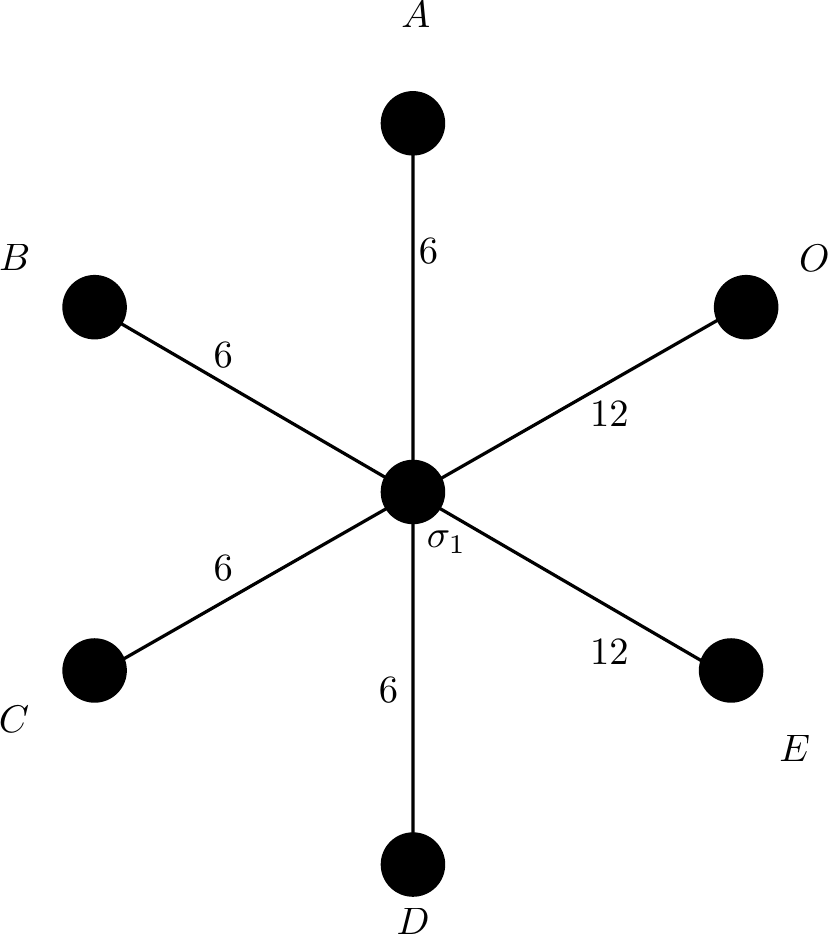}
\caption{\label{fig:ray6}  }
\end{figure}

\begin{table}[H]
\begin{center}
\scriptsize
\begin{tabular}{| c |c | c | c | c | c | c | }
\hline
Type & Target & Contributing Species & Vertex  & Example hyperthread & $e_{\sigma_1 A}$ & $e_{\sigma_1 O}$  \\ 
\hline
\hline
$h_1$ & -1 & $(ABCD)_3(EO)_1$ & $\sigma_1$ &$\{\sigma_1;A;B;C;O)\}$ & 6 & 4 \\
\hline
$h_2$ & 2 & $ABCD,(ABCD)_2EO$ &$\sigma_1$ &$\{\sigma_1;A;B;C;D\}$ & 4 & 6\\
\hline
$h_3$ & 4 & $ABCDEO$ &$\sigma_1$ &$\{\sigma_1;A;B;C;D;E;O\}$ & 1 & 1\\
\hline
\end{tabular}
\end{center}
\label{tab:ray6}
\end{table}

\be
\begin{split}
 e_{\sigma_1 A}: \quad &6h_1+4h_2+h_3=6\\
 e_{\sigma_1 O}: \quad &4h_1+6h_2+h_3=12\\
&h_1 \rightarrow -1, \; h_2 \rightarrow 2, \; h_3 \rightarrow 4
\end{split}
\ee

\begin{figure}[H]
\centering
\begin{tabular}{cc}
\centering
\includegraphics[width=.45\textwidth,page=25]{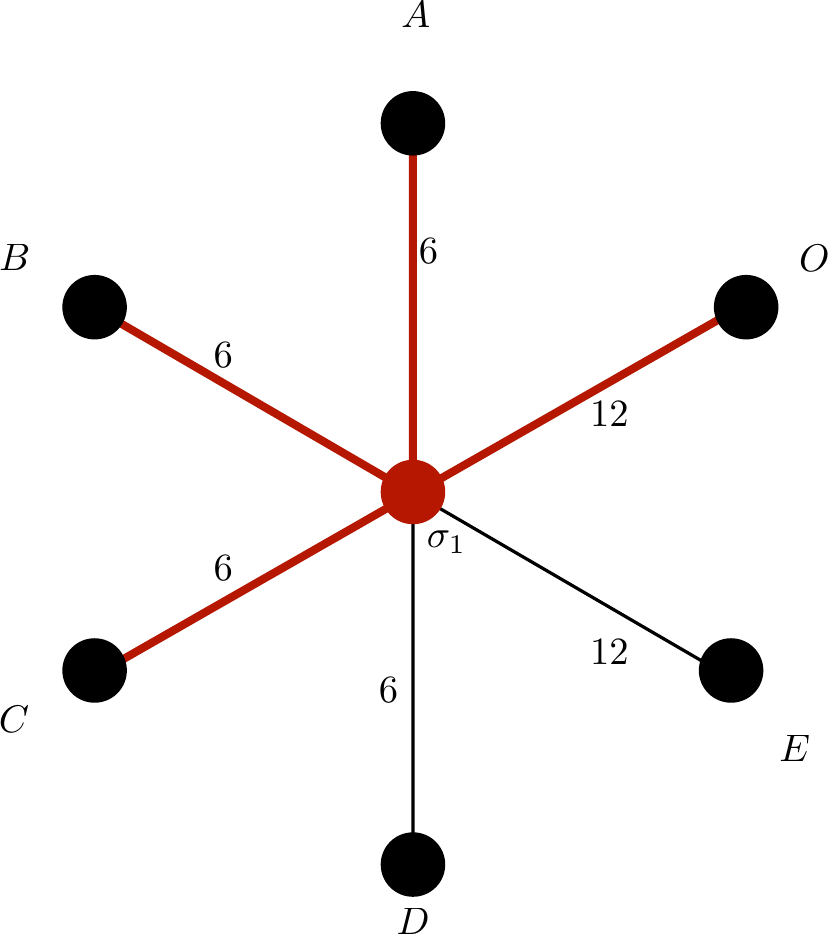}&
\includegraphics[width=.45\textwidth,page=24]{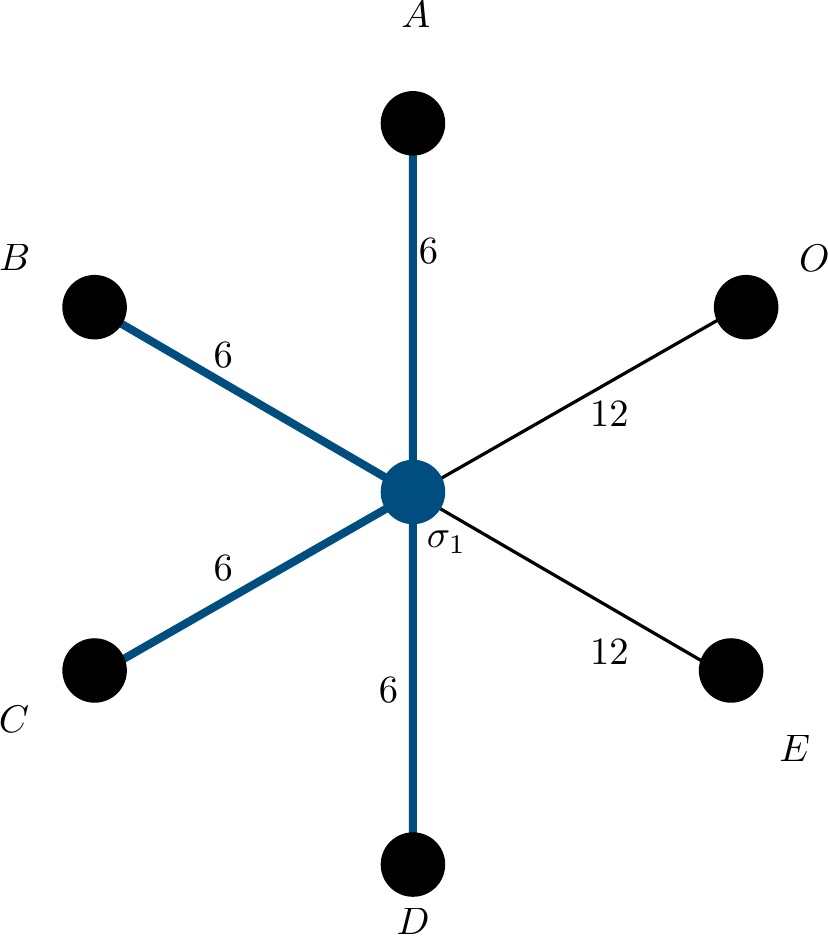}\\
$h_1$ & $h_2$ \\[6pt]
\multicolumn{2}{c}{\includegraphics[width=.45\textwidth,page=23]{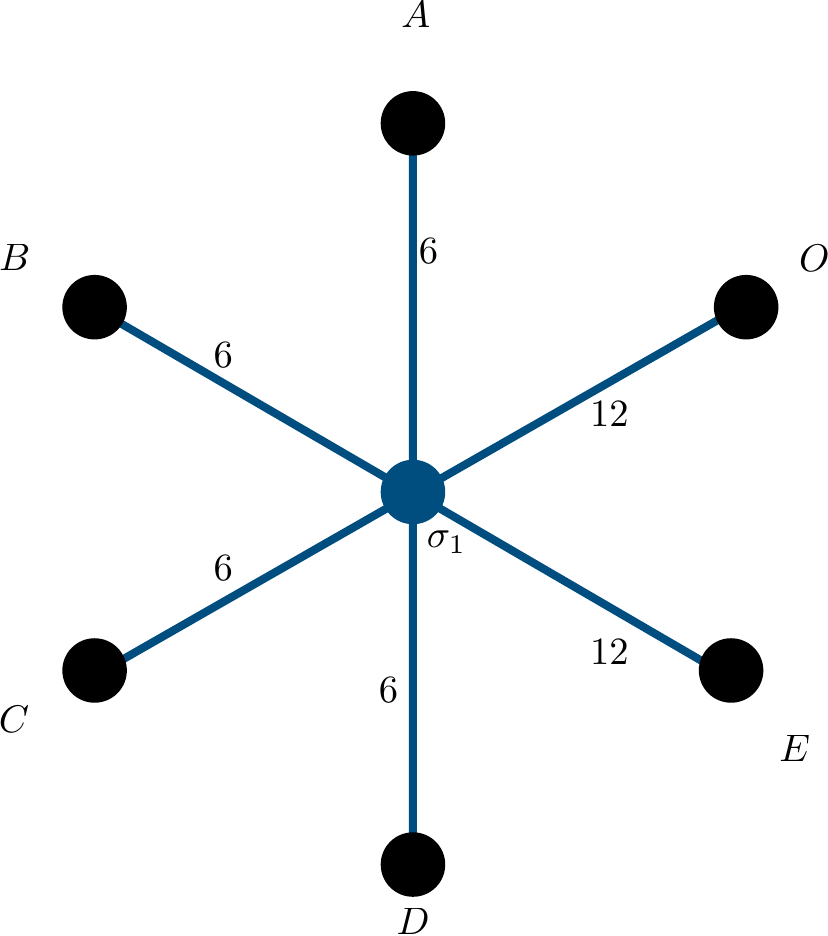}}\\
\multicolumn{2}{c}{$h_3$}
\end{tabular}
\caption{}
\end{figure}

\newpage
\subsection*{Ray 7}
\be\label{eq:ray7S}
\mathcal{S}=6*\{ 11122; 2233233334; 3444454455; 55444; 3\}
\ee
\be\label{eq:ray7K}
\mathcal{K}=\{000000000000000;11-2-211-2114-21144;2\}
\ee
\begin{figure}[H]
\centering
\includegraphics[width=.5\textwidth,page=37]{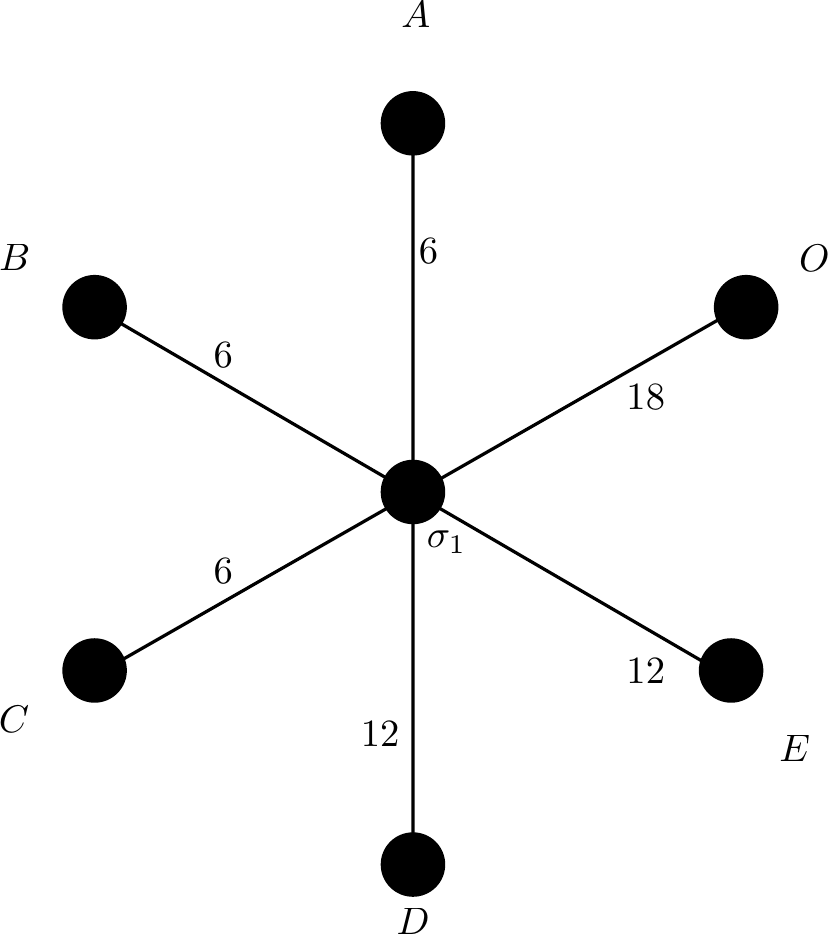}
\caption{\label{fig:ray7}  }
\end{figure}
\begin{table}[H]
\begin{center}
\scriptsize
\begin{tabular}{| c |c | c | c | c | c | c | c | }
\hline
Type & Target & Contributing Species & Vertex  & Example hyperthread & $e_{\sigma_1 A}$ & $e_{\sigma_1 E}$  & $e_{\sigma_1 O}$  \\ 
\hline
\hline
$h_1$ & 1 & $ABC(DE), (ABC)_2(DE)O$ & $\sigma_1$ &$\{\sigma_1;A;B;D;O)\}$ & 6 & 4 & 6 \\
\hline
$h_2$ & -2 & $(ABC)_2DE, ABCO$ &$\sigma_1$ &$\{\sigma_1;A;B;D;E\}$ & 3 & 3 & 1\\
\hline
$h_3$ & 4 & $(ABC)DEO$ &$\sigma_1$ &$\{\sigma_1;A;D;E;O\}$ & 1 & 3 & 3\\
\hline
$h_4$ & 2 & $ABCDEO$ &$\sigma_1$ &$\{\sigma_1;A;B;C;D;E;O\}$ & 1 & 1 & 1\\
\hline
\end{tabular}
\end{center}
\label{tab:ray7}
\end{table}
\be
\begin{split}
 e_{\sigma_1 A}: \quad &6h_1+3h_2+h_3+h_4=6\\
 e_{\sigma_1 E}: \quad &4h_1+3h_2+3h_3+h_4=12\\
 e_{\sigma_1 O}: \quad &6h_1+h_2+3h_3+h_4=18\\
&h_1 \rightarrow 1, \; h_2 \rightarrow -2, \; h_3 \rightarrow 4, \; h_4 \rightarrow 2
\end{split}
\ee
\begin{figure}[H]
\centering
\begin{tabular}{cc}
\centering
\includegraphics[width=.45\textwidth,page=38]{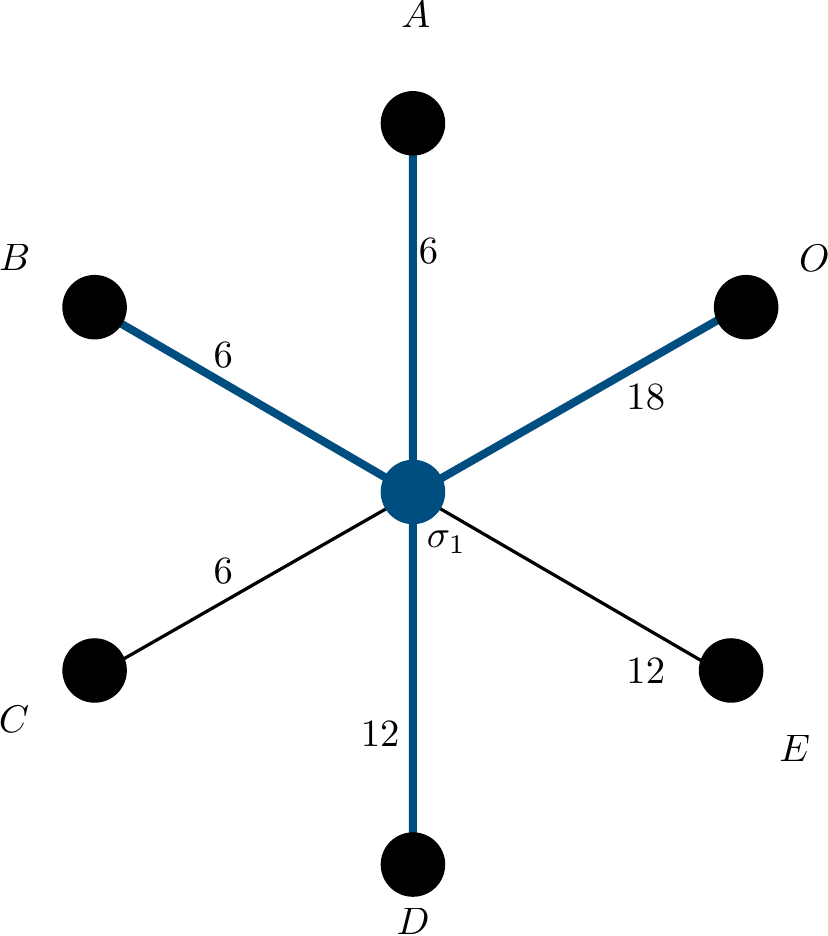}&
\includegraphics[width=.45\textwidth,page=39]{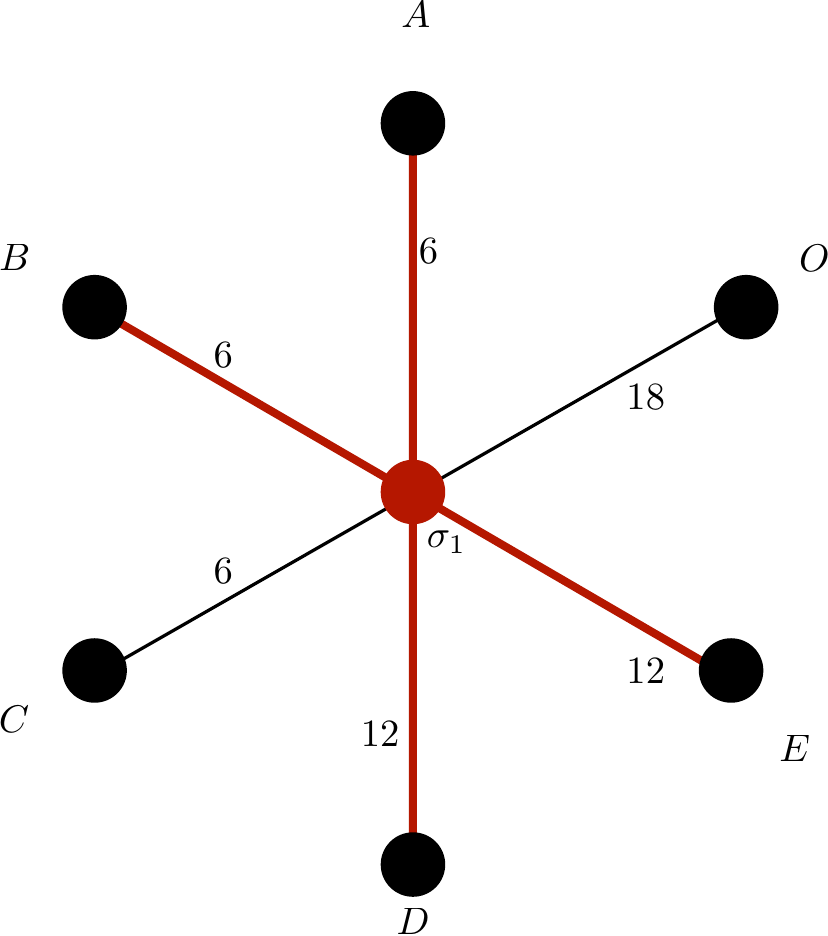}\\
$h_1$ & $h_2$ \\[6pt]
\includegraphics[width=.45\textwidth,page=40]{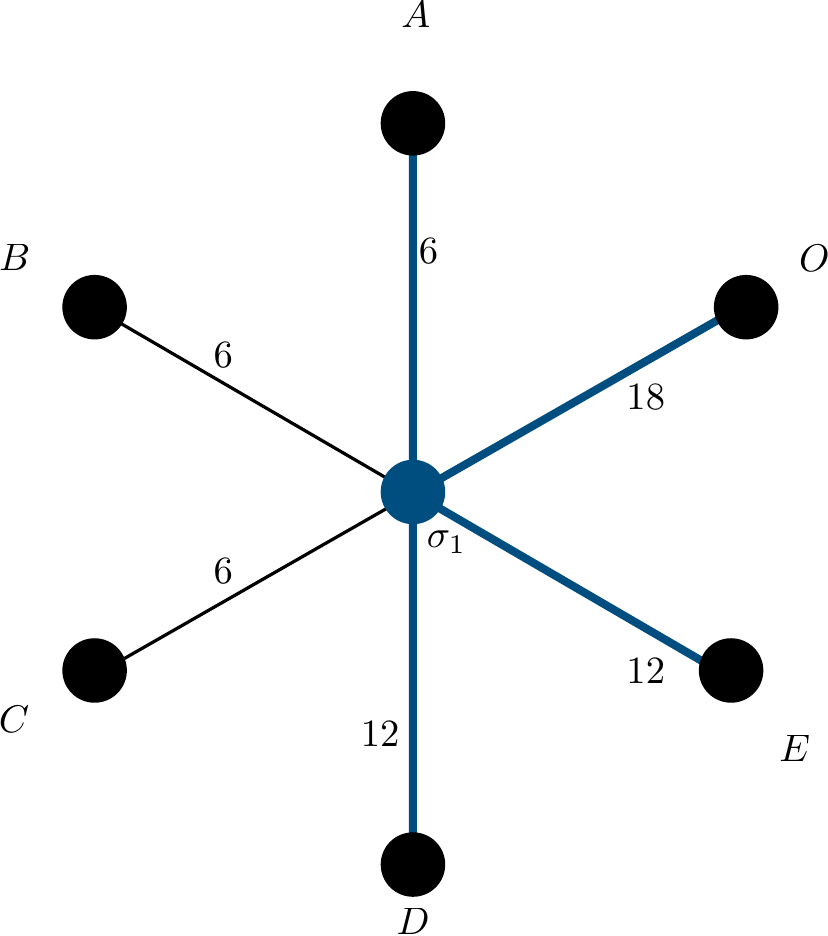}&
\includegraphics[width=.45\textwidth,page=41]{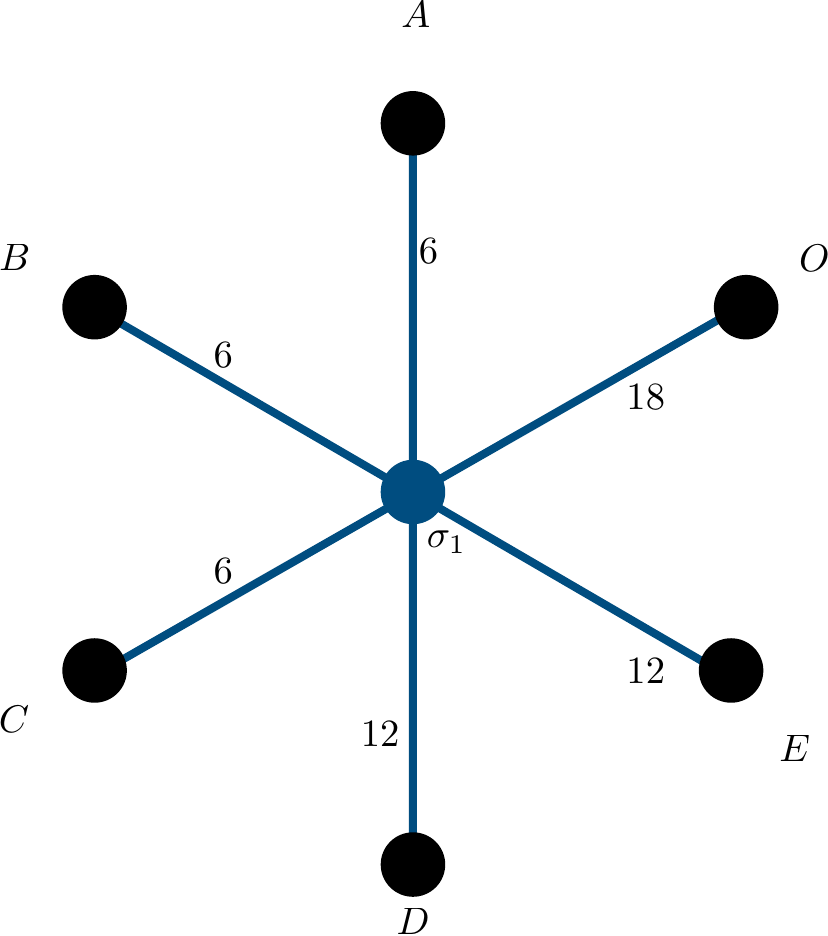}\\
$h_3$ & $h_4$ \\[6pt]
\end{tabular}
\caption{}
\end{figure}

\newpage
\subsection*{Ray 8}
\be\label{eq:ray8S}
\mathcal{S}=6*\{ 11111; 2222222222; 3333333333; 22222; 1\}
\ee
\be\label{eq:ray8K}
\mathcal{K}=\{000000000000000;-1-12-1222-1-1-12-1-1-12;4\}
\ee
\begin{figure}[H]
\centering
\includegraphics[width=.75\textwidth,page=42]{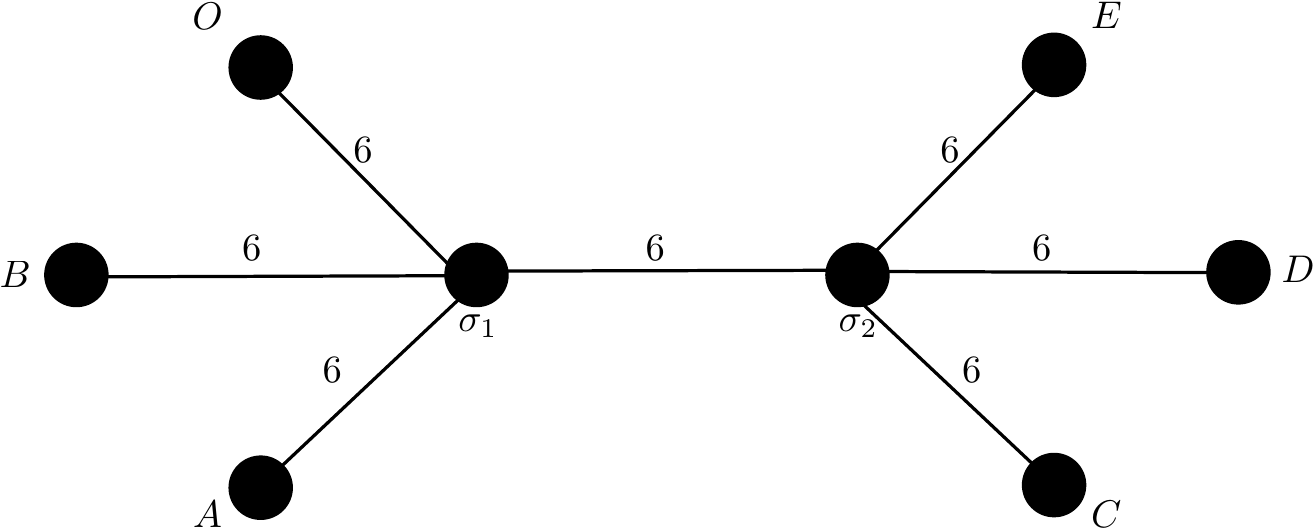}
\caption{\label{fig:ray8}  }
\end{figure}
\begin{table}[H]
\begin{center}
\scriptsize
\begin{tabular}{| c |c | c | c | c | c | c | }
\hline
Type & Target & Contributing Species & Vertex  & Example hyperthread & $e_{\sigma_1 A}$ & $e_{\sigma_1 \sigma_2}$  \\ 
\hline
\hline
$h_1$ & -1 & $(ABO)_2(CDE)_2$ & $\sigma_{1,2}$ &$\{\sigma_1;A;B;\sigma_2C;\sigma_2D)\}$ & 6 & 18 \\
\hline
$h_2$ & 2 & $(ABO)CDE, ABO(CDE)$ &$\sigma_{1,2}$ &$\{\sigma_1;A;B;\sigma_2C;O\}$ & 4 & 6\\
\hline
$h_3$ & 4 & $ABCDEO$ &$\sigma_{1,2}$ &$\{\sigma_1;A;B;\sigma_2C;\sigma_2D;\sigma_2E;O\}$ & 1 & 3\\
\hline
\end{tabular}
\end{center}
\label{tab:ray8}
\end{table}
\be
\begin{split}
 e_{\sigma_1 A}: \quad &6h_1+4h_2+h_3=6\\
 e_{\sigma_1 \sigma_2}: \quad &18h_1+6h_2+3h_3=6\\
&h_1 \rightarrow -1, \; h_2 \rightarrow 2, \; h_3 \rightarrow 4
\end{split}
\ee
\begin{figure}[H]
\centering
\begin{tabular}{cc}
\centering
\includegraphics[width=.5\textwidth,page=43]{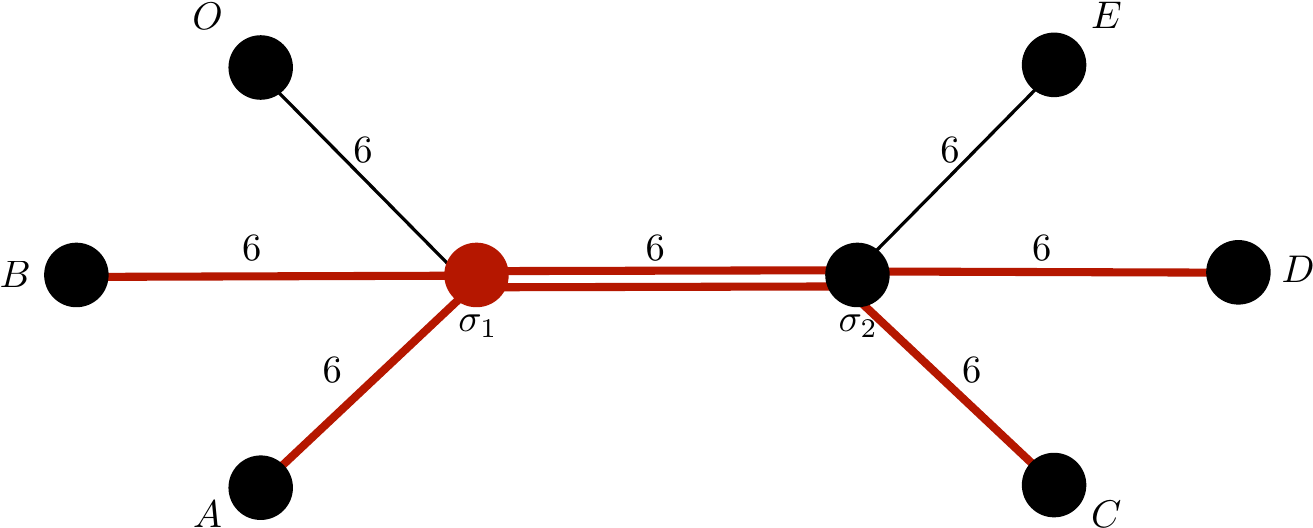}&
\includegraphics[width=.5\textwidth,page=44]{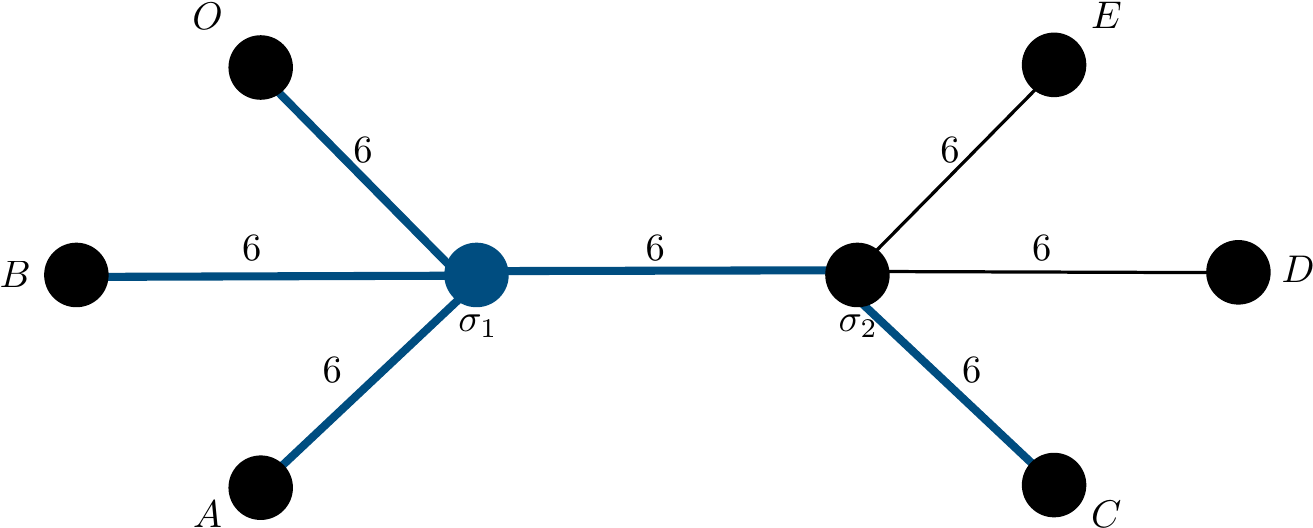}\\
$h_1$ & $h_2$ \\[6pt]
\multicolumn{2}{c}{\includegraphics[width=.5\textwidth,page=45]{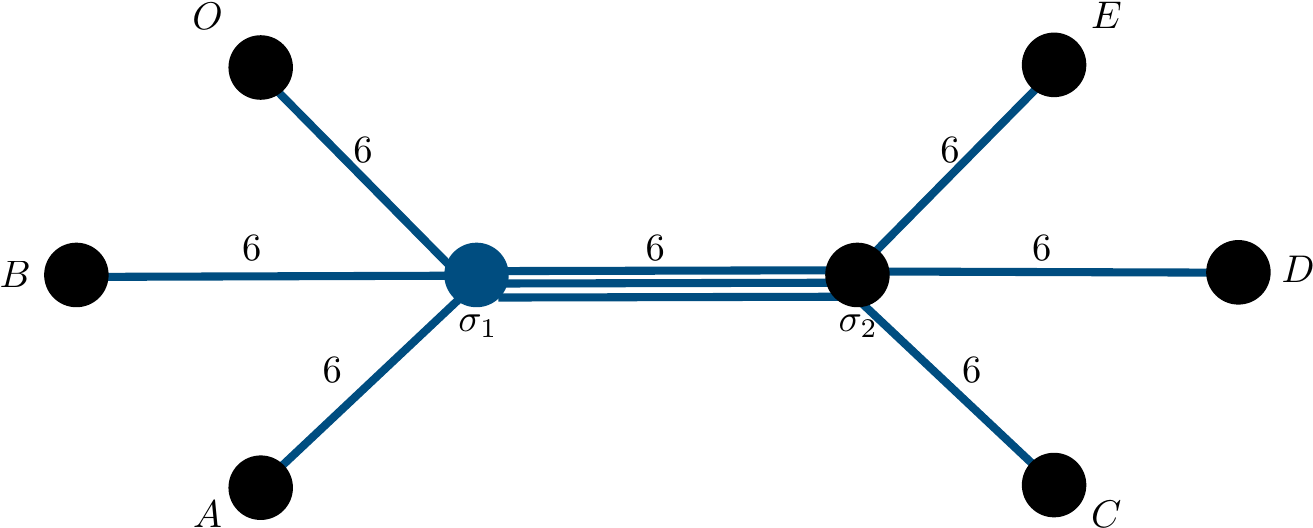}}\\
\multicolumn{2}{c}{$h_3$}
\end{tabular}
\caption{}
\end{figure}

\subsection*{Ray 9}
\be\label{eq:ray9S}
\mathcal{S}=6*\{ 11112; 2223223223; 3343443442; 43333; 2\}
\ee
\be\label{eq:ray9K}
\mathcal{K}=\{000000000000000;1-21-2141-2111-2114;2\}
\ee
\begin{figure}[H]
\centering
\includegraphics[width=.75\textwidth,page=46]{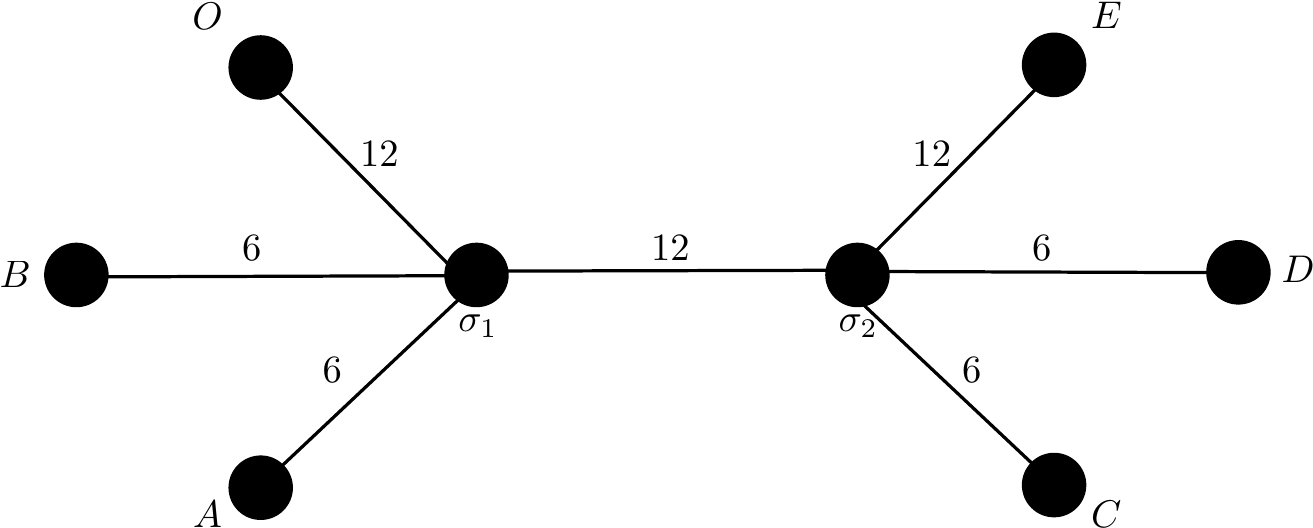}
\caption{\label{fig:ray9}  }
\end{figure}
\begin{table}[H]
\begin{center}
\scriptsize
\begin{tabular}{| c |c | c | c | c | c | c | c | }
\hline
Type & Target & Contributing Species & Vertex  & Example hyperthread & $e_{\sigma_1 A}$ & $e_{\sigma_1O}$ & $e_{\sigma_1 \sigma_2}$  \\ 
\hline
\hline
$h_1$ & 4 & $ABEO,CDEO$ & $\sigma_{1,2}$ &$\{\sigma_1;A;B;\sigma_2E;O)\}$ & 1 & 2 & 2 \\
\hline
$h_2$ & 1 &  \begin{tabular}{@{}c@{}}$AB(CDO)_2,(AB)CDE$,\\$(AB)(CD)EO$\end{tabular}  &$\sigma_{1,2}$ &$\{\sigma_1;A;B;\sigma_2C;\sigma_2D\}$ & 6 & 6 & 14\\
\hline
$h_3$ & -2 & $AB(CD)E,(AB)CDO$ &$\sigma_{1,2}$ &$\{\sigma_1;A;B;\sigma_2C;\sigma_2O\}$ & 3 & 2 & 8\\
\hline
$h_4$ & 2 & $ABCDEO$ &$\sigma_{1,2}$ &$\{\sigma_1;A;B;\sigma_2C;\sigma_2D;\sigma_2E;O\}$ & 1 & 1 & 3\\
\hline
\end{tabular}
\end{center}
\label{tab:ray9}
\end{table}
\be
\begin{split}
 e_{\sigma_1 A}: \quad &h_1+6h_2+3h_3+h_4=6\\
 e_{\sigma_1 O}: \quad &2h_1+6h_2+2h_3+h_4=12\\
 e_{\sigma_1 \sigma_2}: \quad &2h_1+14h_2+8h_3+3h_4=12\\
&h_1 \rightarrow 4, \; h_2 \rightarrow 1, \; h_3 \rightarrow -2, \; h_4 \rightarrow 2
\end{split}
\ee
\begin{figure}[H]
\centering
\begin{tabular}{cc}
\centering
\includegraphics[width=.5\textwidth,page=47]{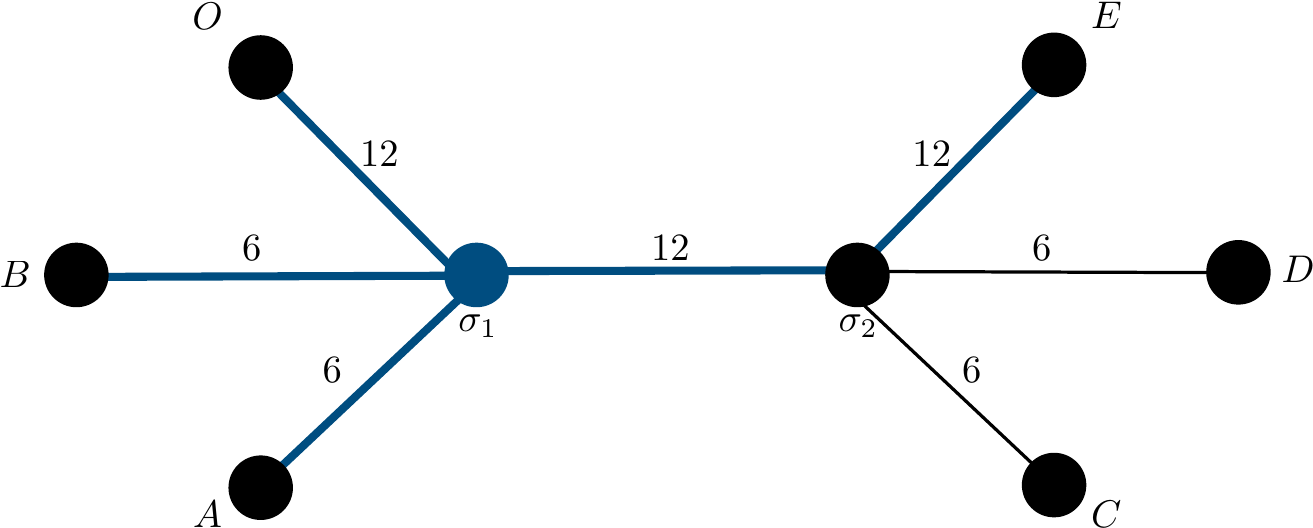}&
\includegraphics[width=.5\textwidth,page=48]{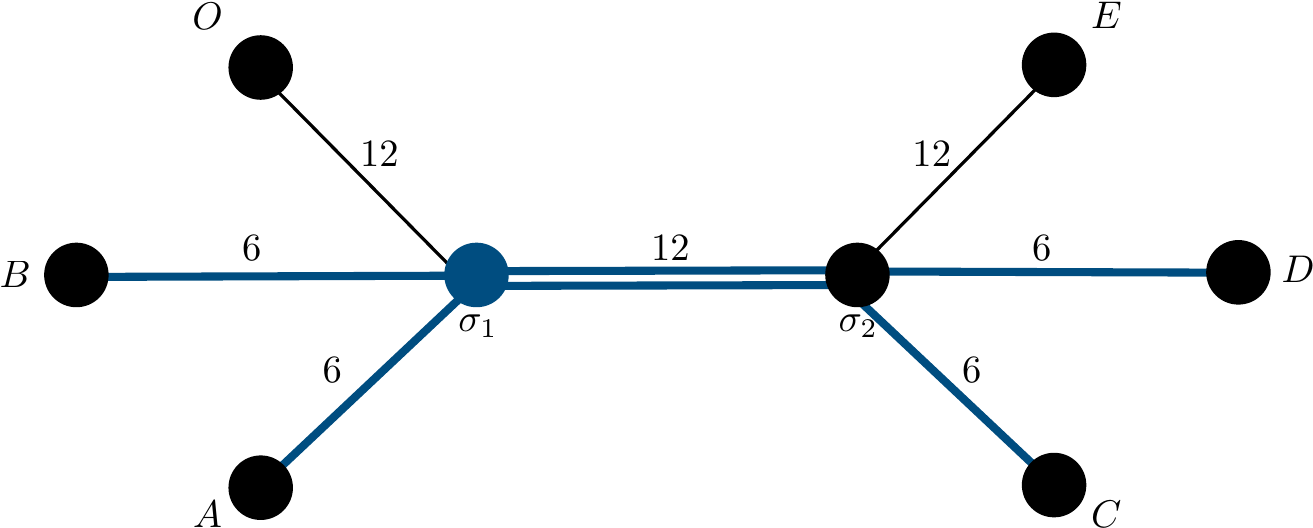}\\
$h_1$ & $h_2$ \\[6pt]
\includegraphics[width=.5\textwidth,page=49]{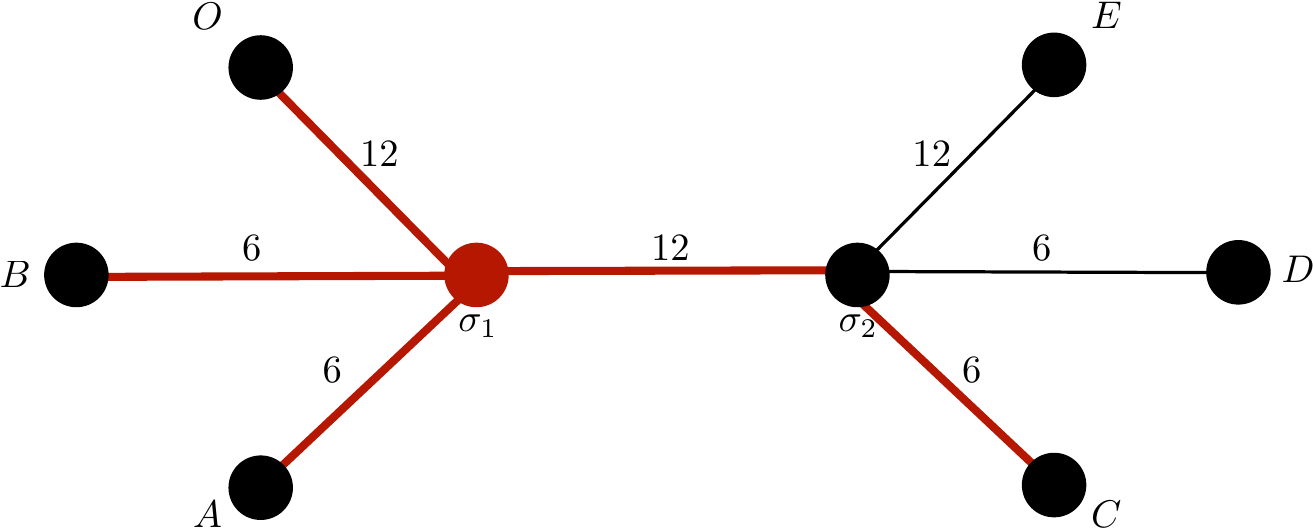}&
\includegraphics[width=.5\textwidth,page=50]{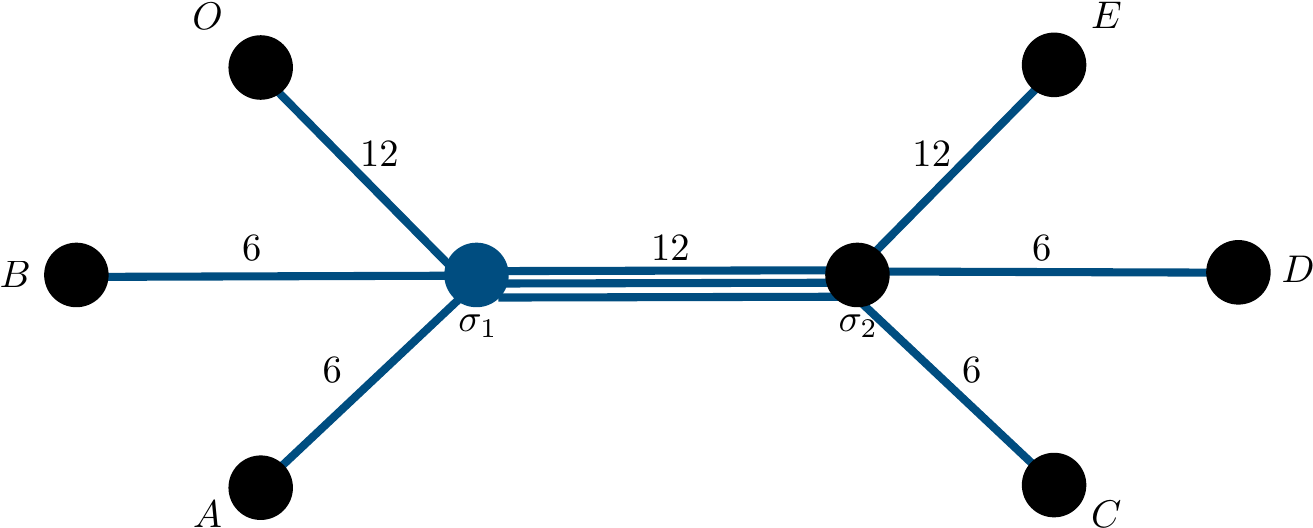}\\
$h_3$ & $h_4$ \\[6pt]
\end{tabular}
\caption{}
\end{figure}

\newpage
\subsection*{Ray 10}
\be\label{eq:ray10S}
\mathcal{S}=8*\{ 11111; 2222222222; 2333332332; 22222; 1\}
\ee
\be\label{eq:ray10K}
\mathcal{K}=\{000000000000000;202-2020-20220-202;4\}
\ee
\begin{figure}[H]
\centering
\includegraphics[width=.75\textwidth,page=56]{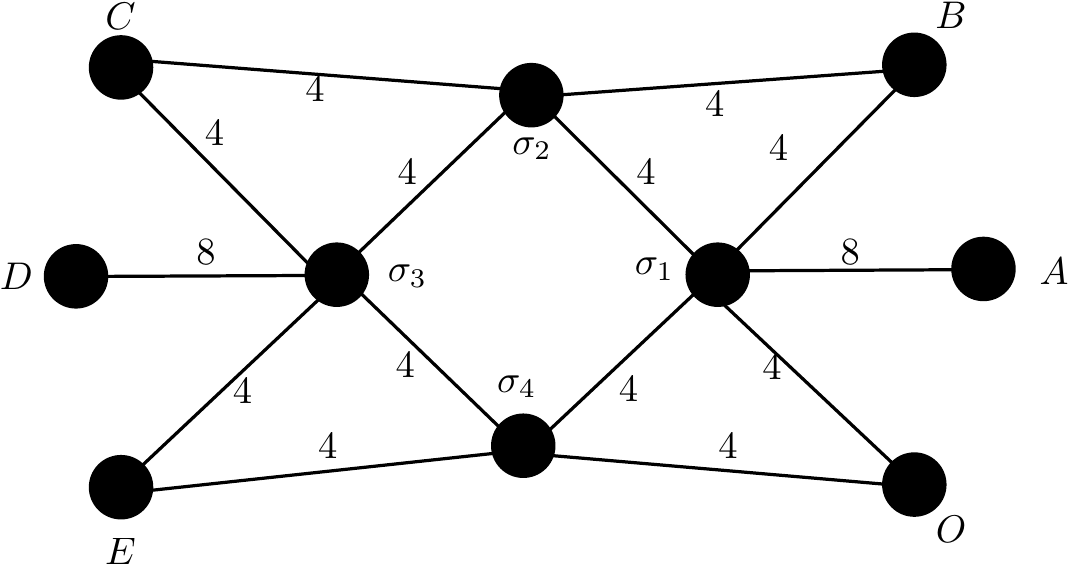}
\caption{\label{fig:ray10}  }
\end{figure}

\begin{table}[H]
\begin{center}
\scriptsize
\begin{tabular}{| c |c | c | c | c | c | c | c | c | }
\hline
Type & Target & Contributing Species & Vertex  & Example hyperthread & $e_{\sigma_1 A}$ & $e_{\sigma_1B}$ & $e_{\sigma_1 \sigma_2}$ & $e_{\sigma_2B}$ \\ 
\hline
\hline
$h_1$ & 2 & $ABCD,ADEO$ & $\sigma_{2,4}$ &$\{\sigma_2;\sigma_1A;B;C;\sigma_3D)\}$ & 2 & 0 & 1 & 1 \\
\hline
$h_2$ & 2 &  $AB(CE)O, (BO)CDE $  &$\sigma_{1,3}$ &$\{\sigma_1;A;B;\sigma_2C;O\}$ & 2 & 2 & 1 & 1\\
\hline
$h_3$ & -2 & $ABDE, ACDO, BCEO$ &$\sigma_{2,4}$ &$\{\sigma_2;\sigma_1A;\sigma_1B;\sigma_3D;\sigma_3E\}$ & 2 & 1 & 2 & 1\\
\hline
$h_4$ & 4 & $ABCDEO$ &$\sigma_{1,3}$ &$\{\sigma_1;A;B;\sigma_2C;\sigma_2\sigma_3D;\sigma_4E;O\}$ & 1 & $\frac{1}{2}$ & 1 & $\frac{1}{2}$\\
\hline
\end{tabular}
\end{center}
\label{tab:ray10}
\end{table}

\be
\begin{split}
 e_{\sigma_1 A}: \quad &2h_1+2h_2+2h_3+h_4=8\\
 e_{\sigma_1 B}: \quad &2h_2+h_3+\frac{1}{2}h_4=4\\
 e_{\sigma_1 \sigma_2}: \quad &h_1+h_2+2h_3+h_4=4\\
  e_{\sigma_2 B}: \quad &h_1+h_2+h_3+\frac{1}{2}h_4=4\\
&h_1 \rightarrow 2, \; h_2 \rightarrow 2, \; h_3 \rightarrow -2, \; h_4 \rightarrow 4
\end{split}
\ee
\begin{figure}[H]
\centering
\begin{tabular}{cc}
\centering
\includegraphics[width=.5\textwidth,page=57]{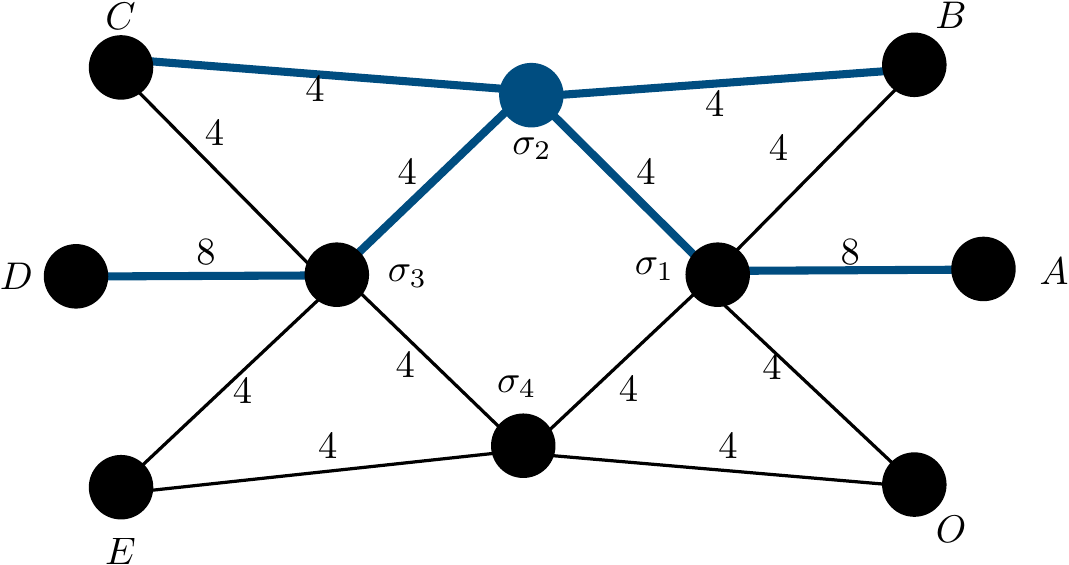}&
\includegraphics[width=.5\textwidth,page=58]{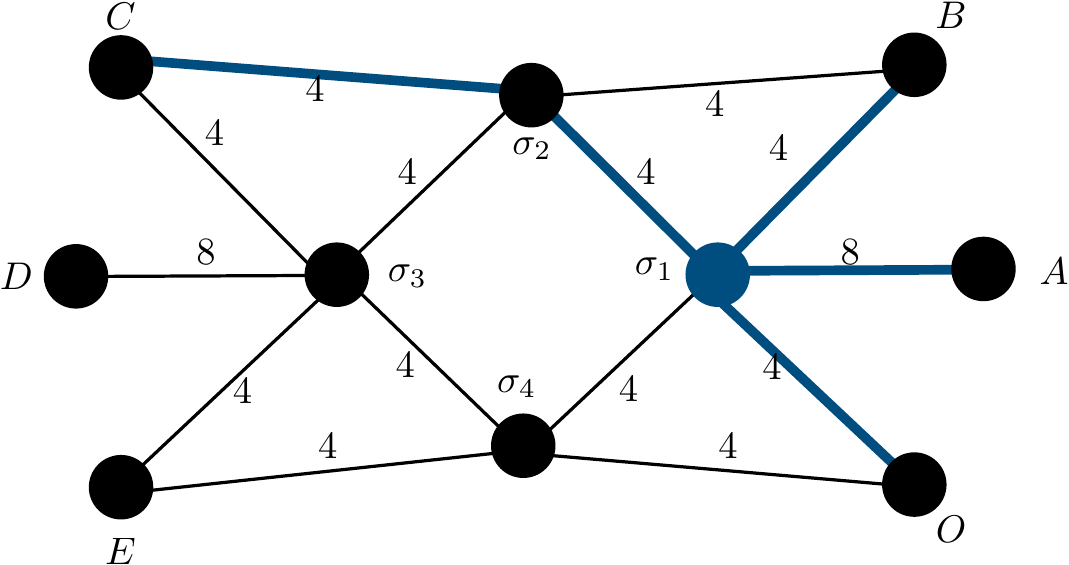}\\
$h_1$ & $h_2$ \\[6pt]
\includegraphics[width=.5\textwidth,page=59]{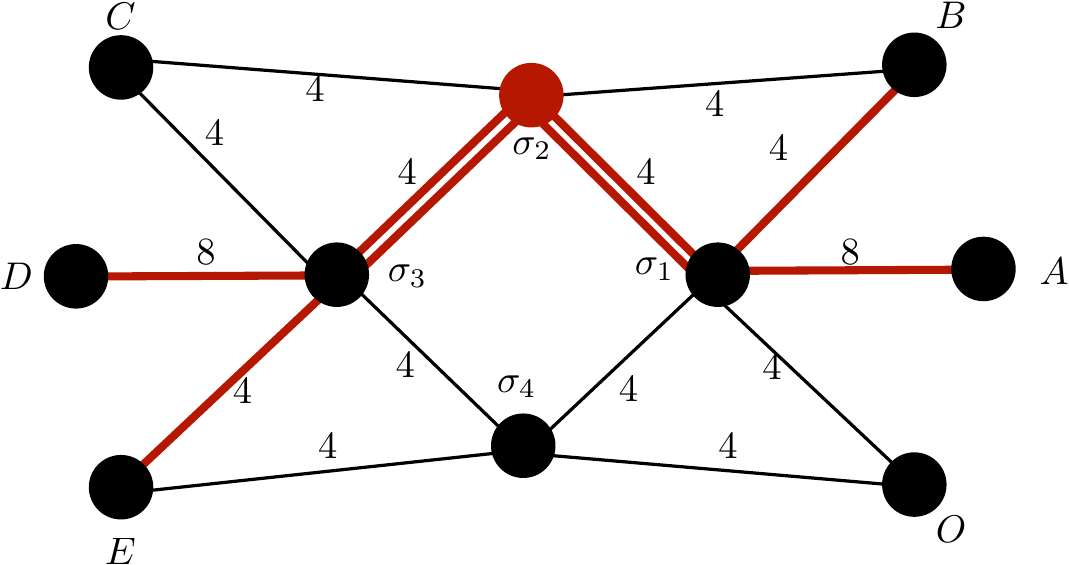}&
\includegraphics[width=.5\textwidth,page=60]{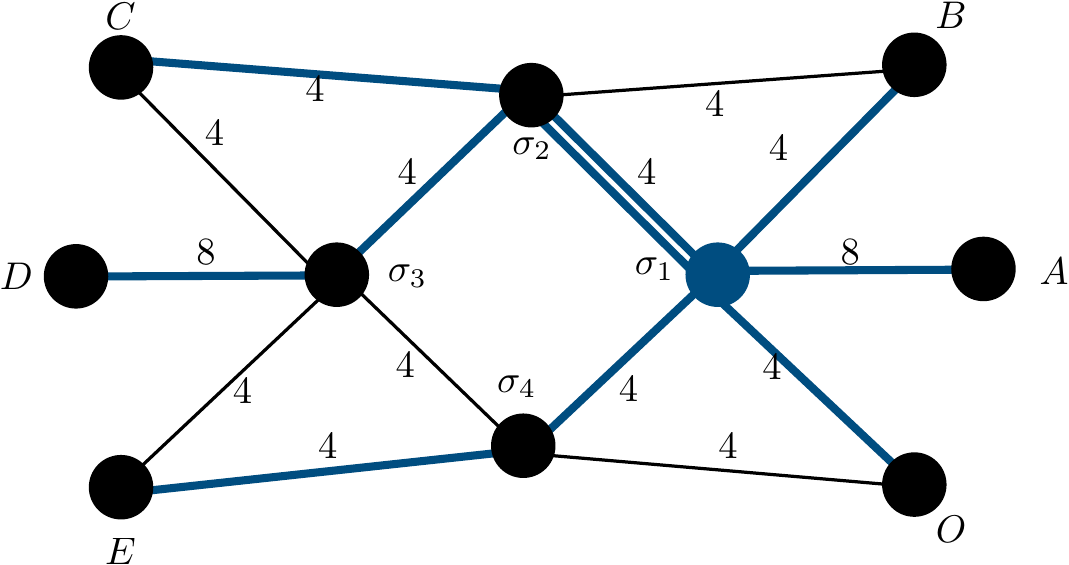}\\
$h_3$ & $h_4$ \\[6pt]
\end{tabular}
\caption{}
\end{figure}

\newpage
\subsection*{Ray 11}
\be\label{eq:ray11S}
\mathcal{S}=6*\{ 11222; 2333333444; 4445535354; 44433; 2\}
\ee

\be\label{eq:ray11K}
\mathcal{K}=\{000000000000000;-21111-211-24114-24;2\}
\ee

\begin{figure}[H]
\centering
\includegraphics[width=.5\textwidth,page=66]{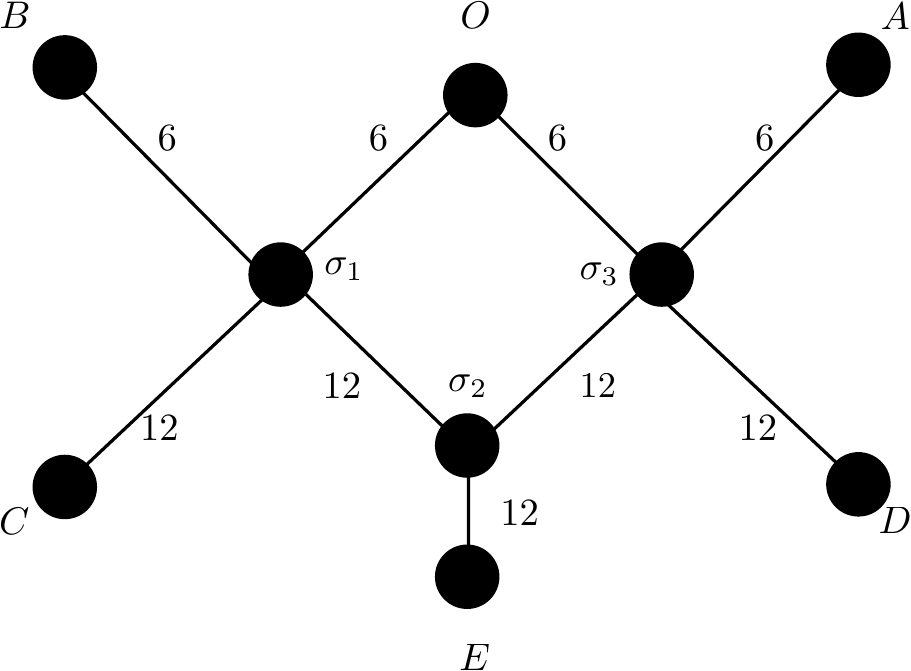}
\caption{\label{fig:ray11}  }
\end{figure}
\begin{adjustbox}{center}
\scriptsize
\begin{tabular}{| c |c | c | c | c | c | c | c | c | c |}
\hline
Type & Target & Contributing Species & Vertex  & Example hyperthread & $e_{\sigma_3 A}$ & $e_{\sigma_3D}$ & $e_{\sigma_3 \sigma_O}$ & $e_{\sigma_2 \sigma_3}$ & $e_{\sigma_2 E}$  \\ 
\hline
\hline
$h_1$ & -2 & $ABCD$ & $\sigma_{1,3}$ &$\{\sigma_3;A;\sigma_2\sigma_1B;\sigma_2\sigma_1C;D\}$ & 1 & 1 & 0 & 2 & 0 \\
\hline
$h_2$ & 4 & $ADEO,BCEO$ & $\sigma_{1,3}$ &$\{\sigma_3;A;D;\sigma_2E;O\}$ & 1 & 1 & 1 & 1 & 2 \\
\hline
$h_3$ & 4 & $CDEO$ & $\sigma_{1,3}$ &$\{\sigma_3;\sigma_2\sigma_1C;D;\sigma_2E;O\}$ & 0 & 1 & $\frac{1}{2}$ & $\frac{3}{2}$ & 1 \\
\hline
$h_4$ & -2 & $ABEO$ & $\sigma_{1,3}$ &$\{\sigma_3;A;\sigma_2\sigma_1B;\sigma_2E;O\}$ & 1 & 0 & $\frac{1}{2}$ & $\frac{3}{2}$ & 1 \\
\hline
$h_5$ & -2 & $ACEO,BDEO$ & $\sigma_{1,3}$ &$\{\sigma_3;A;\sigma_2\sigma_1C;\sigma_2E;O\}$ & 1 & 1 & 1 & 3 & 2 \\
\hline
$h_6$ & 1 & $(ABCD)_3E$ & $\sigma_{1,3}$ &$\{\sigma_3;A;\sigma_2\sigma_1B;D;\sigma_2E\}$ & 3 & 3 & 0 & 6 & 4 \\
\hline
$h_7$ & 1 & $(ABCD)_3O$ & $\sigma_{1,3}$ &$\{\sigma_3;A;\sigma_2\sigma_1B;D;O\}$ & 3 & 3 & 2 & 4 & 0 \\
\hline
$h_8$ & 2 & $ABCDEO$ & $\sigma_{1,3}$ &$\{\sigma_3;A;\sigma_2\sigma_1B;\sigma_2\sigma_1C;D;\sigma_2E;O\}$ & 1 & 1 & $\frac{1}{2}$ & $\frac{5}{2}$ & 1 \\
\hline
\end{tabular}
\label{tab:ray10}
\end{adjustbox}
\be
\begin{split}
 e_{\sigma_3 A}: \quad &h_1+h_2+h_4+h_5+3h_6+3h_7+h_8=6\\
 e_{\sigma_3 D}: \quad &h_1+h_2+h_3+h_5+3h_6+3h_7+h_8=12\\
 e_{\sigma_3 O}: \quad &h_2+\frac{1}{2}h_3+\frac{1}{2}h_4+h_5+2h_7+\frac{1}{2}h_8=6\\
 e_{\sigma_2 \sigma_3}: \quad &2h_1+h_2+\frac{3}{2}h_3+\frac{3}{2}h_4+3h_5+6h_6+4h_7+\frac{5}{2}h_8=12\\
 e_{\sigma_2 E}: \quad &2h_2+h_3+h_4+2h_5+4h_6+h_8=12\\
&h_1 \rightarrow -2, \; h_2 \rightarrow 4, \; h_3 \rightarrow 4, \; h_4 \rightarrow -2\; h_5 \rightarrow -2,\; h_6 \rightarrow 1,\; h_7 \rightarrow 1,\; h_8 \rightarrow 2
\end{split}
\ee
\begin{figure}[H]
\centering
\begin{tabular}{cc}
\centering
\includegraphics[width=.42\textwidth,page=67]{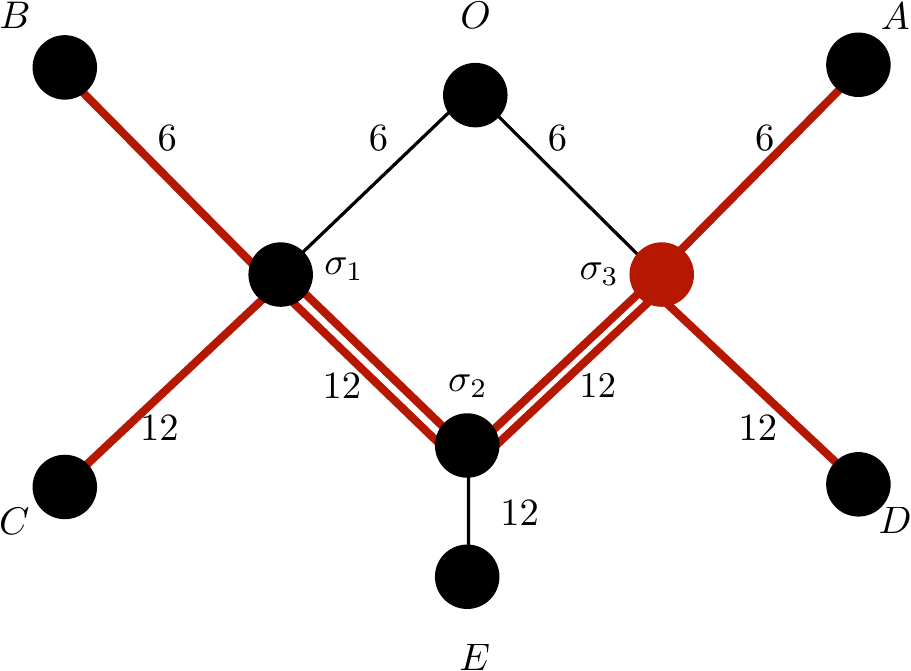}&
\includegraphics[width=.42\textwidth,page=68]{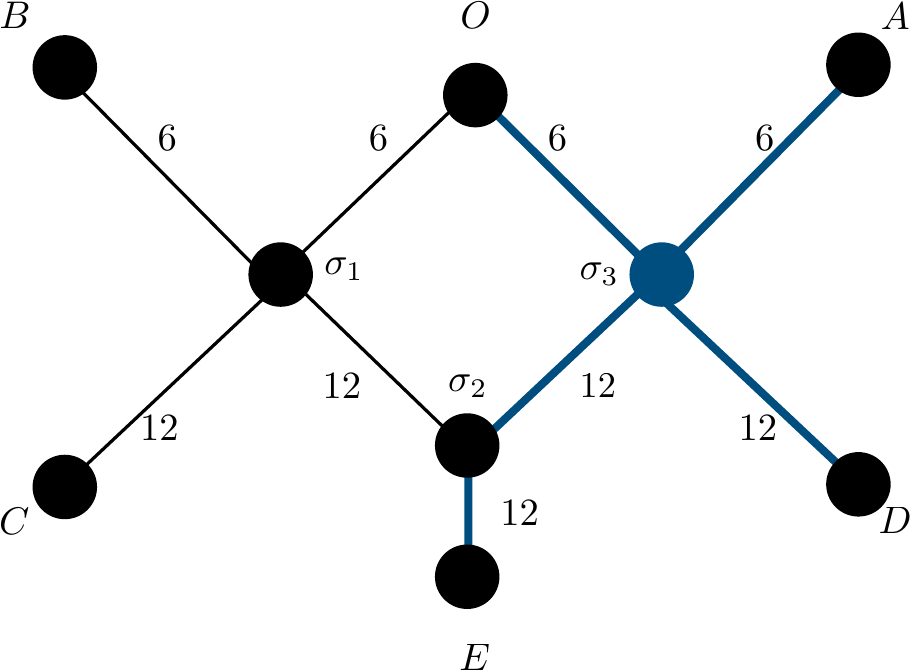}\\
$h_1$ & $h_2$ \\[6pt]
\includegraphics[width=.42\textwidth,page=69]{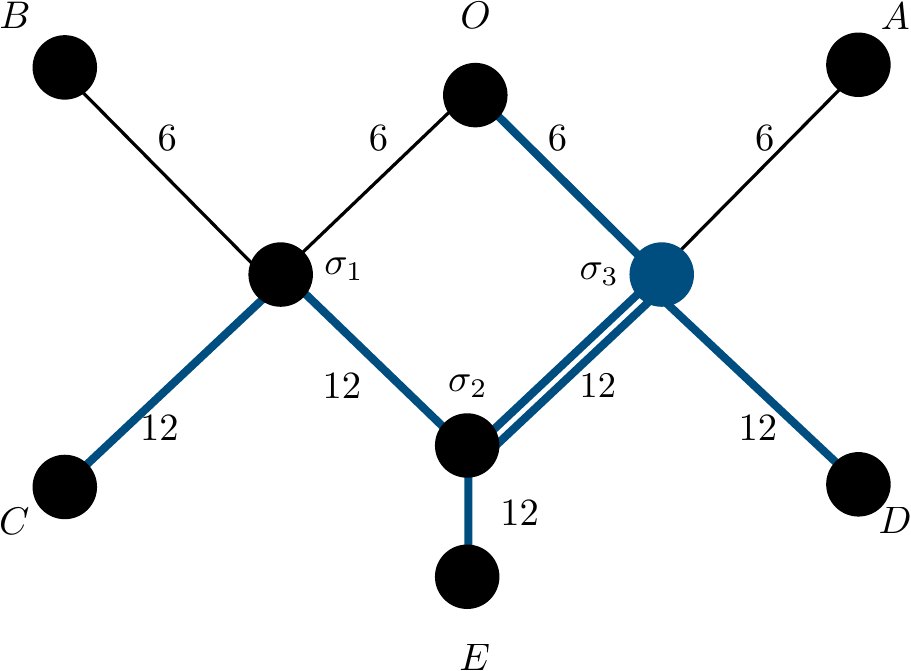}&
\includegraphics[width=.42\textwidth,page=70]{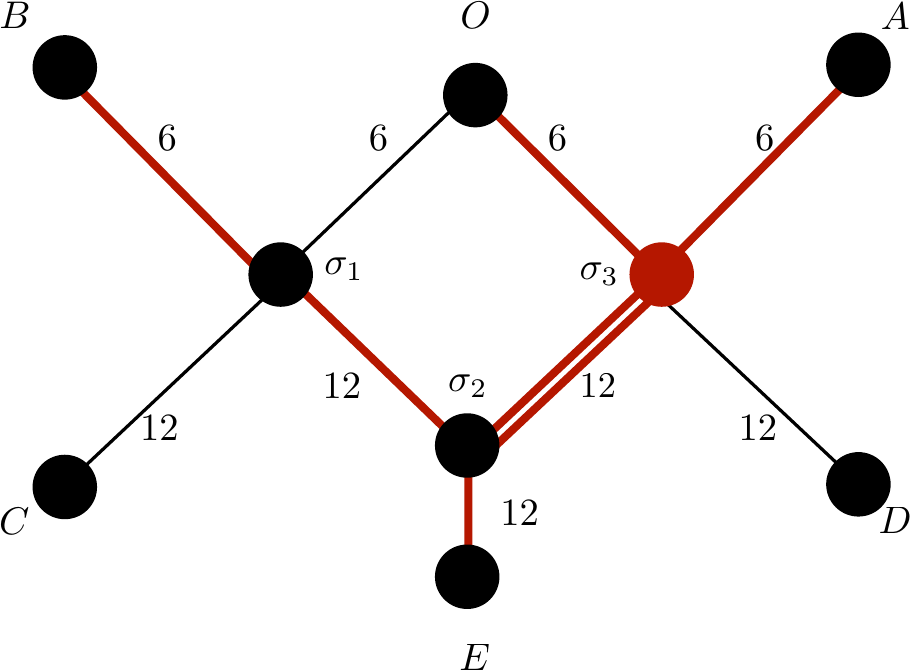}\\
$h_3$ & $h_4$ \\[6pt]
\includegraphics[width=.42\textwidth,page=71]{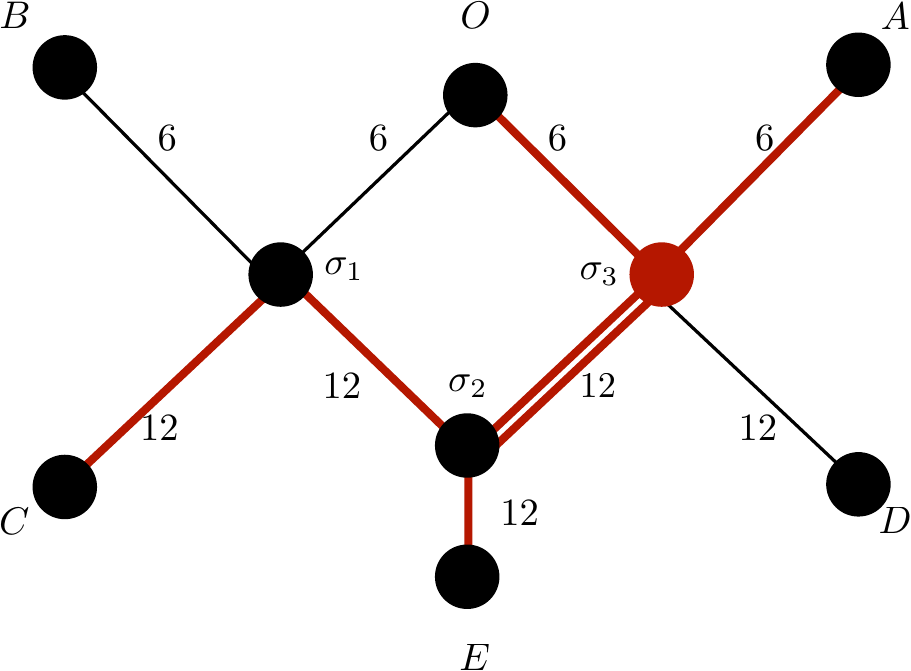}&
\includegraphics[width=.42\textwidth,page=72]{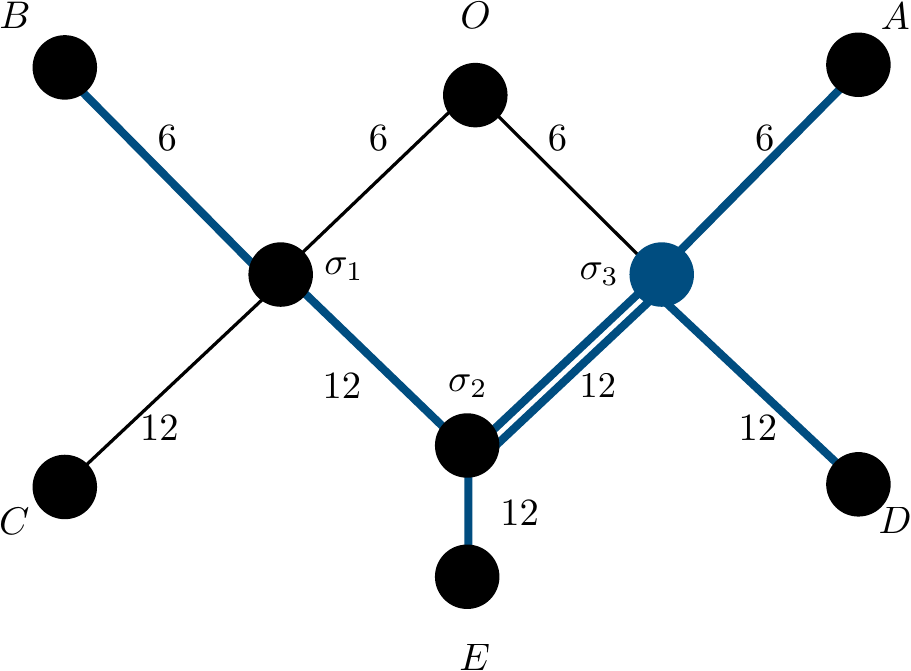}\\
$h_5$ & $h_6$ \\[6pt]
\includegraphics[width=.42\textwidth,page=73]{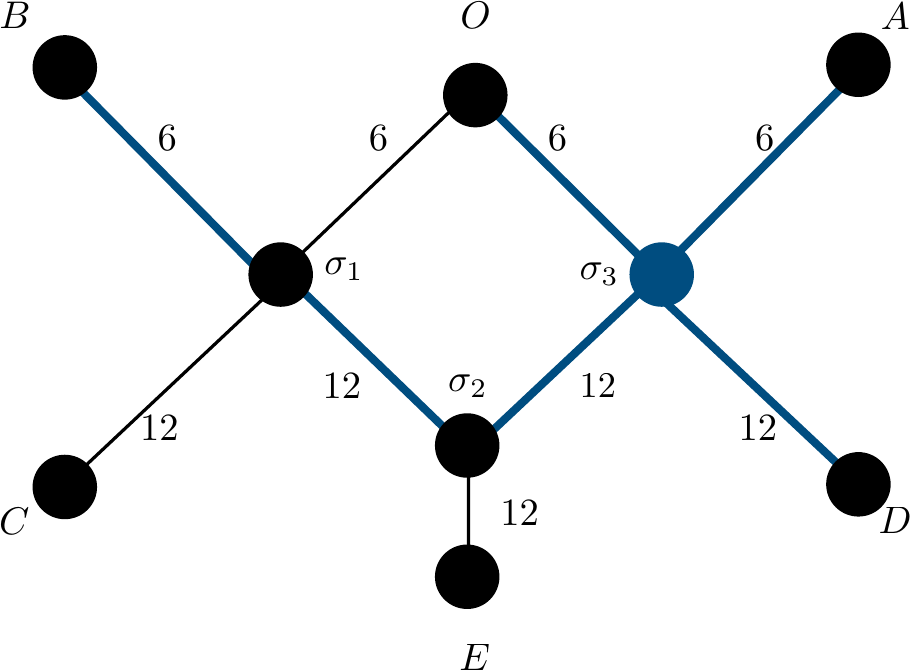}&
\includegraphics[width=.42\textwidth,page=74]{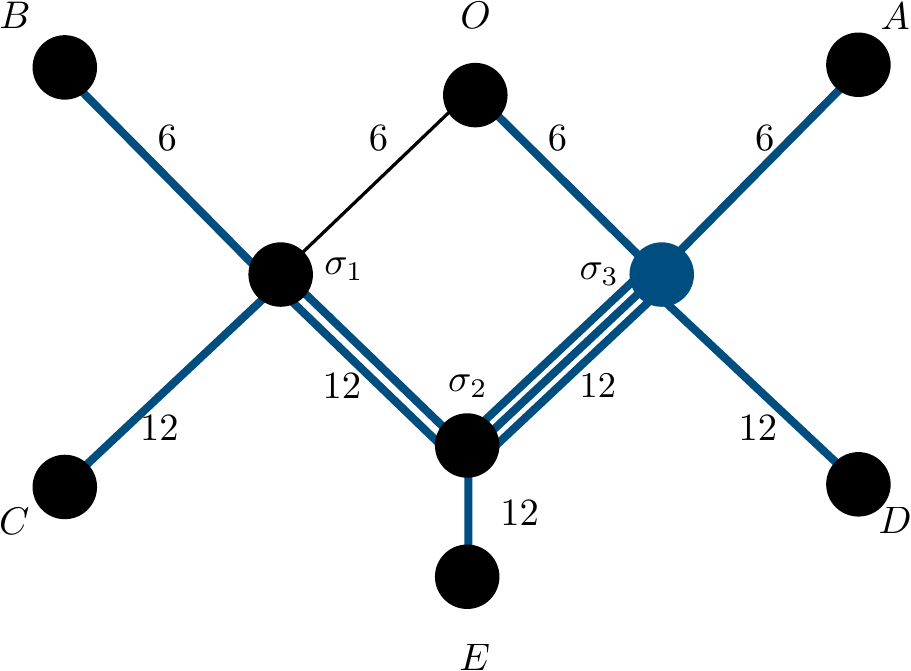}\\
$h_7$ & $h_8$ \\[6pt]
\end{tabular}
\caption{}
\end{figure}

\subsection*{Ray 12}
\be\label{eq:ray12S}
\mathcal{S}=12*\{ 11111; 2222222222; 3323323232; 22222; 1\}
\ee

\be\label{eq:ray12K}
\mathcal{K}=\{000000000000000;-42222222-4222242;4\}
\ee

\begin{figure}[H]
\centering
\includegraphics[width=.5\textwidth,page=26]{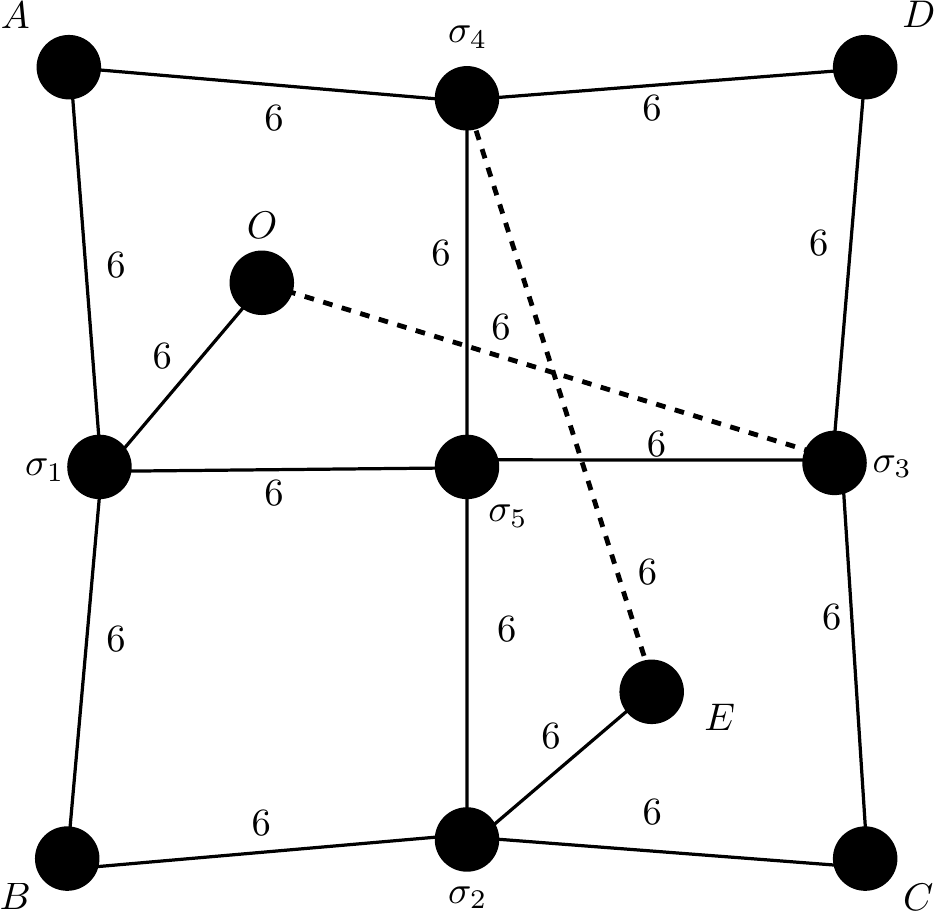}
\caption{\label{fig:ray12}  }
\end{figure}

\begin{table}[H]
\begin{center}
\scriptsize
\begin{tabular}{| c |c | c | c | c | c | c | c |}
\hline
Type & Target & Contributing Species & Vertex  & Example hyperthread & $e_{\sigma_1 A}$ & $e_{\sigma_1 O}$ & $e_{\sigma_1 \sigma_5}$  \\ 
\hline
\hline
$h_1$ & -4 & $ABCD$ & $\sigma_5$ &$\{\sigma_5;\sigma_1A;\sigma_1B;\sigma_3C;\sigma_3D\}$ & $\frac{1}{2}$ & 0 & 1 \\
\hline
$h_2$ & -4 & $ACEO, BDEO$ &$\sigma_5$ &$\{\sigma_5;\sigma_1A;\sigma_2C;\sigma_2E;\sigma_1O\}$ & $\frac{1}{2}$ & 1 & 2\\
\hline
$h_3$ & 2 & All other 4-threads &$\sigma_{1,2,3,4}$ &$\{\sigma_1;A;B;\sigma_5\sigma_3D;O\}$ & 4 & 4 & 6\\
\hline
$h_4$ & 4 & $ABCDEO$ &$\sigma_5$ &$\{\sigma_5;\sigma_1A;\sigma_2B;\sigma_3C;\sigma_4D;\sigma_4E;\sigma_1O\}$ & $\frac{1}{2}$ & $\frac{1}{2}$ & $\frac{3}{2}$\\
\hline
\end{tabular}
\end{center}
\label{tab:ray12}
\end{table}
\be
\begin{split}
 e_{\sigma_1 A}: \quad &\frac{1}{2}h_1+\frac{1}{2}h_2+4h_3+\frac{1}{2}h_4=6\\
 e_{\sigma_1 O}: \quad &h_2+4h_3+\frac{1}{2}h_4=6\\
 e_{\sigma_1 \sigma_5}: \quad &h_1+2h_2+6h_3+\frac{3}{2}h_4=6\\
&h_1 \rightarrow -4, \; h_2 \rightarrow -4, \; h_3 \rightarrow 2, \; h_4\rightarrow 4
\end{split}
\ee
\begin{figure}[H]
\centering
\begin{tabular}{cc}
\centering
\includegraphics[width=.45\textwidth,page=27]{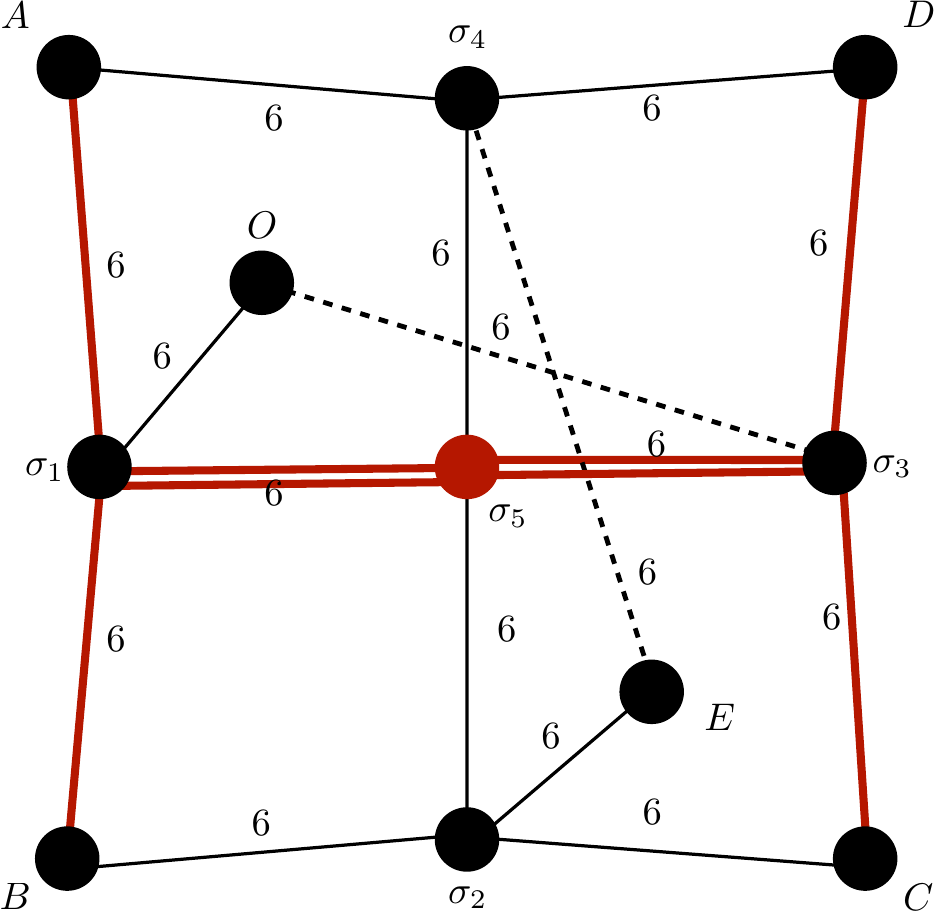}&
\includegraphics[width=.45\textwidth,page=28]{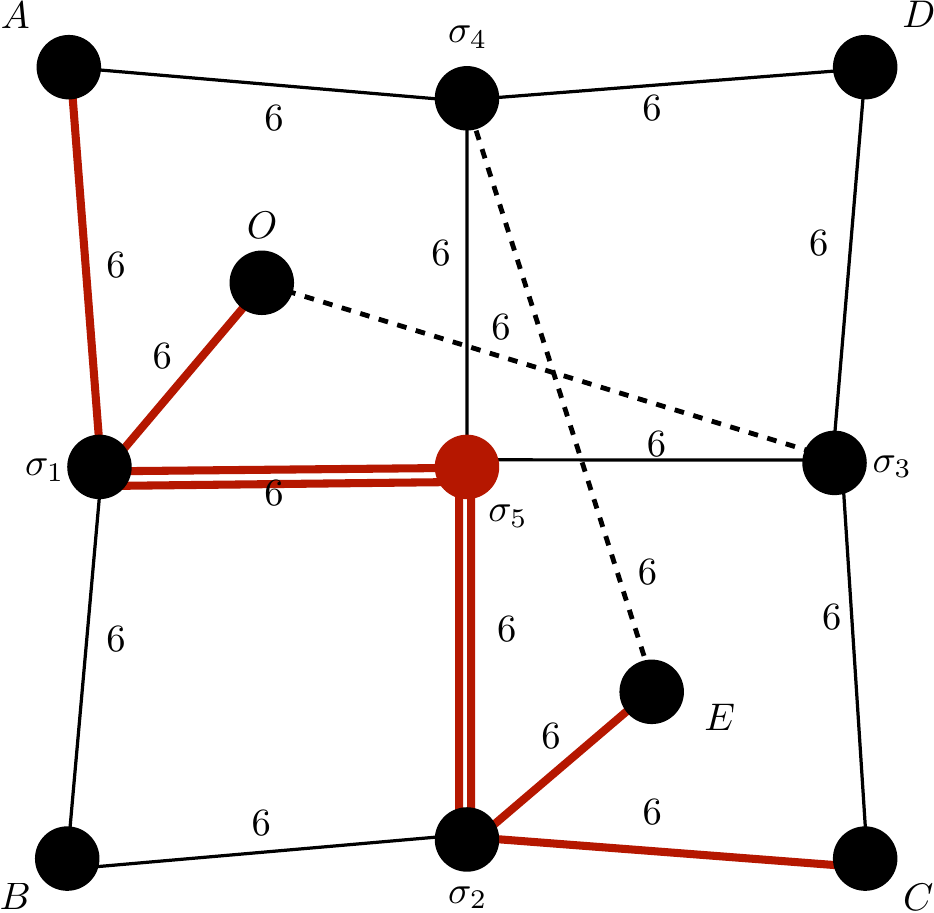}\\
$h_1$ & $h_2$ \\[6pt]
\includegraphics[width=.45\textwidth,page=29]{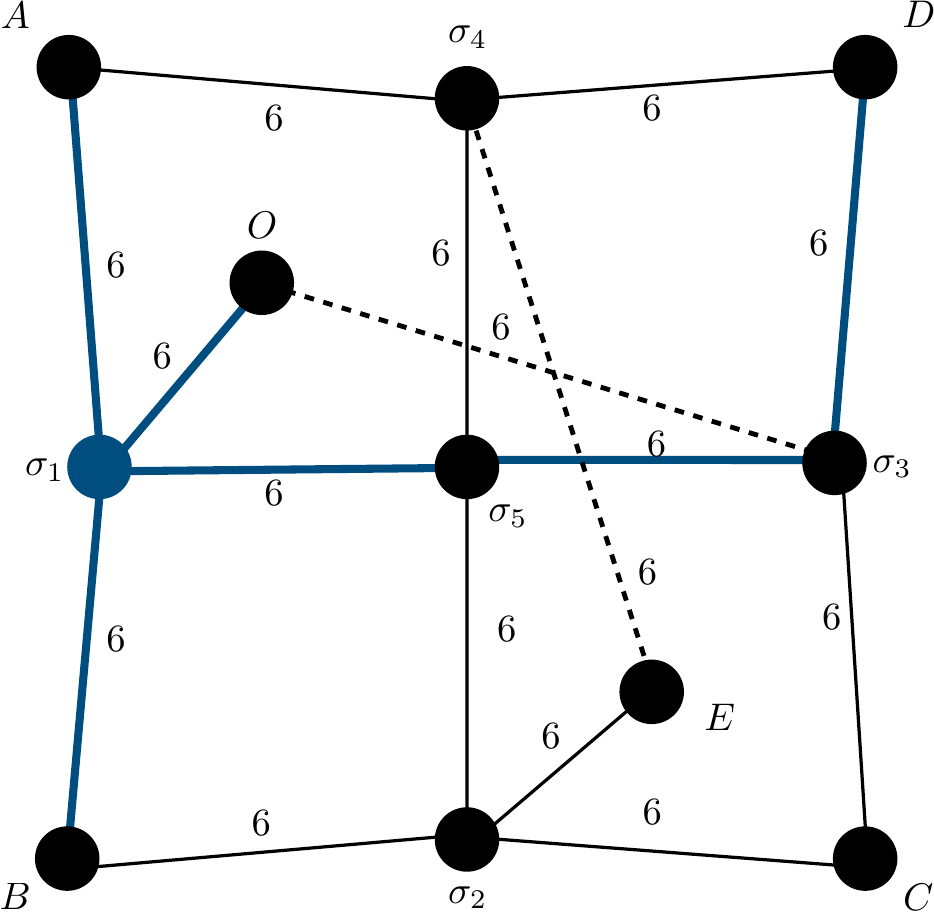}&
\includegraphics[width=.45\textwidth,page=30]{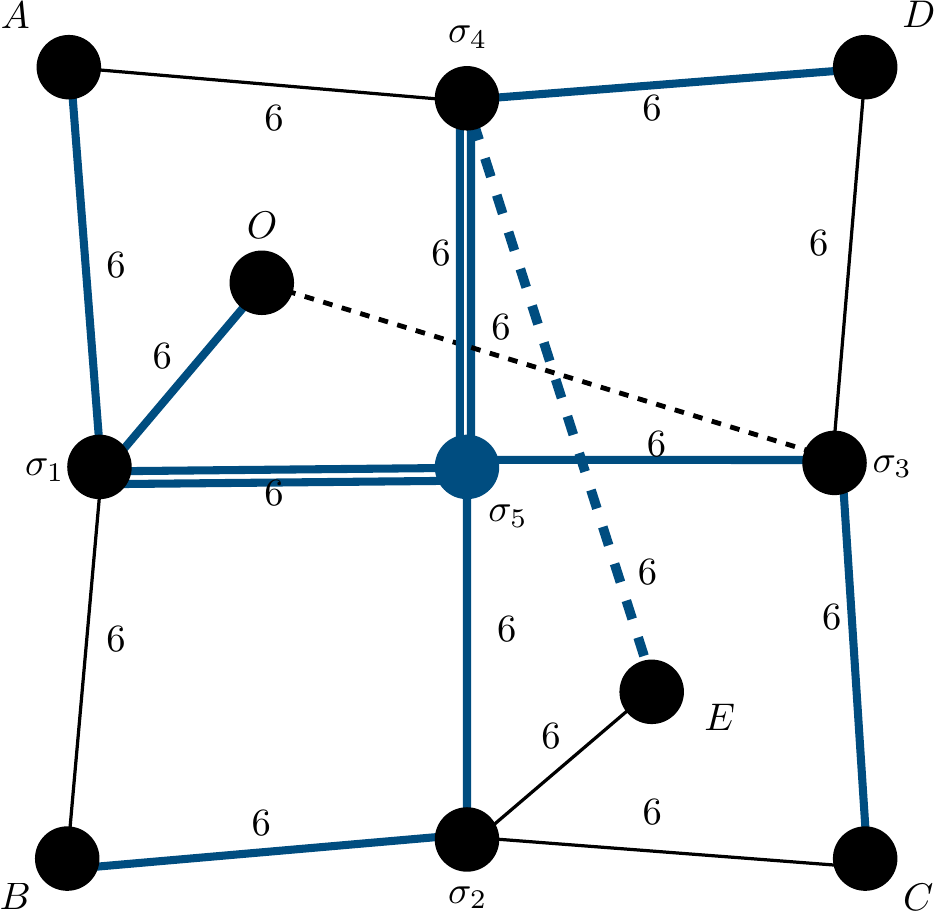}\\
$h_3$ & $h_4$ \\[6pt]
\end{tabular}
\caption{}
\end{figure}

\newpage
\subsection*{Ray 13}
\be\label{eq:ray13S}
\mathcal{S}=2*\{ 22222; 4444444444; 6466666464; 44444; 2\}
\ee
\be\label{eq:ray13K}
\mathcal{K}=\{000000000000000;00002000001-11-11;6\}
\ee
\begin{figure}[H]
\centering
\includegraphics[width=.5\textwidth,page=51]{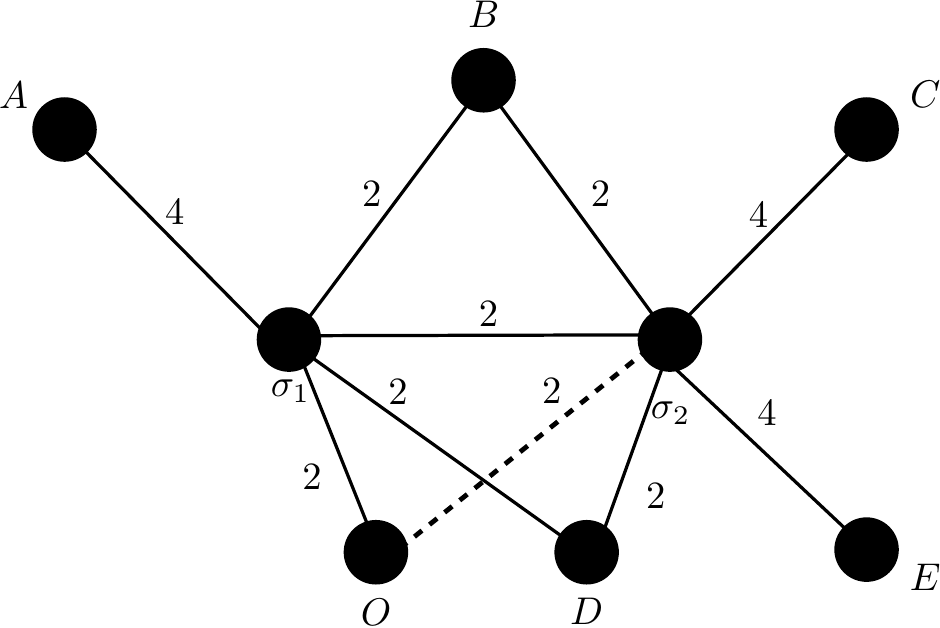}
\caption{\label{fig:ray13}  }
\end{figure}
\begin{table}[H]
\begin{center}
\scriptsize
\begin{tabular}{| c |c | c | c | c | c | c | c | c | c | }
\hline
Type & Target & Contributing Species & Vertex  & Example hyperthread & $e_{\sigma_1 A}$ & $e_{\sigma_1 B}$ & $e_{\sigma_2 B}$  & $e_{\sigma_2 C}$  & $e_{\sigma_1 \sigma_2}$  \\ 
\hline
\hline
$h_1$ & 2 & $ABDO$ & $\sigma_1$ &$\{\sigma_1;A;B;D;O)\}$ & 1 & 1 & 0 & 0 & 0\\
\hline
$h_2$ & -1 & $BDO(CE)$ &$\sigma_2$ &$\{\sigma_2;B;C;D;O\}$ & 0 & 0 & 2 & 1&0\\
\hline
$h_3$ & 1 & $(BDO)_2CE$ &$\sigma_1$ &$\{\sigma_2;B;C;D;E\}$ & 0 & 0 & 2 & 3 & 0\\
\hline
$h_4$ & 2 & $ABCDEO$ &$\sigma_2$ &$\{\sigma_2;\sigma_1A;B;C;D;E;O\}$ & 1 & 0 & 1 & 1 & 1\\
\hline
\end{tabular}
\end{center}
\label{tab:ray13}
\end{table}
\be
\begin{split}
 e_{\sigma_1 A}: \quad &h_1+h_4=4\\
 e_{\sigma_1 B}: \quad &h_1=2\\
 e_{\sigma_2 B}: \quad &2h_2+2h_3+h_4=2\\
 e_{\sigma_2 C}: \quad &h_2+3h_3+h_4=4\\
 e_{\sigma_1\sigma_2 }: \quad &h_4=2\\
&h_1 \rightarrow 2, \; h_2 \rightarrow -1, \; h_3 \rightarrow 1, \; h_4 \rightarrow 2
\end{split}
\ee
\begin{figure}[H]
\centering
\begin{tabular}{cc}
\centering
\includegraphics[width=.45\textwidth,page=52]{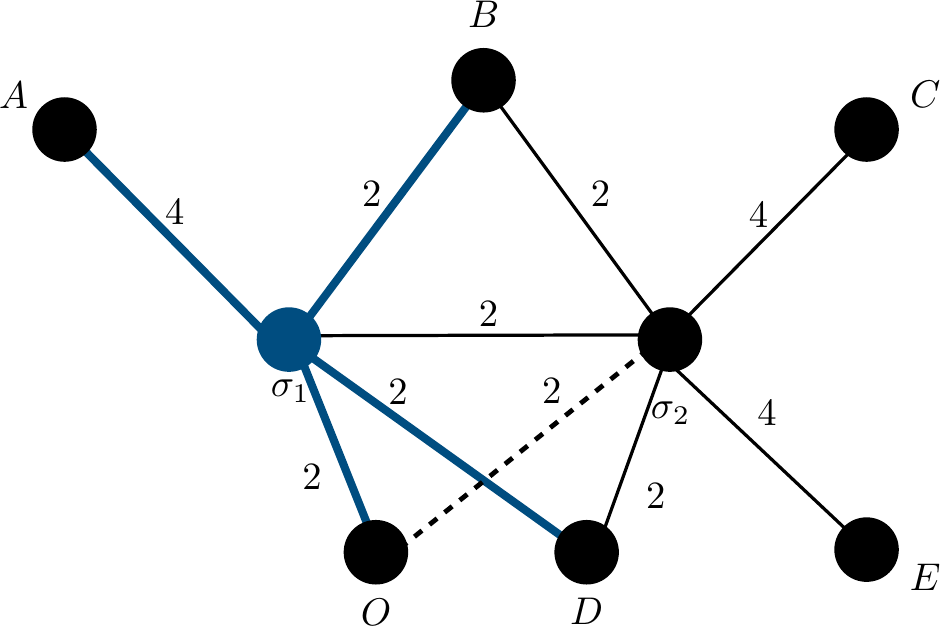}&
\includegraphics[width=.45\textwidth,page=53]{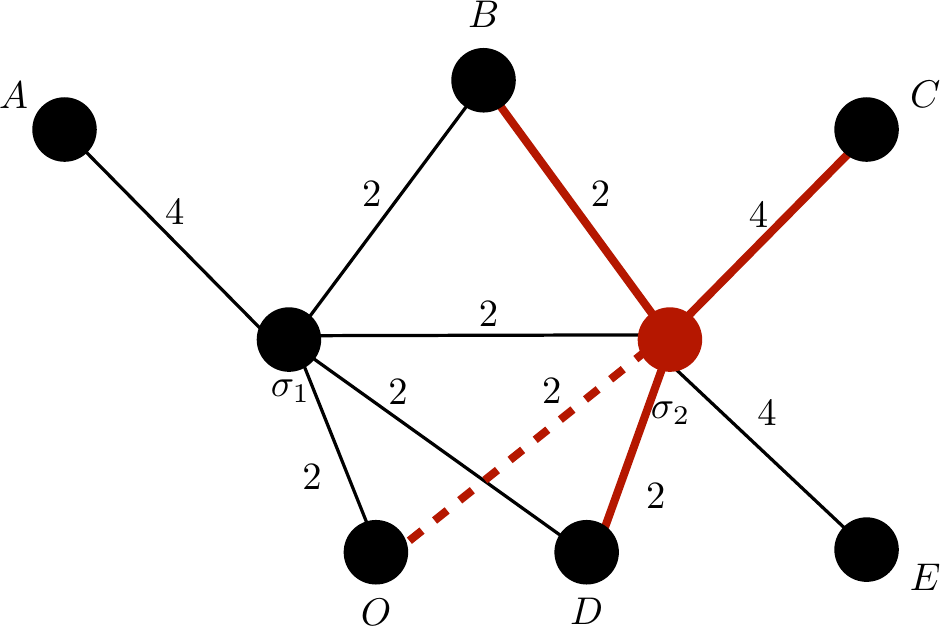}\\
$h_1$ & $h_2$ \\[6pt]
\includegraphics[width=.45\textwidth,page=54]{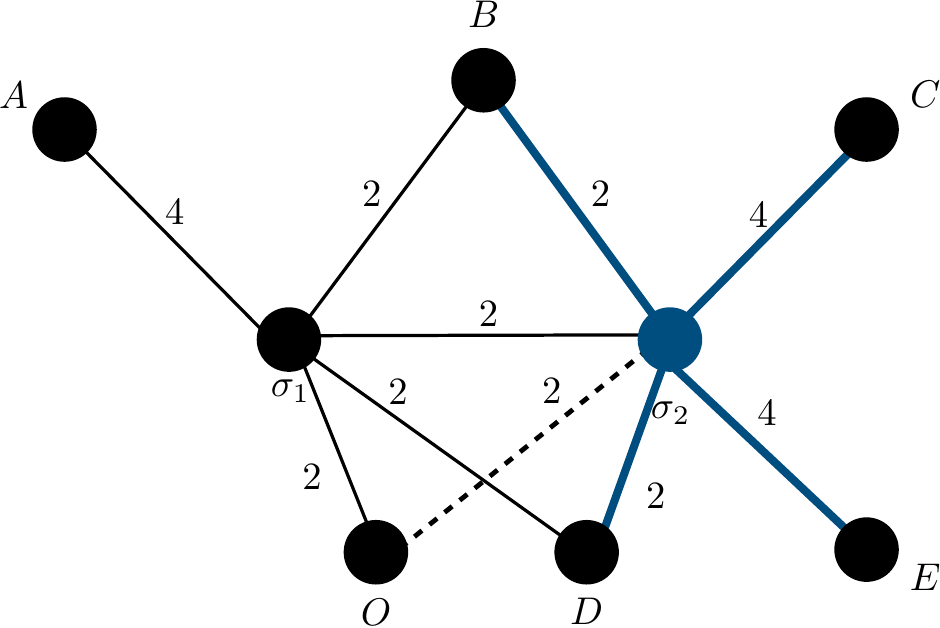}&
\includegraphics[width=.45\textwidth,page=55]{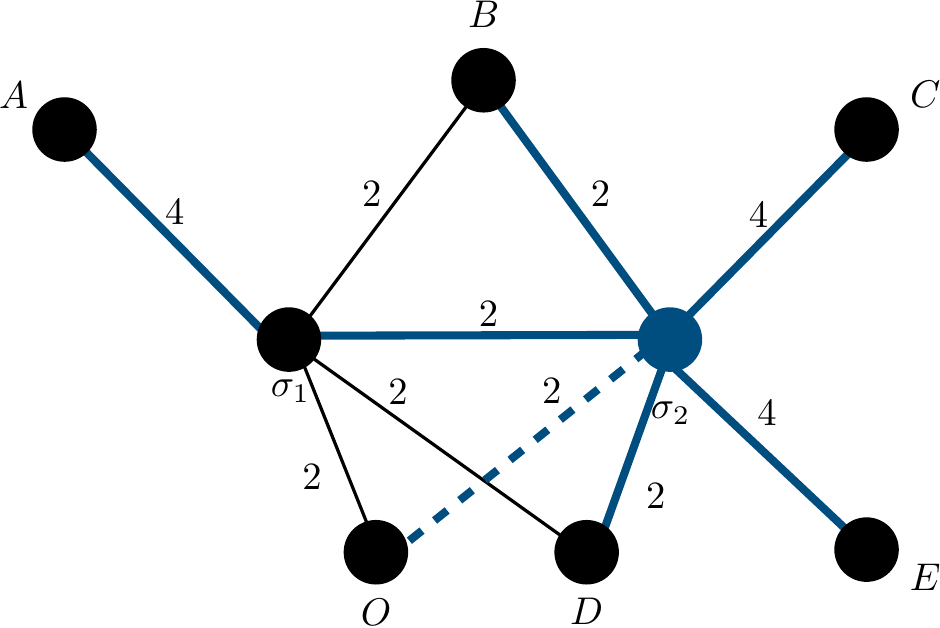}\\
$h_3$ & $h_4$ \\[6pt]
\end{tabular}
\caption{}
\end{figure}

\newpage
\subsection*{Ray 14}
\be\label{eq:ray14S}
\mathcal{S}=12*\{ 22223; 4445445455; 6476776575; 65555; 3\}
\ee
\be\label{eq:ray14K}
\mathcal{K}=\{000000000000000;4-2-2-2104-2-2444-810-210;8\}
\ee

\begin{figure}[H]
\centering
\includegraphics[width=.5\textwidth,page=75]{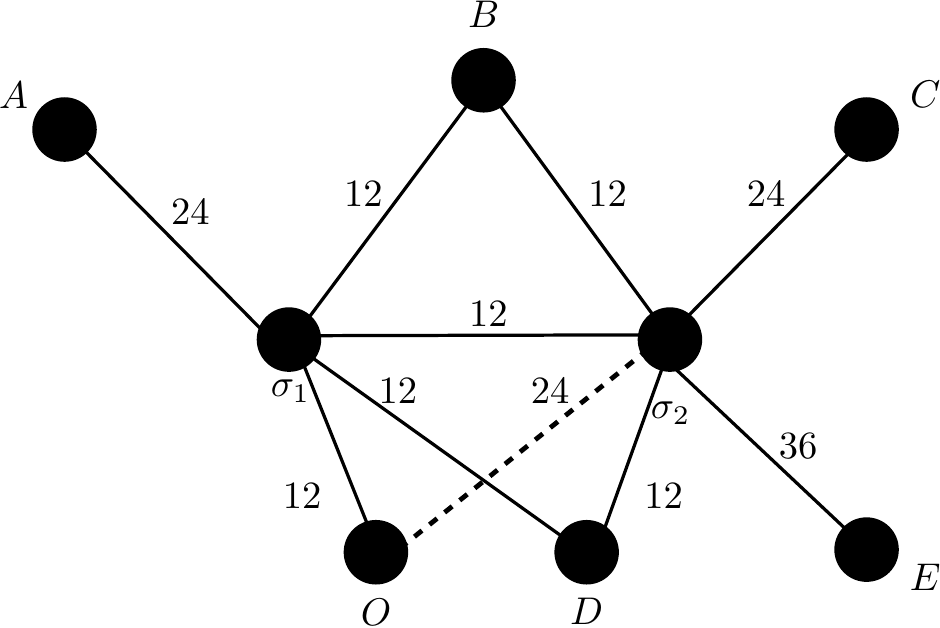}
\caption{\label{fig:ray14}  }
\end{figure}

\begin{adjustbox}{center}
\scriptsize
\begin{tabular}{| c |c | c | c | c | c | c | c | c | c | c | c | c |}
\hline
Type & Target & Contributing Species & Vertex  & Example hyperthread & $e_{\sigma_1 A}$ & $e_{\sigma_1B}$ & $e_{\sigma_1 O}$ & $e_{\sigma_1 \sigma_2}$ & $e_{\sigma_2 B}$  & $e_{\sigma_2 O}$ & $e_{\sigma_2 C}$ & $e_{\sigma_2 E}$  \\ 
\hline
\hline
$h_1$ & 10 & $ABDO$ & $\sigma_1$ &$\{\sigma_1;A;B;D;O\}$ & 1 & 1 & 1 & 0 & 0 & 0 & 0 & 0 \\
\hline
$h_2$ & 10 & $BCEO,CDEO$ & $\sigma_2$ &$\{\sigma_2;B;C;E;O\}$ & 0 & 0 & 0 & 0 & 1 & 2 & 2 & 2 \\
\hline
$h_3$ & -8 & $BCDO$ & $\sigma_2$ &$\{\sigma_2;B;C;D;O\}$ & 0 & 0 & 0 & 0 & 1 & 1 & 1 & 0 \\
\hline
$h_4$ & -2 & $BDEO$ & $\sigma_2$ &$\{\sigma_2;B;D;E;O\}$ & 0 & 0 & 0 & 0 & 1 & 1 & 0 & 1 \\
\hline
$h_5$ & -2 & $ABCE,ACDE$ & $\sigma_2$ &$\{\sigma_2;\sigma_1A;B;C;E\}$  & 2 & 0 & 0 & 2 & 1 & 0 & 2 & 2 \\
\hline
$h_6$ & -2 & $ABCO,ACDO$ & $\sigma_{1,2}$ &$\{\sigma_2;\sigma_1A;B;C;O\}$ & 2 & $\frac{1}{2}$ & 1 & 2 & $\frac{1}{2}$ & 1 & 2 & 0 \\
\hline
$h_7$ & -2 & $ABDE$ & $\sigma_{1,2}$ &$\{\sigma_1;A;B;\sigma_1D;\sigma_1E\}$ & 1 & $\frac{1}{2}$ & 0 & 2 & $\frac{1}{2}$ & 0 & 0 & 1 \\
\hline
$h_8$ & 4 & $ABCD$ & $\sigma_{1,2}$ &$\{\sigma_2;\sigma_1A;B;C;D\}$ & 1 & $\frac{1}{2}$ & 0 & 1 & $\frac{1}{2}$ & 0 & 1 & 0 \\
\hline
$h_9$ & 4 & $ABEO,ADEO$ & $\sigma_{1,2}$ &$\{\sigma_2;\sigma_1A;B;E;O\}$ & 2 & $\frac{1}{2}$ & 1 & 2 & $\frac{1}{2}$ & 1 & 0 & 2 \\
\hline
$h_{10}$ & 4 & $ACEO$ & $\sigma_2$ &$\{\sigma_2;\sigma_1A;C;E;O\}$ & 1 & 0 & 0 & 1 & 0 & 1 & 1 & 1 \\
\hline
$h_{11}$ & 4 & $BCDE$ & $\sigma_2$ &$\{\sigma_2;B;C;D;E\}$ & 0 & 0 & 0 & 0 & 1 & 0 & 1 & 1 \\
\hline
$h_{12}$ & 8 & $ABCDEO$ & $\sigma_2$ &$\{\sigma_2;\sigma_1A;B;C;D;E;O\}$ & 1 & 0 & 0 & 1 & 1 & 1 & 1 & 1 \\
\hline
\end{tabular}
\label{tab:ray14}
\end{adjustbox}
\be
\begin{split}
 e_{\sigma_1 A}: \quad &h_1+2h_5+2h_6+h_7+h_8+2h_9+h_{10}+h_{12}=24\\
 e_{\sigma_1 B}: \quad &h_1+\frac{1}{2}h_6+\frac{1}{2}h_7+\frac{1}{2}h_8=12\\
 e_{\sigma_1 O}: \quad &h_1+h_6+h_9=12\\
 e_{\sigma_1\sigma_2}: \quad &2h_5+2h_6+2h_7+h_8+2h_9+h_{10}+h_{12}=12\\
 e_{\sigma_2 B}: \quad &h_2+h_3+h_4+h_5+\frac{1}{2}h_6+\frac{1}{2}h_7+\frac{1}{2}h_8+\frac{1}{2}h_9+h_{11}+h_{12}=12\\
 e_{\sigma_2 O}: \quad &2h_2+h_3+h_4+h_6+h_9+h_{10}+h_{12}=24\\
 e_{\sigma_2 C}: \quad &2h_2+h_3+2h_5+2h_6+h_8+h_{10}+h_{11}+h_{12}=24\\
 e_{\sigma_2 E}: \quad &2h_2+h_4+2h_5+h_7+2h_9+h_{10}+h_{11}+h_{12}=36\\
&h_1 \rightarrow 10, \; h_2 \rightarrow 10, \; h_3 \rightarrow -8, \; h_4 \rightarrow -2 \; h_5 \rightarrow -2, \; h_6 \rightarrow -2, \\ &h_7 \rightarrow -2, \; h_8 \rightarrow 4 \; h_9 \rightarrow 4, \; h_{10} \rightarrow 4, \; h_{11} \rightarrow 4, \; h_{12} \rightarrow 8
\end{split}
\ee

\begin{figure}[H]
\centering
\begin{tabular}{cc}
\centering
\includegraphics[width=.45\textwidth,page=76]{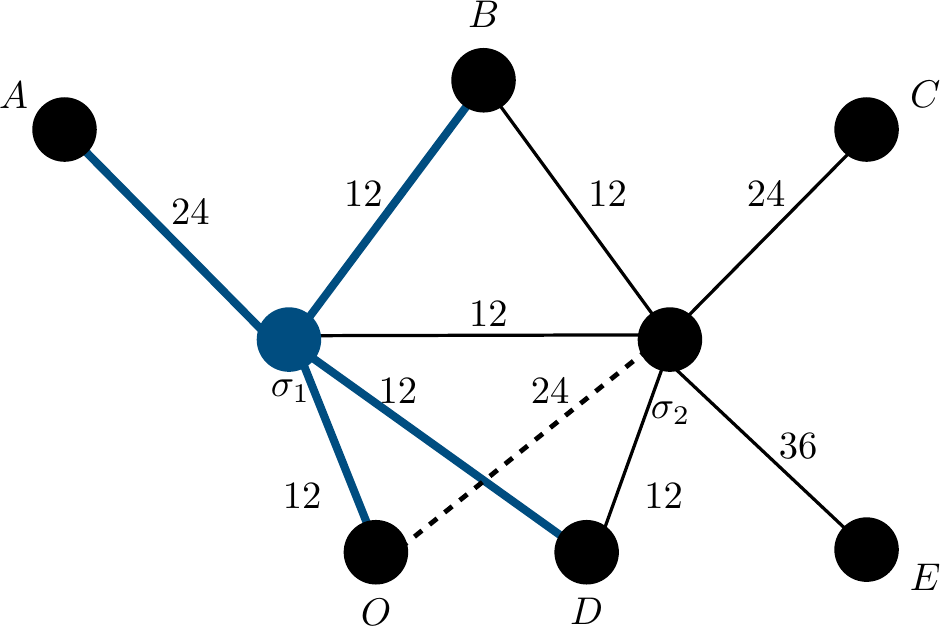}&
\includegraphics[width=.45\textwidth,page=77]{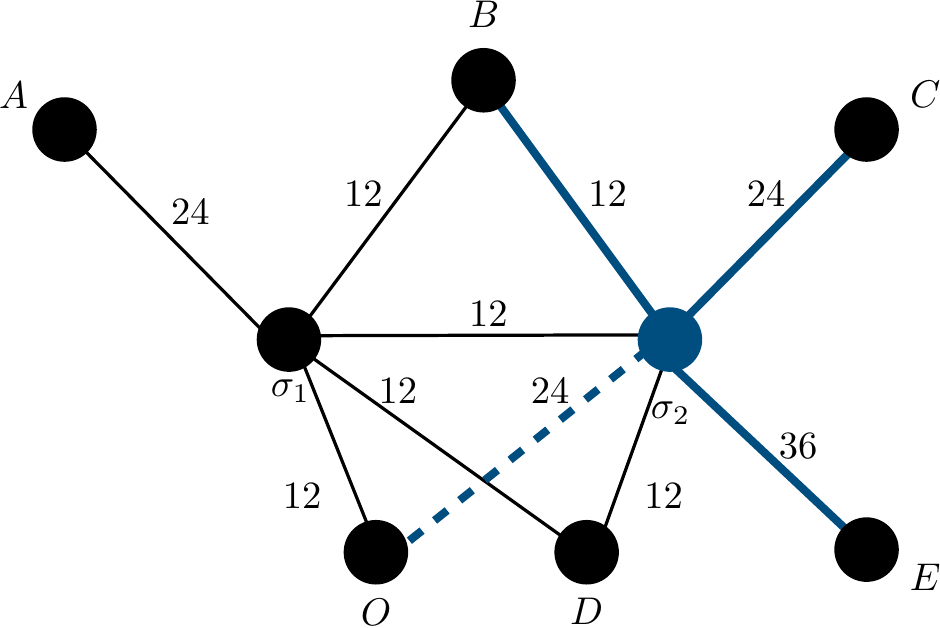}\\
$h_1$ & $h_2$ \\[6pt]
\includegraphics[width=.45\textwidth,page=78]{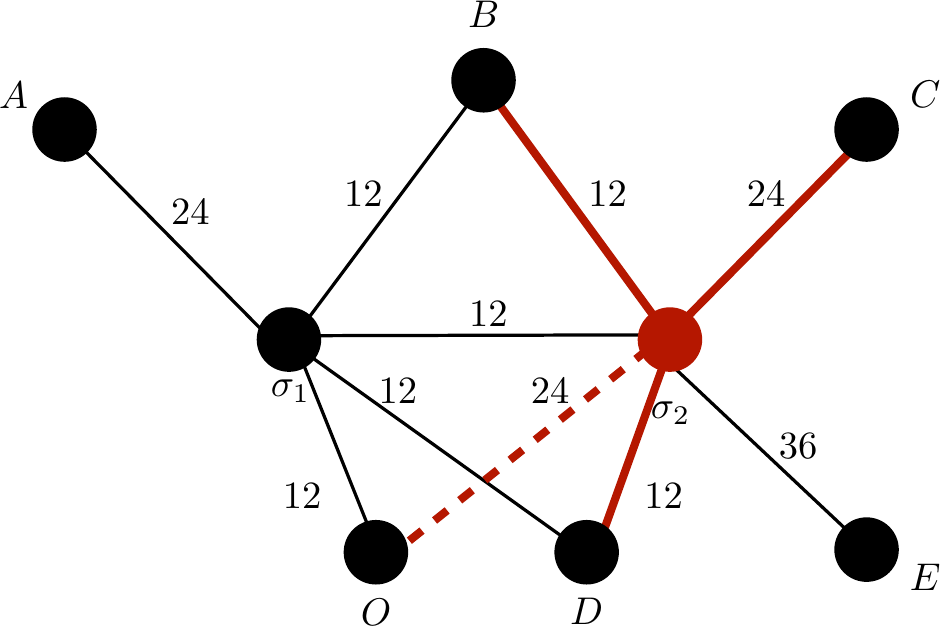}&
\includegraphics[width=.45\textwidth,page=79]{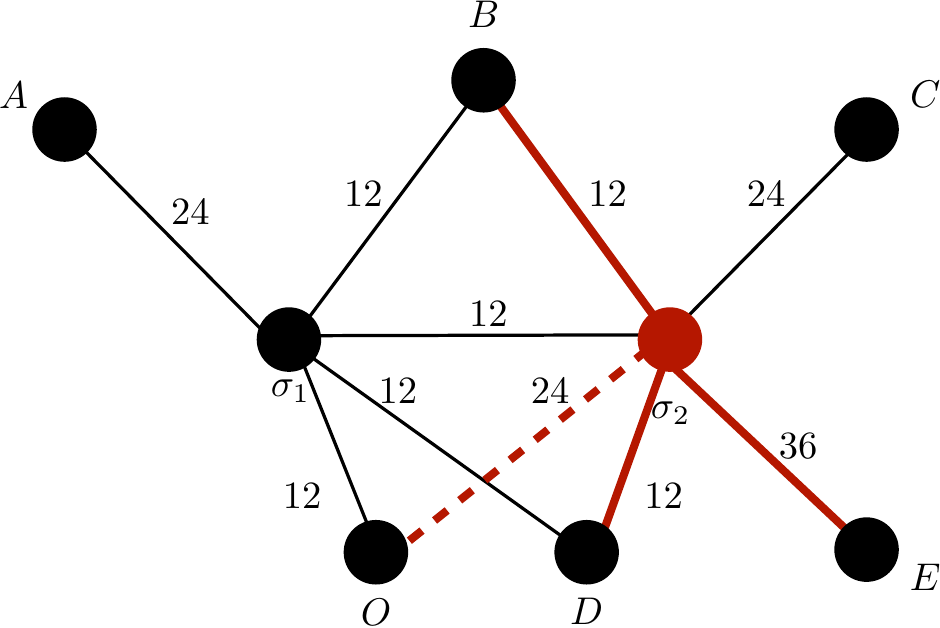}\\
$h_3$ & $h_4$ \\[6pt]
\includegraphics[width=.45\textwidth,page=80]{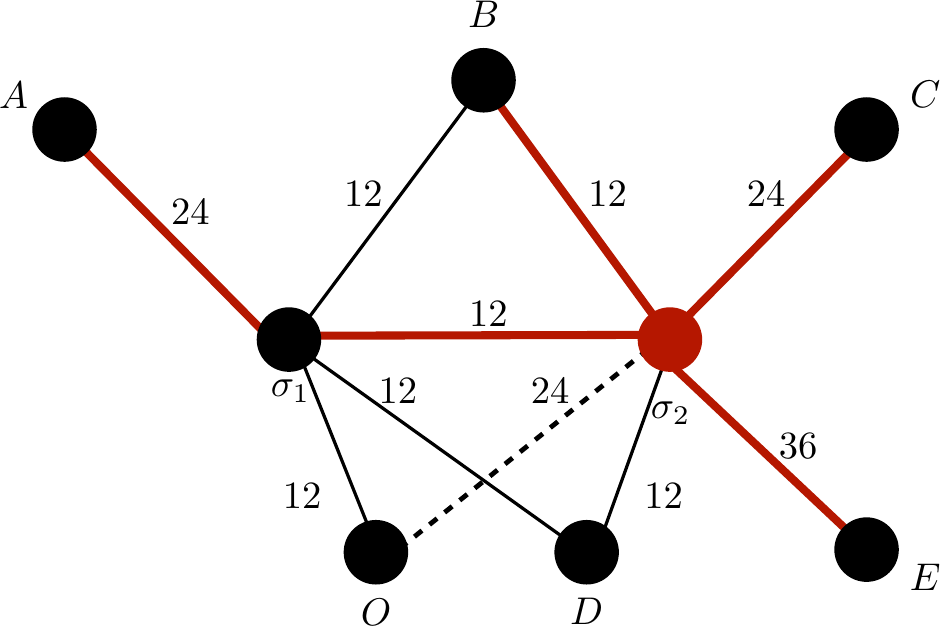}&
\includegraphics[width=.45\textwidth,page=81]{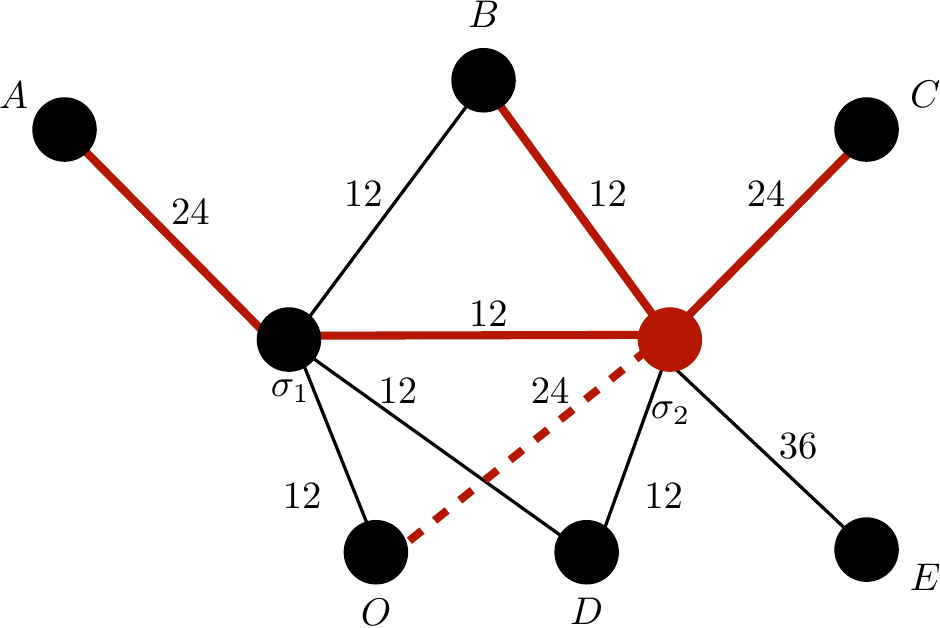}\\
$h_5$ & $h_6$ \\[6pt]
\end{tabular}
\caption{}
\end{figure}

\begin{figure}[H]
\centering
\begin{tabular}{cc}
\centering
\includegraphics[width=.45\textwidth,page=82]{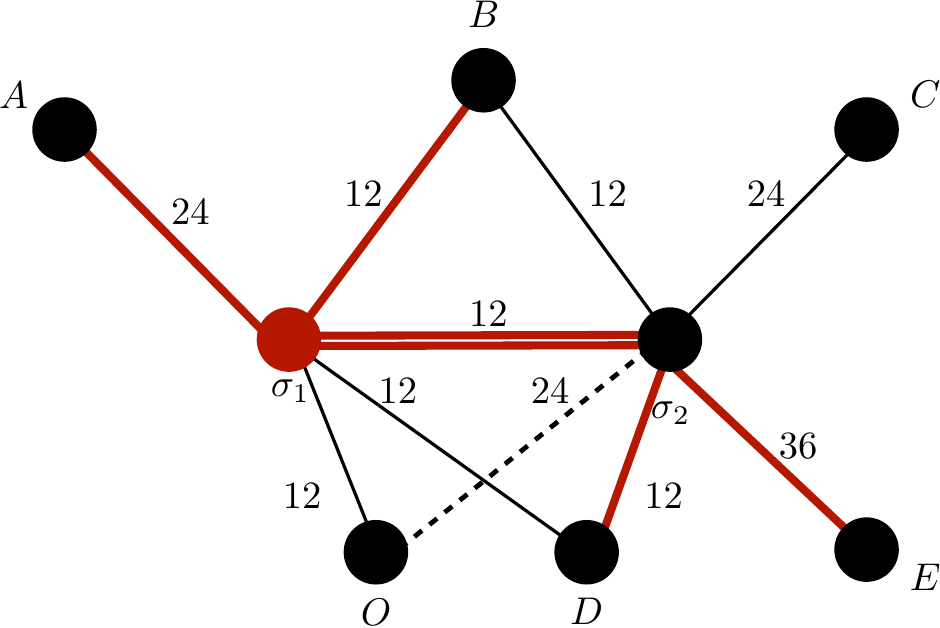}&
\includegraphics[width=.45\textwidth,page=83]{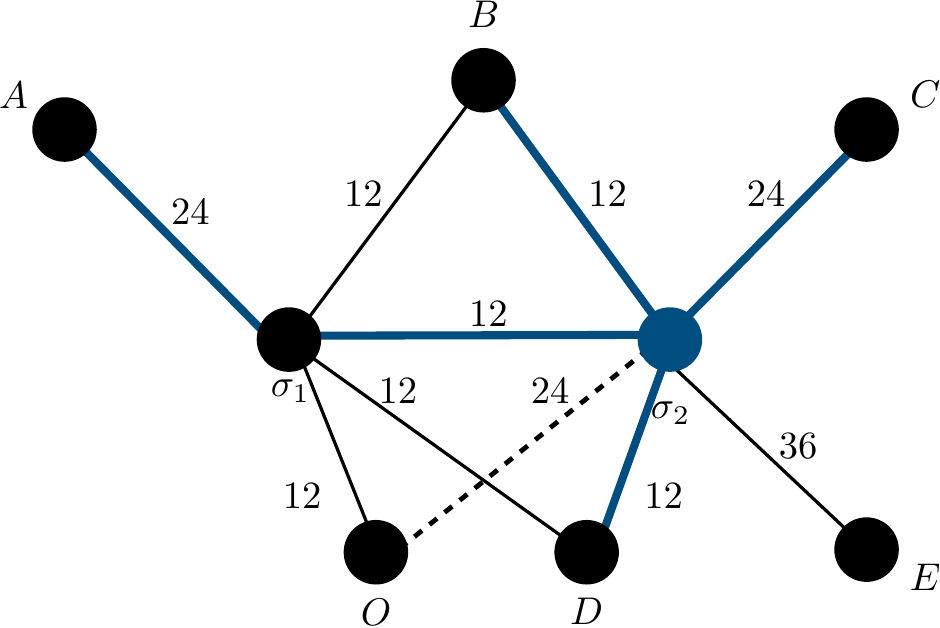}\\
$h_7$ & $h_8$ \\[6pt]
\includegraphics[width=.45\textwidth,page=84]{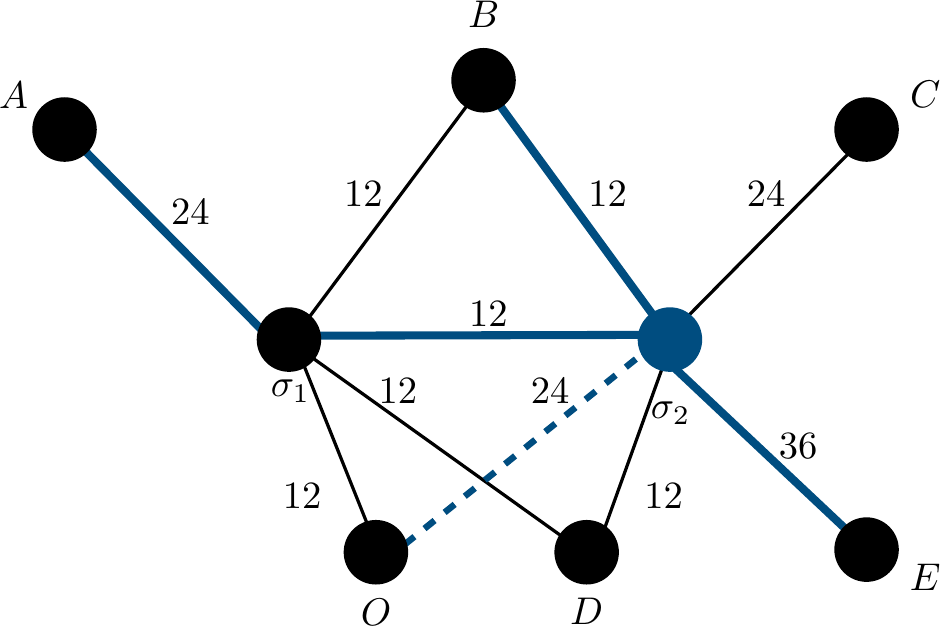}&
\includegraphics[width=.45\textwidth,page=85]{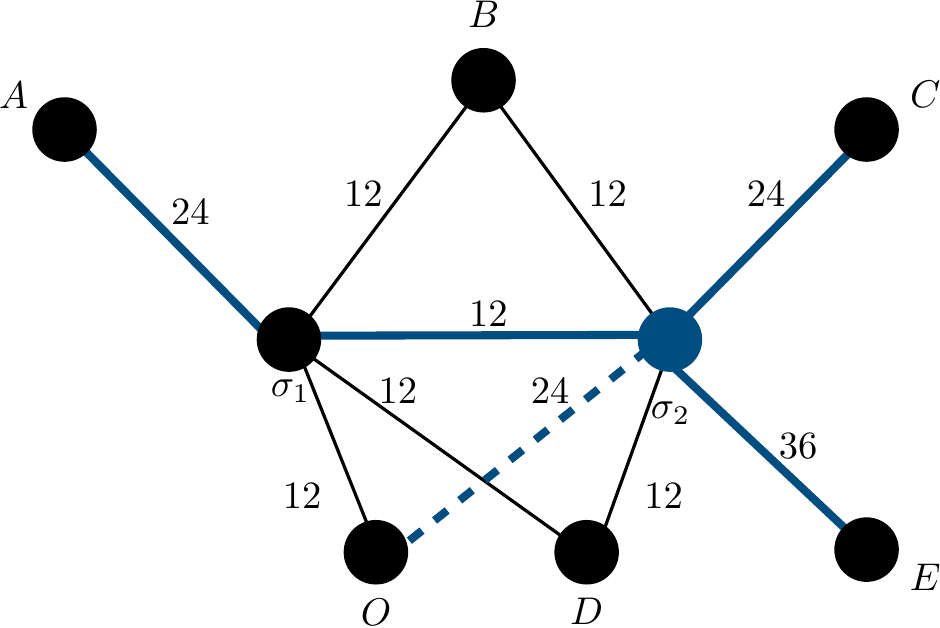}\\
$h_9$ & $h_{10}$ \\[6pt]
\includegraphics[width=.45\textwidth,page=86]{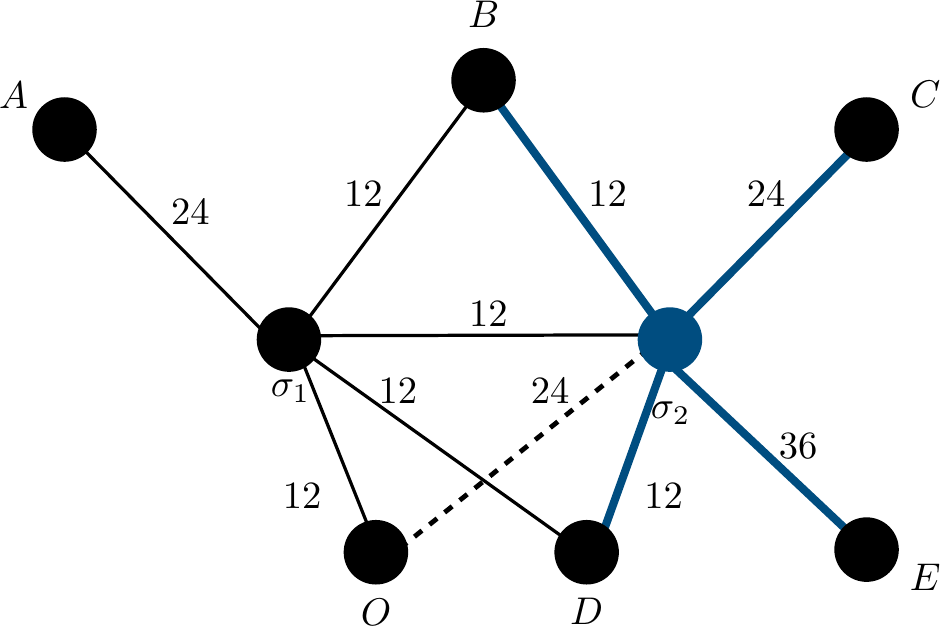}&
\includegraphics[width=.45\textwidth,page=87]{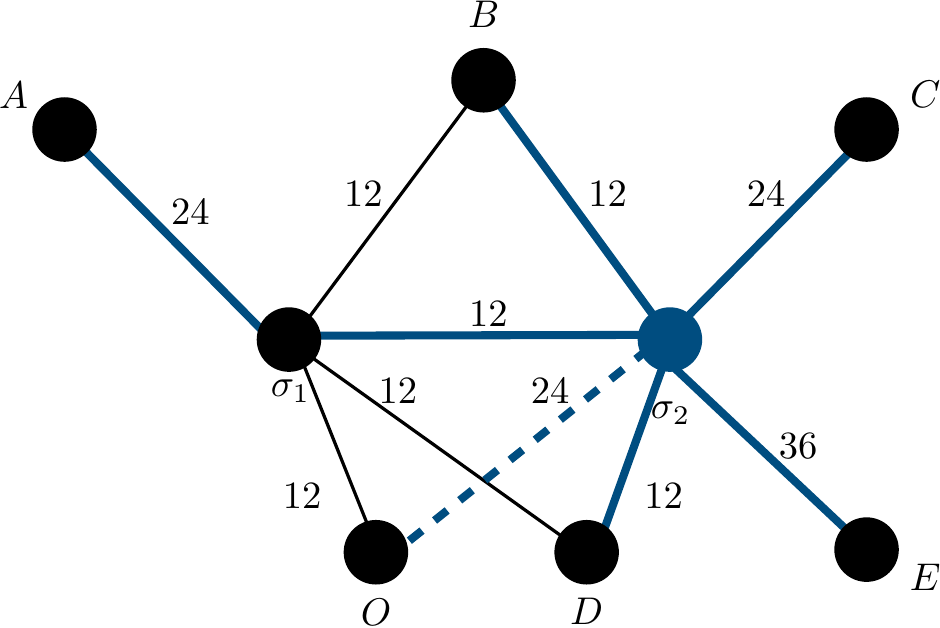}\\
$h_{11}$ & $h_{12}$ \\[6pt]
\end{tabular}
\caption{}
\end{figure}

\newpage
\subsection*{Ray 15}
\be\label{eq:ray15S}
\mathcal{S}=2*\{ 33333; 6666666666; 7759779999; 66666; 3\}
\ee
\be\label{eq:ray15K}
\mathcal{K}=\{000000000000000;0202000000-22002;2\}
\ee
\begin{figure}[H]
\centering
\includegraphics[width=.5\textwidth,page=61]{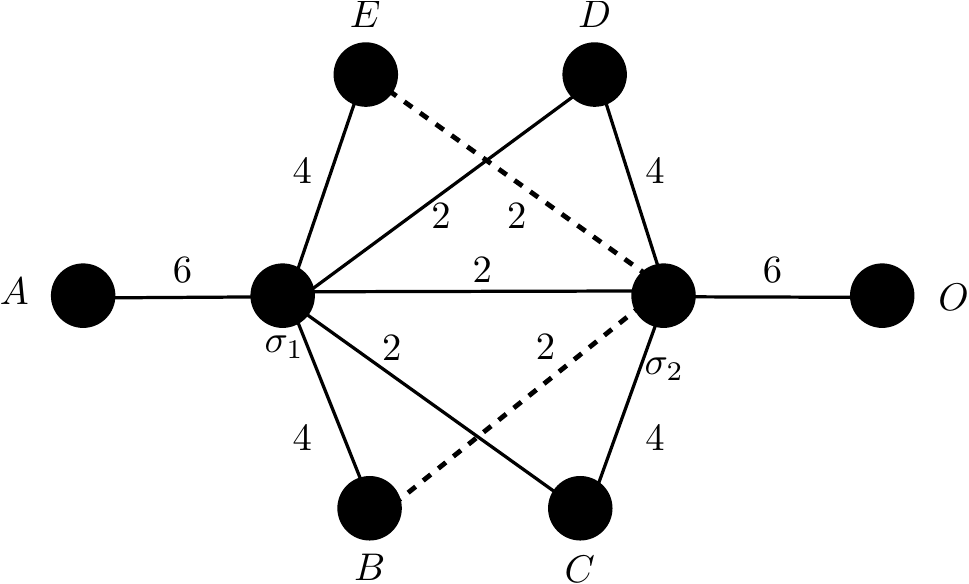}
\caption{\label{fig:ray15}  }
\end{figure}
\begin{table}[H]
\begin{center}
\scriptsize
\begin{tabular}{| c |c | c | c | c | c | c | c | c |  }
\hline
Type & Target & Contributing Species & Vertex  & Example hyperthread & $e_{\sigma_1 A}$ & $e_{\sigma_1 D}$ & $e_{\sigma_1 E}$  & $e_{\sigma_1 \sigma_2}$    \\ 
\hline
\hline
$h_1$ & 2 & $AB(CD)E$ & $\sigma_1$ &$\{\sigma_1;A;B;C;E)\}$ & 2 & 1 & 2 & 0 \\
\hline
$h_2$ & 2 & $(BE)CDO$ &$\sigma_2$ &$\{\sigma_2;B;C;D;O\}$ & 0 & 0 & 0 & 0\\
\hline
$h_3$ &-2 & $BCDE$ &$\sigma_{1,2}$ &$\{\sigma_1;B;C;D;E\}$ & 0 & $\frac{1}{2}$ & $\frac{1}{2}$ & 0\\
\hline
$h_4$ & 2 & $ABCDEO$ &$\sigma_{1,2}$ &$\{\sigma_1;\sigma_1A;B;C;D;E;\sigma_2O\}$ & 1 & $\frac{1}{2}$ & $\frac{1}{2}$ & 1\\
\hline
\end{tabular}
\end{center}
\label{tab:ray15}
\end{table}
\be
\begin{split}
 e_{\sigma_1 A}: \quad &2h_1+h_4=6\\
 e_{\sigma_1 D}: \quad &h_1+\frac{1}{2}h_3+\frac{1}{2}h_4=2\\
 e_{\sigma_1 E}: \quad &2h_1+\frac{1}{2}h_3+\frac{1}{2}h_4=4\\
 e_{\sigma_1\sigma_2 }: \quad &h_4=2\\
 \text{Symmetry:} \quad &h_1=h_2\\
&h_1 \rightarrow 2, \; h_2 \rightarrow 2, \; h_3 \rightarrow -2, \; h_4 \rightarrow 2
\end{split}
\ee
\begin{figure}[H]
\centering
\begin{tabular}{cc}
\centering
\includegraphics[width=.45\textwidth,page=62]{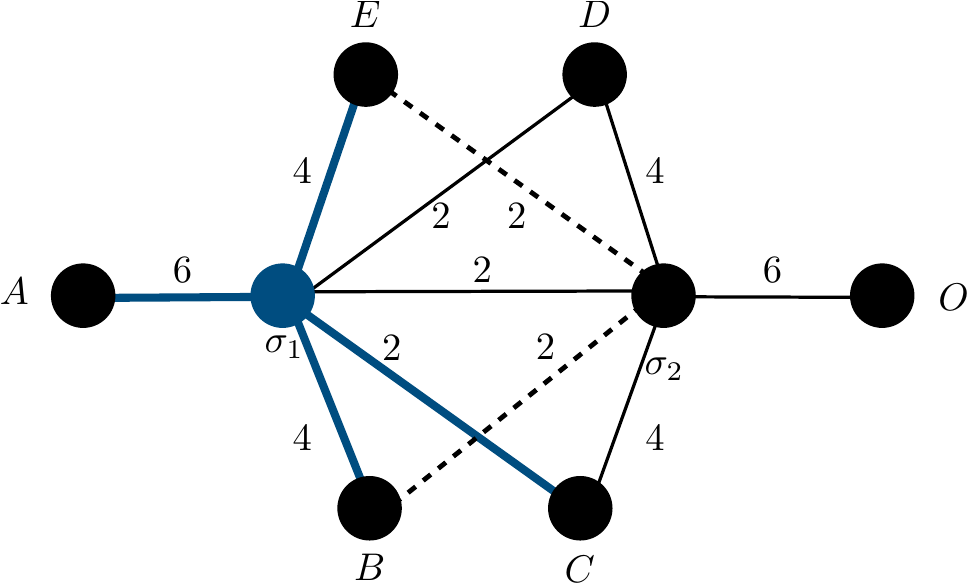}&
\includegraphics[width=.45\textwidth,page=63]{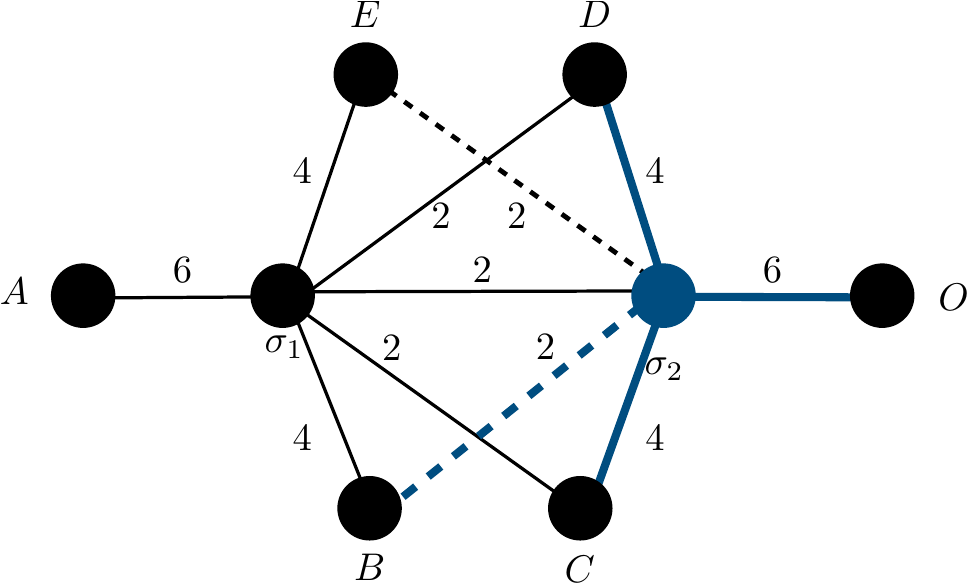}\\
$h_1$ & $h_2$ \\[6pt]
\includegraphics[width=.45\textwidth,page=64]{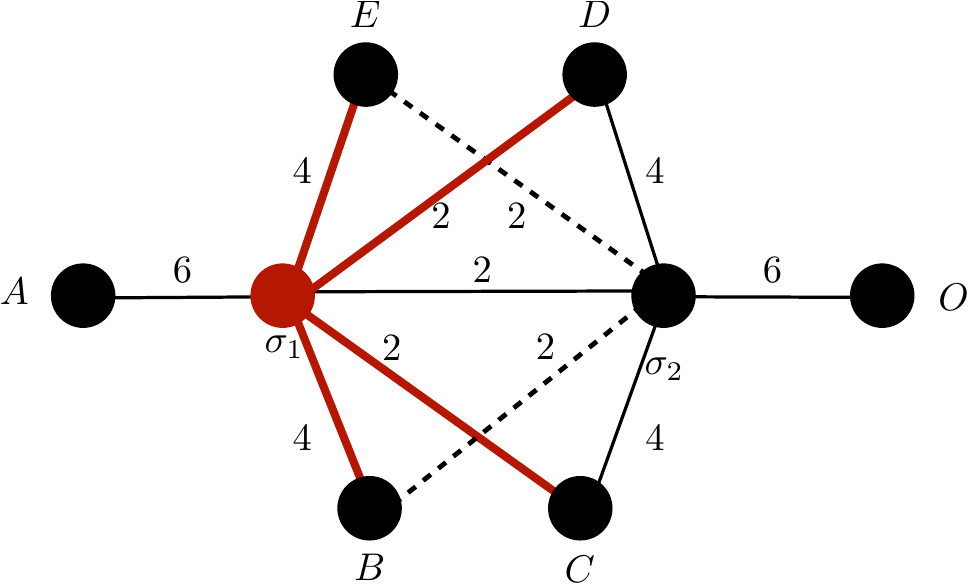}&
\includegraphics[width=.45\textwidth,page=65]{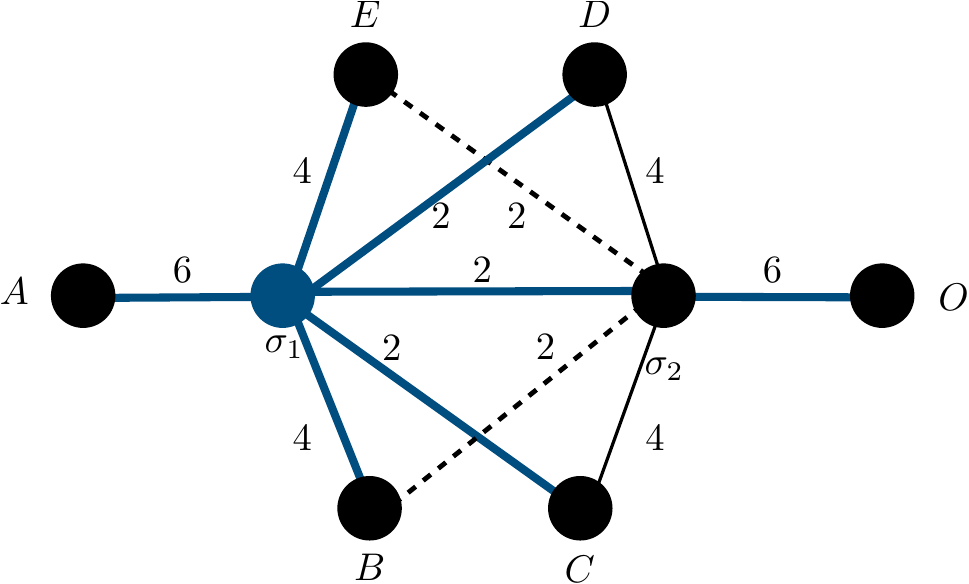}\\
$h_3$ & $h_4$ \\[6pt]
\end{tabular}
\caption{}
\end{figure}

\bibliographystyle{JHEP}
\bibliography{main}

\providecommand{\href}[2]{#2}\begingroup\raggedright\begin{thebibliography}{10}

\bibitem{2006JHEP...08..045R}
S.~{Ryu} and T.~{Takayanagi}, \emph{{Aspects of holographic entanglement
  entropy}},
  \href{http://dx.doi.org/10.1088/1126-6708/2006/08/045}{\emph{Journal of High
  Energy Physics} {\bfseries 2006} (Aug., 2006) 045},
  [\href{https://arxiv.org/abs/hep-th/0605073}{{\ttfamily hep-th/0605073}}].

\bibitem{2017CMaPh.352..407F}
M.~{Freedman} and M.~{Headrick}, \emph{{Bit Threads and Holographic
  Entanglement}},
  \href{http://dx.doi.org/10.1007/s00220-016-2796-3}{\emph{Communications in
  Mathematical Physics} {\bfseries 352} (May, 2017) 407--438},
  [\href{https://arxiv.org/abs/1604.00354}{{\ttfamily 1604.00354}}].

\bibitem{2018CQGra..35j5012H}
M.~{Headrick} and V.~E. {Hubeny}, \emph{{Riemannian and Lorentzian flow-cut
  theorems}}, \href{http://dx.doi.org/10.1088/1361-6382/aab83c}{\emph{Classical
  and Quantum Gravity} {\bfseries 35} (May, 2018) 105012},
  [\href{https://arxiv.org/abs/1710.09516}{{\ttfamily 1710.09516}}].

\bibitem{r2}
N.~{Bao}, S.~{Nezami}, H.~{Ooguri}, B.~{Stoica}, J.~{Sully} and M.~{Walter},
  \emph{{The holographic entropy cone}},
  \href{http://dx.doi.org/10.1007/JHEP09(2015)130}{\emph{Journal of High Energy
  Physics} {\bfseries 2015} (Sept., 2015) 130},
  [\href{https://arxiv.org/abs/1505.07839}{{\ttfamily 1505.07839}}].

\bibitem{2019ForPh..6700011H}
V.~E. {Hubeny}, M.~{Rangamani} and M.~{Rota}, \emph{{The Holographic Entropy
  Arrangement}},
  \href{http://dx.doi.org/10.1002/prop.201900011}{\emph{Fortschritte der
  Physik} {\bfseries 67} (Apr., 2019) 1900011},
  [\href{https://arxiv.org/abs/1812.08133}{{\ttfamily 1812.08133}}].

\bibitem{2018ForPh..6600067H}
V.~E. {Hubeny}, M.~{Rangamani} and M.~{Rota}, \emph{{Holographic Entropy
  Relations}},
  \href{http://dx.doi.org/10.1002/prop.201800067}{\emph{Fortschritte der
  Physik} {\bfseries 66} (Nov., 2018) 1800067},
  [\href{https://arxiv.org/abs/1808.07871}{{\ttfamily 1808.07871}}].

\bibitem{r3}
S.~{Hern{\'a}ndez Cuenca}, \emph{{Holographic entropy cone for five regions}},
  \href{http://dx.doi.org/10.1103/PhysRevD.100.026004}{\emph{Phys. Rev. D}
  {\bfseries 100} (July, 2019) 026004},
  [\href{https://arxiv.org/abs/1903.09148}{{\ttfamily 1903.09148}}].

\bibitem{2007PhRvD..76j6013H}
M.~{Headrick} and T.~{Takayanagi}, \emph{{Holographic proof of the strong
  subadditivity of entanglement entropy}},
  \href{http://dx.doi.org/10.1103/PhysRevD.76.106013}{\emph{Phys. Rev. D}
  {\bfseries 76} (Nov., 2007) 106013},
  [\href{https://arxiv.org/abs/0704.3719}{{\ttfamily 0704.3719}}].

\bibitem{2013PhRvD..87d6003H}
P.~{Hayden}, M.~{Headrick} and A.~{Maloney}, \emph{{Holographic mutual
  information is monogamous}},
  \href{http://dx.doi.org/10.1103/PhysRevD.87.046003}{\emph{Phys. Rev. D}
  {\bfseries 87} (Feb., 2013) 046003},
  [\href{https://arxiv.org/abs/1107.2940}{{\ttfamily 1107.2940}}].

\bibitem{Cui:2018aa}
S.~X. {Cui}, P.~{Hayden}, T.~{He}, M.~{Headrick}, B.~{Stoica} and M.~{Walter},
  \emph{{Bit Threads and Holographic Monogamy}},
  \href{http://dx.doi.org/10.1007/s00220-019-03510-8}{\emph{Communications in
  Mathematical Physics} {\bfseries 376} (July, 2019) 609--648},
  [\href{https://arxiv.org/abs/1808.05234}{{\ttfamily 1808.05234}}].

\bibitem{N6rays}
D.~{Avis} and S.~{Hern{\'a}ndez-Cuenca}, \emph{{The Six-Party Holographic
  Entropy Cone, {Work in progress}}},  2022.

\bibitem{2022PhRvD.105h6008F}
M.~{Fadel} and S.~{Hern{\'a}ndez-Cuenca}, \emph{{Symmetrized holographic
  entropy cone}},
  \href{http://dx.doi.org/10.1103/PhysRevD.105.086008}{\emph{Phys. Rev. D}
  {\bfseries 105} (Apr., 2022) 086008},
  [\href{https://arxiv.org/abs/2112.03862}{{\ttfamily 2112.03862}}].

\bibitem{2021arXiv210207535A}
D.~{Avis} and S.~{Hern{\'a}ndez-Cuenca}, \emph{{On the foundations and extremal
  structure of the holographic entropy cone}}, {\emph{arXiv e-prints} (Feb.,
  2021) arXiv:2102.07535}, [\href{https://arxiv.org/abs/2102.07535}{{\ttfamily
  2102.07535}}].

\bibitem{2020JHEP...07..245H}
T.~{He}, V.~E. {Hubeny} and M.~{Rangamani}, \emph{{Superbalance of holographic
  entropy inequalities}},
  \href{http://dx.doi.org/10.1007/JHEP07(2020)245}{\emph{Journal of High Energy
  Physics} {\bfseries 2020} (July, 2020) 245},
  [\href{https://arxiv.org/abs/2002.04558}{{\ttfamily 2002.04558}}].

\bibitem{2022arXiv220400075H}
S.~{Hern{\'a}ndez-Cuenca}, V.~E. {Hubeny} and M.~{Rota}, \emph{{The holographic
  entropy cone from marginal independence}}, {\emph{arXiv e-prints} (Mar.,
  2022) arXiv:2204.00075}, [\href{https://arxiv.org/abs/2204.00075}{{\ttfamily
  2204.00075}}].

\bibitem{2021arXiv211200763C}
B.~{Czech} and S.~{Shuai}, \emph{{Holographic Cone of Average Entropies}},
  {\emph{arXiv e-prints} (Dec., 2021) arXiv:2112.00763},
  [\href{https://arxiv.org/abs/2112.00763}{{\ttfamily 2112.00763}}].

\bibitem{r1}
T.~{He}, M.~{Headrick} and V.~E. {Hubeny}, \emph{{Holographic entropy relations
  repackaged}}, \href{http://dx.doi.org/10.1007/JHEP10(2019)118}{\emph{Journal
  of High Energy Physics} {\bfseries 2019} (Oct., 2019) 118},
  [\href{https://arxiv.org/abs/1905.06985}{{\ttfamily 1905.06985}}].

\bibitem{boyd2004convex}
S.~Boyd, \emph{Convex optimization}.
\newblock Cambridge University Press, Cambridge, UK New York, 2004.

\bibitem{locking}
M.~{Headrick}, J.~{Held} and J.~{Herman}, \emph{{Crossing versus locking: Bit
  threads and continuum multiflows}}, {\emph{arXiv e-prints} (Aug., 2020)
  arXiv:2008.03197}, [\href{https://arxiv.org/abs/2008.03197}{{\ttfamily
  2008.03197}}].

\bibitem{2021JHEP...09..118H}
J.~{Harper}, \emph{{Hyperthreads in holographic spacetimes}},
  \href{http://dx.doi.org/10.1007/JHEP09(2021)118}{\emph{Journal of High Energy
  Physics} {\bfseries 2021} (Sept., 2021) 118},
  [\href{https://arxiv.org/abs/2107.10276}{{\ttfamily 2107.10276}}].

\bibitem{doi:10.1137/S0895480195287723}
A.~Frank, A.~V. Karzanov and A.~Sebo, \emph{On integer multiflow maximization},
  \href{http://dx.doi.org/10.1137/S0895480195287723}{\emph{SIAM Journal on
  Discrete Mathematics} {\bfseries 10} (1997) 158--170},
  [\href{https://arxiv.org/abs/https://doi.org/10.1137/S0895480195287723}{{\ttfamily
  https://doi.org/10.1137/S0895480195287723}}].

\bibitem{2020JHEP...04..208A}
C.~{Akers} and P.~{Rath}, \emph{{Entanglement wedge cross sections require
  tripartite entanglement}},
  \href{http://dx.doi.org/10.1007/JHEP04(2020)208}{\emph{Journal of High Energy
  Physics} {\bfseries 2020} (Apr., 2020) 208},
  [\href{https://arxiv.org/abs/1911.07852}{{\ttfamily 1911.07852}}].

\bibitem{2021JHEP...10..047H}
P.~{Hayden}, O.~{Parrikar} and J.~{Sorce}, \emph{{The Markov gap for geometric
  reflected entropy}},
  \href{http://dx.doi.org/10.1007/JHEP10(2021)047}{\emph{Journal of High Energy
  Physics} {\bfseries 2021} (Oct., 2021) 47},
  [\href{https://arxiv.org/abs/2107.00009}{{\ttfamily 2107.00009}}].

\bibitem{2018NatPh..14..573U}
K.~{Umemoto} and T.~{Takayanagi}, \emph{{Entanglement of purification through
  holographic duality}},
  \href{http://dx.doi.org/10.1038/s41567-018-0075-2}{\emph{Nature Physics}
  {\bfseries 14} (Mar., 2018) 573--577},
  [\href{https://arxiv.org/abs/1708.09393}{{\ttfamily 1708.09393}}].

\bibitem{2018JHEP...01..098N}
P.~{Nguyen}, T.~{Devakul}, M.~G. {Halbasch}, M.~P. {Zaletel} and B.~{Swingle},
  \emph{{Entanglement of purification: from spin chains to holography}},
  \href{http://dx.doi.org/10.1007/JHEP01(2018)098}{\emph{Journal of High Energy
  Physics} {\bfseries 2018} (Jan., 2018) 98},
  [\href{https://arxiv.org/abs/1709.07424}{{\ttfamily 1709.07424}}].

\bibitem{2019arXiv190500577D}
S.~{Dutta} and T.~{Faulkner}, \emph{{A canonical purification for the
  entanglement wedge cross-section}}, {\emph{arXiv e-prints} (May, 2019)
  arXiv:1905.00577}, [\href{https://arxiv.org/abs/1905.00577}{{\ttfamily
  1905.00577}}].

\end{thebibliography}\endgroup
\end{document}